\documentclass[11pt]{article}
\pdfoutput=1
\usepackage{jheppubme}
\usepackage[table]{xcolor}
\usepackage{graphicx, psfrag, dsfont}
\usepackage{float}
\usepackage{caption}
\usepackage[mathscr]{euscript}
\usepackage{dynkin-diagrams}
\usepackage{appendix}
\usepackage{coolstr}
\usepackage{hyperref}
\usepackage[normalem]{ulem}
\hypersetup{pdfauthor={Name}}
\usepackage{color}
\usepackage{chngcntr}
\counterwithin{table}{subsection}
\usepackage{booktabs,caption}
\usepackage{lscape}
\definecolor{Mygrey}{gray}{0.8}
\definecolor{Mywhite}{gray}{1.0}

\newcommand{\be}{\begin{equation}}
\newcommand{\ee}{\end{equation}}
\newcommand{\bea}{\begin{eqnarray}}
\newcommand{\eea}{\end{eqnarray}}
\usepackage{amsmath,amssymb,epsfig,jheppubme}
\usepackage{multirow,longtable,enumerate,bm}
\usepackage[flushleft]{threeparttable}

\newcommand\half{\frac12}

\newcommand\bi{\begin{itemize}}
\newcommand\ei{\end{itemize}}

\newcommand\tl{{\tilde \ell}}

\newcommand\tc{{\tilde c}}

\newcommand\tilh{{\tilde h}}
\newcommand\tchi{{\tilde \chi}}

\newcommand\btau{{\bar \tau}}

\newcommand\cG{{\cal G}}

\newcommand\cW{{\cal W}}

\newcommand\cN{{\cal N}}

\newcommand\cC{{\cal C}}
\newcommand\tcW{{\widetilde{\cal W}}}
\newcommand\tcC{{\tilde{\cal C}}}

\newcommand{\cH}{{\cal H}}

\newcommand\bchi{{\overline \chi}}

\newcommand\tD{{\tilde D}}

\newcommand\shalf{{\textstyle\frac12}}

\newcommand\ZZ{\hbox{Z\kern-.4emZ}}
\newcommand\sZZ{\hbox{\sevenfont Z\kern-.4emZ}}

\newcommand{\cM}{{\cal M}}

\newcommand{\cE}{{\cal E}}
\newcommand{\cI}{{\cal I}}
\newcommand{\eref}[1]{Eq.\,(\ref{#1})}
\newcommand{\Comment}[1]{{}}

\newcommand{\mfg}{{\mathfrak g}}
\newcommand{\mfh}{{\mathfrak h}}
\newcommand{\tmfh}{{\tilde{\mathfrak h}}}

\definecolor{Mygrey}{gray}{0.8}
\definecolor{Mywhite}{gray}{1.0}

\colorlet{dred}{red!70!black!100!}

\def\IB{\relax{\rm I\kern-.18em B}}

\def\ID{\relax{\rm I\kern-.18em D}}
\def\IE{\relax{\rm I\kern-.18em E}}
\def\IF{\relax{\rm I\kern-.18em F}}
\def\II{\relax{\rm I\kern-.18em I}}

\def\Id{\relax{1\kern-.32em 1}}
\def\IG{\relax\hbox{$\inbar\kern-.3em{\rm G}$}}
\def\IR{\relax{\rm I\kern-.18em R}}

\title{Meromorphic Cosets and the Classification of Three-Character CFT} 
\author[a]{Arpit Das,}
\author[b,c]{Chethan N. Gowdigere} 
\author[d,e]{and Sunil Mukhi\footnote{Adjunct Professor, ICTS-TIFR, Bengaluru.}}

\affiliation[a]{Centre for Particle Theory, Department of Mathematical Sciences,\\
Durham University, South Road, Durham DH1 3LE, United Kingdom}
\affiliation[b]{National Institute of Science Education and Research Bhubaneshwar,\\
P.O. Jatni, Khurdha, 752050, India}
\affiliation[c]{Homi Bhabha National Institute, Training School Complex,\\
Anushakti Nagar, Mumbai 400094, India}
\affiliation[d]{School of Natural Sciences, Institute for Advanced Study,\\
1, Einstein Drive, Princeton NJ 08540, USA}
\affiliation[e]{Indian Institute of Science Education and Research,\\ 
Homi Bhabha Rd, Pune 411008, India}

\emailAdd{arpit.das@durham.ac.uk}
\emailAdd{chethan.gowdigere@niser.ac.in}
\emailAdd{sunil.mukhi@gmail.com}
    
\abstract{We investigate the admissible vector-valued modular forms having three independent characters and vanishing Wronskian index and determine which ones correspond to genuine 2d conformal field theories. This is done by finding bilinear coset-type relations that pair them into meromorphic characters with central charges 8, 16, 24, 32 and 40. Such pairings allow us to identify some characters with definite CFTs and rule out others. As a key result we classify all unitary three-character CFT with vanishing Wronskian index, excluding $c=8,16$. The complete list has two infinite affine series $B_{r,1},D_{r,1}$ and 45 additional theories. As a by-product, at higher values of the total central charge we also find constraints on the existence or otherwise of meromorphic theories. We separately list  several cases that potentially correspond to Intermediate Vertex Operator Algebras.}

\preprint{}


\begin{document}

\maketitle

\section{Introduction}
\label{Introduction}

A rational 2d conformal field theory has a finite set of holomorphic characters $\chi_i(\tau)$ and a partition function of the form:
\be\label{eq1}
Z(\tau,\btau)=\sum_{i,j=0}^{n-1}M_{ij}\,\bchi_i(\btau)\chi_j(\tau) = |\chi_0|^2 + \sum_{i=1}^{n-1}Y_i|\chi_i|^2
\ee
Here the integer $n$ is the number of linearly independent characters, which is less than or equal to the number of independent primaries, which we denote by $p$ and refer to as the ``rank''. It is possible for multiple primaries to have the same character \footnote{This occurs in particular whenever a primary is complex, since its complex conjugate has the same character, but there are also more general cases of this phenomenon.}. The positive integers $Y_i$ in Eq.(\ref{eq1}) are the multiplicities of the characters, and the number of primaries is given in terms of these by $p=1+\sum_{i=1}^{n-1}Y_i$.  When $n=1$, the only character is the identity character, and since the vacuum state is unique and real we also have $p=1$. In this case we will refer to the resulting theory as a meromorphic CFT \footnote{Some authors restrict the word ``meromorphic'' to those CFT whose character is exactly modular invariant without a phase, and hence $c$ is a multiple of 24. However in this work we will use the term for all one-character CFT, whose central charges can be any positive integral multiple of 8.}.

A classification programme initiated in \cite{Mathur:1988rx,Mathur:1988na,Mathur:1988gt} and pursued by both mathematicians and physicists in more recent times \cite{Naculich:1988xv,Kiritsis:1988kq,Bantay:2005vk,Mason:2007,Mason:2008,Bantay:2007zz,Tuite:2008pt,Bantay:2010uy,Marks:2011,Gannon:2013jua,Kawasetsu:2014,Hampapura:2015cea,Franc:2016,Gaberdiel:2016zke,Hampapura:2016mmz,Arike:2016ana,Tener:2016lcn,Mason:2018,Harvey:2018rdc,Chandra:2018pjq,Chandra:2018ezv,Bae:2018qfh,Bae:2018qym,Mason:c8c16,franc2020classification,Bae:2020xzl,Mukhi:2020gnj,Kaidi:2020ecu,Das:2020wsi,Kaidi:2021ent,Das:2021uvd,Bae:2021mej,Duan:2022ltz}, is based on the fact that characters are vector-valued modular forms (VVMF) of weight 0:
\be
\chi_i(\gamma\tau)=\sum_{j=0}^{n=1}\varrho_{ij}(\gamma)\chi_j(\tau)
\ee
where:
\be
\gamma=\begin{pmatrix}
a & b\\
c & d
\end{pmatrix}
\in \hbox{SL(2,$\mathbb{Z}$)},\quad
\gamma\tau\equiv \frac{a\tau+b}{c\tau+d}, \qquad \tau\in\mathbb{H}
\ee
and $\mathbb{H}=\{\tau\in\mathbb{C} \, \, | \, \, \text{Im}(\tau)>0\}$ is the upper half plane.

For the partition function in \eref{eq1} to be modular invariant, we must have:
\be
\varrho^\dagger{\rm diag}(1,Y_i)\varrho={\rm diag}(1,Y_i)
\label{rhoY}
\ee

Characters that transform in this way under the modular transformations can be shown to solve modular linear differential equations (MLDE) \cite{Mathur:1988na, Mathur:1988gt}. Such equations have finitely many parameters and these can be varied to scan for solutions that satisfy the basic criteria to be those of a conformal field theory. These criteria correspond to the fact that each character is holomorphic in $q=e^{2\pi i\tau}$ (except as $q\to 0$), and have an expansion of the form:
\be
\chi_i(\tau)=q^{\alpha_i}\sum\limits_{s\geq 0}^{\infty}a_{i,s} \, q^{s}, \qquad s\in\mathbb{Z}
\label{charexp}
\ee
If the VVMF correspond to a genuine CFT then these critical exponents, $\alpha_i$s, can be identified with the central charge and (chiral) conformal dimensions as:
\be
\alpha_i=-\frac{c}{24}+h_i
\label{critexp}
\ee
with $h_0=0\to \alpha_0=-\frac{c}{24}$ corresponding to the identity character of the CFT.

The coefficients $a_{i,s},s\ge 1$ should be non-negative integers for some choice of positive integer $a_{i,0}$ that provides the overall normalisation of each character. To satisfy the axioms of CFT we must choose $a_{0,0}=1$ (non-degeneracy of the vacuum), while for each $i\ne 0$ we define the integer $D_i=a_{i,0}$. Since the MLDE from which characters are obtained is homogeneous, the degeneracies are not uniquely determined without some additional input. One tentatively chooses the minimum integral $D_i$ that make the coefficients $a_{i,s}$, $s\ge 1$, of each character, non-negative integers and then checks for consistency. We discuss this point in some detail in Section \ref{background}.

In \cite{Chandra:2018pjq} character sets with the above properties were called ``admissible''. For any admissible character $\chi_i$, we define $m_1=a_{0,1}$, the degeneracy of the first excited state in the identity character. For a CFT, this corresponds to the number of spin-1 generators in the chiral algebra.  The integers $m_1,D_i,Y_i$ will be important in what follows.

In general, admissible characters do not correspond to a CFT, as we discuss in  detail below. While much of the literature cited above has focused on classifying admissible characters, from the CFT point of view the result should be interpreted as a ``superset'' of candidates of which actual CFTs form a subset. The problem of identifying this subset has been addressed in varying degrees of detail, for small numbers of characters, in \cite{Mathur:1988gt, Schellekens:1992db, Gaberdiel:2016zke, Hampapura:2016mmz, Chandra:2018ezv, Tener:2016lcn, connor2018classification, Mason:c8c16, franc2020classification, Bae:2021mej, Duan:2022ltz, Mukhi:2022bte}. In the present work we take this goal forward by completing the classification of three-character CFT with vanishing Wronskian index (explained below) for any central charge, but excluding central charge $=8,16$ where the classification of admissible characters is itself problematic \cite{Mason:2018, Kaidi:2021ent, Das:2021uvd, Bae:2021mej}. The significance of our result is that we decisively rule in, or out, every admissible character as being a CFT by making an exhaustive list of bilinear pairings. In a  different context, some recent work where the distinction between consistent partition functions and actual CFTs is highlighted is \cite{Dymarsky:2020qom, Dymarsky:2022kwb}.

An important quantity in the classification procedure is the number of zeroes of the Wronskian determinant of the characters in moduli space. Because the torus moduli space has cusps, the number of zeroes can be fractional in units of $\frac16$. Hence we define the Wronskian index $\ell$ \cite{Mathur:1988na} to be an integer such that the number of zeroes is $\frac{\ell}{6}$. Certain values of $\ell$ can be ruled out -- we have $\ell\ne 1$ in general, $\ell$ even for $n=2$ \cite{Naculich:1988xv}, $\ell$ a multiple of 3 for $n=3$, and again $\ell$ even for $n=4$ \cite{Kaidi:2021ent}. 

Our focus in this work is on admissible characters with $(n,\ell)
=(3,0)$. Progress in classifying these was made in \cite{Mathur:1988gt,Tuite:2008pt,franc2020classification, Gaberdiel:2016zke} and more recently in three independent works: \cite{Kaidi:2021ent, Das:2021uvd, Bae:2021mej} which all found a set of seven new solutions that had previously been missed. Of these, the work of Kaidi, Lin and Parra-Martinez \cite{Kaidi:2021ent} was able to complete the classification of admissible characters using a method based on \cite{Mathur:1989pk}. In view of their proof, the classification in \cite{Das:2021uvd} (originally restricted to $c \le 96$) is likewise complete. In the rest of this work we will closely follow the notation of this paper. There is one caveat to the above statements: there are infinitely many admissible $(3,0)$ characters at $c=8,16$ \cite{Mason:c8c16, Kaidi:2021ent, Das:2021uvd, Bae:2021mej} that are harder to classify and would need to be considered separately.

In the present work we start with the complete set of admissible characters (excluding those with $c=8$ and 16) and make use of the coset construction \cite{Goddard:1984vk,frenkel1992vertex,Gaberdiel:2016zke} to complete the classification of $(n,\ell)=(3,0)$ CFT. The cosets we consider are in the spirit of \cite{Gaberdiel:2016zke} where the numerator is a meromorphic CFT with $c=8N$ with $N\in\mathbb{N}$. However we go far beyond this work by exhaustively tabulating all possible bilinear pairings with a total central charge of $c=8,16,24,32,40$. Notably, even at $c=24$ we find interesting classes of pairings that were not considered in \cite{Gaberdiel:2016zke}.

A significant spinoff of our coset pairings is that we can use them to predict several non-lattice meromorphic CFT at arbitrary high values of $c=8N$. The results have been presented in \cite{Das:2022slz} and here they are placed in a larger context. Moreover we will also rule {\em out} certain classes of meromorphic theories at $c=32,40$. 

Returning to the three-character case, the restriction on Wronskian index makes this in one sense a weaker classification than that of \cite{Mukhi:2022bte} for two primaries, where there was no restriction on the Wronskian index, but in one sense also stronger since the present work has no restriction on the central charge. This should finally bring closure to a programme for the ``simplest'' three-character theories (those with vanishing Wronskian index) that was initiated over three decades ago in \cite{Mathur:1988gt}. By contrast, the analogous problem for two characters and vanishing Wronskian index was simple enough to solve in a single paper \cite{Mathur:1988na} with completeness being rigorously proved more recently \cite{Mason:2018}.

Apart from the fact that we restrict the Wronskian index but not the central charge, the approach in the present work has some other important differences from \cite{Mukhi:2022bte}. Here we start from a given {\em finite} set of admissible characters, then look for bilinear coset-type relations for them based on their $q$-expansion. Thereafter we use embedding techniques to identify one of these as a CFT if the other one is known, We also allow any number of primaries as long as the number of characters (dimension of the VVMF) is three, while the rank (number of primaries or ``simple modules'') can be larger. We do not impose unitarity, but always work with the unitary presentation of the characters (the most singular term is treated as defining the central charge). 

In Section \ref{background} we start by describing the methodology used and provide a list of VVMFs that potentially describe three-character CFTs but were so far uncharacterised. Thereafter we summarise some relevant facts about embeddings, extensions of chiral algebras and bilinear or ``gluing'' relations. We also review a class of admissible characters  that have formally negative fusion rules (as computed from the Verlinde formula \cite{Verlinde:1988sn}, after extending if necessary the modular $S$-matrix to have the same rank as the number of primaries). Some of these have been identified as ``Intermediate Vertex Operator Algebras'' \cite{Kawasetsu:2014}. Section \ref{cosetpairs} is devoted to the detailed presentation of our results, with tables detailing the coset pairs at the level of VVMFs and descriptions of the tables that explain how individual entries are either identified with definite CFTs or ruled out. We summarise our results and discuss significant general features of our classification in Section \ref{discussion}. At the very end we abstract a complete table of unitary CFTs with three characters and zero Wronskian index (excluding $c=8,16$ as  mentioned above). The reader who is only interested in the results may skip directly  to Section \ref{discussion}.

While this work was in progress we came to know of \cite{Duan:2022ltz} which has significant overlap with Table \ref{T7} of our paper which positively identifies 6 of the 41 previously uncharacterised solutions. However, in the present work we are also able to unambiguously categorise all the remaining 35 solutions, separating them into 20 that are of IVOA type and 15 that we can rule out as CFTs, completing the classification process. This process makes use of most of the remaining 20 tables in subsections 3.1 -- 3.5. Also, as mentioned above, we find both positive and negative predictions for classes of meromorphic theories at $c>24$.

\section{Methodology and background}

\label{background}

\subsection{MLDE and coset construction}

As explained in the Introduction, the starting point of the classification procedure in which we are working is the construction of admissible characters using MLDEs. Here we explain some important subtleties in this construction and then go on to discuss the coset construction which we employ in the present work to characterise which admissible characters correspond to CFTs.

Below \eref{critexp} we defined the degeneracy $D_i$ of each non-identity character $\chi_i$ as the minimum integer such that the $q$-series for the corresponding character has non-negative integral coefficients. This assigns a tentative normalisation to each non-identity character. As explained in \cite{Mathur:1988gt}, the test of having found correct degeneracies $D_i$ is that the $S$-matrix in a basis of primaries is unitary. Note also that for an affine theory (WZW model), the degeneracy $D_i$ for a given $\chi_i$ is the dimension of the representation of the finite-dimensional Lie algebra in which the $i^{\text{th}}$ primary transforms, so in this case it is uniquely determined. 

In view of these observations, at some stage it may be needed to change the degeneracy of a primary from the initially determined one to a multiple of itself. However the possibility of such a change is subject to a constraint. 
Suppose we have a solution to a given MLDE where the degeneracies $D_i$ as well as the multiplicities $Y_i$ have been tentatively determined (the $Y_i$s can be computed for MLDE solutions using the procedure given in \cite{Mathur:1988gt}). If we redefine the $D_i$ by multiplying by an integer factor, the $Y_i$ will also change in such a way that the product $Y_iD_i^2$ remains fixed. This can be seen by writing the partition function as:
\begin{align}
    Z(\tau,\btau) = |\chi_0|^2 + \sum\limits_{i=1}^{n-1} Y_i D_i^2\left|\left(1 + \frac{a_{i,1}}{a_{i,0}} \, q + \frac{a_{i,2}}{a_{i,0}} \, q^2 + \ldots\right)\right|^2, \label{exp1}
\end{align}
where everything except $Y_iD_i^2$ is uniquely determined by the MLDE. Then modular invariance uniquely determines the $Y_iD_i^2$ for each $i$.
Thus the change $D_i\to \delta_i D_i$ leads to the scaling $Y_i\to \frac{Y_i}{\delta_i^2}$. The new $Y_i$ will be integer only if the old one was divisible by $\delta_i^2$. This is a stringent constraint -- for any given pair $Y_i,D_i$, rescaling of $D_i$ is only allowed if the original $Y_i$ are divisible by the square of an integer. This point is illustrated in considerable detail in the discussion of Table \ref{T7}.

In fact there are MLDE solutions for which both $D_i$ and $Y_i$ cannot simultaneously be made integral. These cannot be CFTs and are marked with a ``strikethrough'' in Table \ref{T0} (thus they appear as \sout{${\bf III}$} or \sout{${\bf V}$}). We note that none of these solutions appears in \cite{Kaidi:2021ent}, who presumably eliminated them at the outset for the above reasons, however some of them do appear in \cite{Bae:2021mej}. Interestingly even these VVMFs satisfy bilinear relations, and for completeness we display these in our subsequent tables where they continue to be marked with a ``strikethrough''. Though they are inconsistent as CFTs, it is still striking that they satisfy bilinear pairings at all, and this might prove useful for the general understanding of VVMFs.

Next we describe one of our main tools, the coset construction \cite{Goddard:1984vk, Goddard:1986ee, frenkel1992vertex}. This is a general class of relations among CFTs, and we will only use the class of cosets where the numerator factor of the coset is a meromorphic CFT, as we explain below \footnote{This is the form studied in the physics literature in \cite{Gaberdiel:2016zke, Hampapura:2016mmz, Duan:2022ltz, Mukhi:2022bte} and in the mathematics literature in, for example, \cite{Dong:1994, Hoehn:Baby8, Hoehn:thesis}.}. Pick a set of admissible characters $\chi_i, i=0,1,\ldots n-1$ and collectively denote it by $\cW$. Suppose this set has Wronskian index $\ell$, central charge $c$ and conformal dimensions $h_i,i=1,2,\cdots n-1$. $\cW$ will be said to have a ``bilinear relation'' with another set of admissible characters $\tchi_i$, collectively denoted $\tcW$, with $i$ running over the same range and having Wronskian index $\tl$, central charge $\tc$ and conformal dimensions $\tilh_i$ if the following holomorphic identity holds:
\be
\chi_0(\tau)\tchi_0(\tau)+\sum_{i=1}^{n-1} d_i \, \chi_i(\tau)\tchi_i(\tau) = \chi^{\mathcal{H}}(\tau)
\label{bilinrel}
\ee
where $\chi^{\mathcal{H}}(\tau)$ is a polynomial in the Klein $j$-invariant times possible factors of  $j(q)^{\frac{1}{3}}$ or $j(q)^{\frac{2}{3}}$, such that the result has non-negative integral coefficients in a power series in $q\equiv e^{2\pi i \tau}$. Such a relation can only hold if $\tchi_i(\tau)$ transforms the same way as the complex conjugate $\bchi_i(\btau)$ under modular transformations. 
Then the $d_i$ are positive integers satisfying:
\be
\varrho^\dagger{\rm diag}(1,d_i)\varrho=
{\rm diag}(1,d_i)
\label{rhod}
\ee
where $\rho$ is the representation under which the $\chi_i$ transform.

From its properties, $\chi^{\mathcal{H}}(\tau)$ is also an admissible character. It may potentially correspond to a meromorphic CFT of central charge $c+\tc$, but it is not necessary that such a CFT exists. For example at $c+\tc=24$ we have an infinite family of admissible characters but only a finite number correspond to CFT's \cite{Schellekens:1992db}. Bilinear pairings are also known to hold for quasi-characters \cite{Chandra:2018pjq,Mukhi:2020gnj} which are integral but not admissible due to negativity of some coefficients.

Comparing \eref{rhod} with \eref{rhoY} we see that we must have $d_i=Y_i$. Physically this is because on the one hand the modular transformations of
$\cW,\overline{\cW}$ are conjugate to each other (where $\overline{\cW}$ is the complex conjugate VVMF to $\cW$ with characters $\bchi_i(\btau)$), so that the partition function is invariant. On the other hand the modular transformations of $\cW,\tcW$ are also mutually conjugate, so that the bilinear relation is modular invariant -- A slight subtlety here is that the bilinear relation can acquire a phase under modular transformations if $c^\cH =24n+8,24n+16$ with $n$ a non-negative integer. However this phase can be absorbed into the transformations of $\tchi$ and it is still true that $d_i=Y_i$. 

Note that if the degeneracies of one of the members of the pair ($D_i$ or $\tD_i$) are not the correct ones then we may not find $d_i=Y_i$. This will be a useful diagnostic in what follows. However there is another condition under which it is possible to have $(d_1,d_2)\ne(Y_1,Y_2)$, that arises when the dual pair is made up of affine theories of the type $D_{4n,1}$. In such cases the representation of SL(2,$\mathbb{Z}$) on the characters is reducible (see v1 of \cite{franc2020classification}) and as a result there are multiple ways to combine the characters into a modular invariant. This will be explained in more detail in Section \ref{s3}.

The bilinear relation \eref{bilinrel} 
does not imply that any of $\chi(\tau),\tchi(\tau),\chi^{\mathcal{H}}(\tau)$ correspond to a genuine CFT. 
However, if $\chi,\tchi$ and $\cH$ are all CFTs, denoted $\cC,\tcC$ and $\cH$ respectively, 
then the bilinear relation is equivalent to the coset relation:
\be
\tilde {\cal C}=\frac{\cH}{\cal C} 
\label{chc}
\ee 
This means that the chiral algebra of $\tcC$ is the commutant of the embedding of the chiral algebra of $\cC$ in that of $\cH$. The representations of the commutant algebra also follow from this embedding, hence the coset completely defines a CFT. 

If both $\cC$ and $\cH$ correspond to CFT's whose stress tensor is given by the Sugawara construction in terms of Kac-Moody currents, then by embedding the currents of $\cal C$ in those of $\cH$ one defines the stress tensor of the coset theory $\tilde {\cal C}$. This will provide a relatively easy way to prove the existence of a coset relation \cite{Gaberdiel:2016zke}. However it is also possible for \eref{bilinrel} to be satisfied when $\cH$ does not 
have any Kac-Moody currents (an example is the Monster Module \cite{Frenkel:1988xz,Borcherds:1}). In this case the coset construction of \cite{Goddard:1984vk,Goddard:1986ee} does not strictly apply, but the more general one of \cite{frenkel1992vertex} does. In these cases it is easier to verify the bilinear relation rather than compute the commutant of $\cC$ in $\cH$.  One such example, studied in the context of MLDE and holomorphic bilinear relations \cite{Hampapura:2016mmz} arises when $\cC$ is the Ising model and $\tcC$ is the Baby Monster CFT \cite{Hoehn:Baby8}.

The existence of bilinear relations between an admissible solution $\cW$, another admissible solution $\tcW$ and an invariant (up to a phase) character $\chi^\cH$ provides us a number of ways to decide whether given admissible characters do or do not correspond to CFT. These are as follows:

\begin{itemize}

\item
When $\cW$ and $\chi^\cH$ are both known CFTs $\cC,\cH$, the bilinear relation suggests that $\tcW$ may correspond to one or more CFTs $\tcC$. This can then be accurately confirmed by checking for the existence of one or more suitable embeddings of $\cC$ in $\cH$ that would define $\tcC$.

\item When $\cW,\tcW$ are both known CFTs $\cC,\tcC$,  we may conclude that the character $\chi^\cH$ corresponds to a CFT $\cH$ that can be called the ``gluing'' of $\cC,\tcC$ \footnote{Rigorously this is true for CFTs with up to 4 primaries, for which the Modular Tensor Category is unique given the modular transformations of the characters \cite{rowell2009classification}.}.
Several new meromorphic CFT were recently discovered in this way in \cite{Das:2022slz} \footnote{However, again for cases involving $D_{4n,1}$, there can be ways to pair $\cC,\tcC$ that do not lead to a meromorphic theory because the coefficients $d_i$ in the pairing are not integral. We will remark on these as they are encountered.}. 

\item When a bilinear relation exists and $\cW$ is a CFT $\cC$, but the character $\chi^\cH$ is known {\em not} to correspond to a CFT, the bilinear partner $\tcW$ cannot be a CFT. For if it were, then the bilinear relation would predict that $\chi^\cH$ is a CFT, resulting in a contradiction. 

\item When a bilinear relation holds and $\cW$ corresponds to a CFT $\cC$ but $\tcW$ is known  {\em not} to correspond to any CFT, it can sometimes be argued that $\chi^\cH$ does not describe a CFT. The naive reasoning is that if $\chi^\cH$ were a CFT $\cH$, then by taking the coset $\cH/\cC$ we would define a CFT $\tcC$ corresponding to the admissible character $\tcW$,  resulting in a contradiction. However a certain condition needs to be satisfied in this case, so we will explain the statement more precisely in the discussion on Table \ref{T17} where it is implemented for the first time. 

\end{itemize}

We see that the bilinear relation is a powerful diagnostic tool for relating admissible characters to CFT or ruling them out as being CFT. 

Let us note here that the recent work \cite{Duan:2022ltz} also makes use of the coset construction to identify some admissible three-character solutions as CFT, however there are some differences in the criteria used. We will comment on the cases of overlap as we go along. 

In \cite{Gaberdiel:2016zke} the following relation between the data of characters $\chi_i$ and their coset dual $\tchi_i$ was derived:
\be
\ell+\tl = n^2+\left(\frac{c+\tc}{4}-1\right)n-6\sum_{i=1}^{n-1}(h_i+\tilh_i) 
\label{datareln}
\ee
Here we are interested in the case $n=3$. Because $c+\tc$ must be a multiple of 8, we write it as $8N$ where $N$ is an integer \footnote{In \cite{Gaberdiel:2016zke} the convention was to write $c+\tc=24N$ where $N$ is a multiple of $\frac13$.}. Since the right hand side of the bilinear relation is a character with integer dimensions (up to an overall power of $q$), we must have $h_i+\tilh_i=n_i$, an integer $\ge 1$, for each $i$. Thus the above relation can be written:
\be
\ell+\tl= 6\Big(N+1-\sum_{i=1}^{n-1}n_i\Big)
\label{peq3reln}
\ee
As both $\ell,\tl\ge 0$ we have the bound:
\be
\sum_{i=1}^{n-1}n_i\le N+1
\label{sumbound}
\ee
If this bound is saturated it means $\ell=\tl=0$ and we have the possibility of dual $(3,0)$ pairs. Thus we will proceed by listing all possible values $n_i$ that saturate the bound for each $N$, and then classifying dual pairs with these $n_i$. This technical point is of importance because it seems to have been missed in much of the previous literature, starting with \cite{Gaberdiel:2016zke} that only considered a special sub-class of cosets where each $n_i\ge 2$. More general cosets of meromorphic theories were studied recently, and apparently for the first time, in \cite{Mukhi:2022bte} in the context of theories with exactly two primaries.

The values of $n_i$ have considerable significance for the structure of the bilinear pair, which we now explain. Suppose a bilinear relation holds between CFTs $\cC,\tcC$ with Kac-Moody algebras $\mfh,\tmfh$, and they pair up to a CFT $\cH$ with Kac-Moody algebra $\mfg$. Then $\tmfh$ must be the commutant of $\mfh$ in $\mfg$. Now suppose that for any of $i=1,2$ we have $n_i=1$. This means that some spin-1 currents in the theory $\cH$ arise as ``composites'' of primaries in $\cC,\tcC$. This in turn means the total number of spin-1 currents of $\cC,\tcC$ is strictly smaller than that of $\cH$, in other words $\dim \mfh+\dim\tmfh<\dim\mfg$, so the embedding of $\mfh$ in $\mfg$ is a non-trivial one -- typically $\mfh$ is embedded into a simple factor of $\mfg$. Such cases were discussed for the case of two primaries in \cite{Mukhi:2022bte}. On the other hand whenever all $n_i\ge 2$, no currents of $\cH$ can arise as composites of primaries of $\cC,\tcC$. Therefore we have $\dim \mfh+\dim\tmfh=\dim\mfg$. This can only happen if $\mfg$ is non-simple and $\mfh$ corresponds to one or more of its simple factors. Such cases were first studied in \cite{Gaberdiel:2016zke}, and they are simpler because the coset merely ``deletes'' the simple factors of $\mfg$ corresponding to $\mfh$ leaving behind the remaining simple factors as the chiral algebra $\tmfh$ of the coset theory.

At this point it is useful to briefly describe how the concept of ``fusion rules'' applies to VVMFs even before they are identified with CFT. In the MLDE approach to classification of RCFT, one first finds admissible character solutions that transform covariantly under SL(2,Z) and only later addresses their identification with CFT. Thus we can calculate their modular $S$ and $T$ matrices at the outset. Inserting the $S$-matrix into the Verlinde formula \cite{Verlinde:1988sn} one can then compute the following quantities \footnote{This can only be done once a unitary $S$-matrix has been found. In general $S$ does not come out to be unitary, this problem arises when multiple primaries have the same character. In that case the space of primaries has to be manually enlarged and the $S$-matrix recomputed in that space, as explained in \cite{Mathur:1988gt}.}:
\be
N_{ij}^k=\sum_l \frac{S_{il}S_{jl}S^{-1}_{kl}}{S_{0l}}
\label{Verform}
\ee
As long as the $S_{ij}$ are only a property of  admissible characters, the quantities $N_{ij}^k$ have no particular physical meaning. But once the characters are identified with CFTs then these quantities necessarily become the fusion rules of that theory. Hence by abuse of notation we will refer to $N_{ij}^k$ as ``fusion rules'' even when no CFT interpretation has so far been assigned to the corresponding characters. An important point that will come up below is that sometimes one or more of the $N_{ij}^k$ is a negative, rather than positive, integer. We refer to such characters as being of Intermediate Vertex Operator Algebra (IVOA) type, following \cite{Kawasetsu:2014}. 

We now give a short summary of the complete classification of admissible VVMF's with three characters and $\ell=0$ (the characters are extracted from the most recent papers \cite{Kaidi:2021ent, Das:2021uvd, Bae:2021mej} and expressed in the notation of \cite{Das:2021uvd}), referring the reader to the original references for more details. The admissible character sets fall into five categories, labelled $\mathbf{I},\mathbf{II},\ldots,\mathbf{V}$.  Let us briefly review what the various categories mean.

\noindent {\bf Category I:} The admissible VVMFs belonging to this category are all 3-character theories that are affine or tensor products of affine theories, together with the Ising CFT $\mathcal{M}(4,3)$ and the unitary presentations\footnote{By unitary presentation we mean the choice of the most singular character as the identity. However, this does not imply there is a unitary theory.} of $\mathcal{M}(7,2)$ and $\mathcal{M}(5,2)^{\otimes 2}$.

\noindent{\bf Category II:} Most of these are admissible 2-character solutions together with  an ``unstable'' character (or sometimes an admissible 1-character solution together with two ``unstable'' characters). By unstable, we mean that this character has rational coefficients in its $q$-series that cannot be made integral by any choice of normalisation. Such a case was first discussed in \cite{Mathur:1988na} and more general examples were found in \cite{Das:2021uvd}. There are also some type II cases where the conformal dimensions degenerate -- two of them become equal -- and in this case the MLDE has a logarithmic solution. Due to these reasons, type-{\bf II} VVMFs are not genuine 3-character solutions and we do not explore them further.

\noindent{\bf Category III:} The admissible VVMFs belonging to this category are those solutions of the (3,0) MLDE which appeared in \cite{franc2020classification} but not in \cite{Gaberdiel:2016zke} and hence were not previously categorised as CFT. In this category, there exists special infinite sets of solutions, at $c=8$ and $c=16$ that we explain in Appendix \ref{Inhomo}. We will not attempt to include these in our classification, though some of the known cases will appear in our tables.

\noindent{\bf Category IV:} The admissible VVMFs belonging to this category are those solutions of the (3,0) MLDE which appeared in \cite{Gaberdiel:2016zke, Hampapura:2016mmz} where they were precisely characterised as CFTs via the coset construction.

\noindent{\bf Category V:} There are seven admissible VVMFs in this category, these were independently discovered in \cite{Das:2021uvd, Kaidi:2021ent, Bae:2021mej} and not known previously.

We see that all entries in categories $\mathbf{I},\mathbf{II},\mathbf{IV}$ have already been identified as CFT's or else shown to be inconsistent \cite{Das:2021uvd}. Thus we need to focus on the characterisation of classes $\mathbf{III}$ and $\mathbf{V}$ which so far have not been identified as CFTs. To characterise them, we will study their bilinear relations with solutions in category {\bf I} and {\bf IV} (and amongst themselves). 

In Table \ref{T0} we have listed all solutions in categories $\mathbf{III}$ \footnote{Note that ${\bf III_{27}}$ in Table \ref{T0} is actually $E_{7.5}^{\otimes 2}$ and was identified in \cite{Das:2021uvd} as a category ${\bf I}$ solution. However it has a negative fusion rule and therefore is of IVOA type. Here we include it in category ${\bf III}$ as it will pair up with other IVOA-type characters in this category.} (except for the infinite sets having $c = 8$ and $c = 16$ noted above) and $\mathbf{V}$. The subscripts label the set in order of increasing central charge, thus for example $\mathbf{V}_{18}$ ($c=12$) lies between  $\mathbf{III_{17}}$ ($c=12$) and $\mathbf{III_{19}}$ ($c=\frac{25}{2}$). As explained below \eref{charexp}, the integer $m_1$ is the dimension of the weight-1 space in the identity character, while $D_i, i=1,2$ are the ground-state degeneracies of the non-identity characters.

In the last column  of table \ref{T0}, labelled ``sign(fusion)", we list the signature of the fusion coefficients of the concerned VVMF, computed using \eref{Verform}. However we do not bother to compute these for solutions of \sout{${\bf III}$}, \sout{${\bf V}$} type. Also, as noted earlier this computation requires us to enlarge the matrix in cases where there are more than three primaries, and this rapidly becomes tedious. So we restrict  this calculation to solutions that have at most four primaries. The notation `$\cdots$' in the last column of the table denotes that we did not compute the fusion coefficients of these solutions for one of the reasons above. Fortunately these will also not be needed. In the remaining cases a `$+$' sign in the last column denotes that all the fusion coefficients are non-negative while a `$-$' sign denotes that {\em at least one} coefficient is negative. The latter will be called IVOA-type solutions, and we discuss them in more detail in Sec. \ref{ivoa}. 

In Table \ref{T00}, we list the category {\bf III} infinite sets of admissible character solutions at $c = 8$ and $c = 16$ . More details on these infinite sets are in Appendix \ref{Inhomo}.


\setlength\LTleft{-8pt}
\setlength\LTright{0pt}
\rowcolors{2}{Mygrey}{Mywhite}
\begin{longtable}{c||ccccc||c}
\hline
\hline
\rowcolor{Mywhite}\# & $c$ & $(h_1,h_2)$ & $m_1$ & $(D_1,D_2)$ & $(Y_1,Y_2)$ & sign(fusion)  \\
\hline
\hline
${\bf III_{1}}$ & $\frac{12}{7}$ & $(\frac27, \frac37)$ &  $6$  & $(3,2)$ & (1,1) & $-$ \\ 
${\bf III_{2}}$ & $\frac{12}{5}$ & $(\frac15, \frac35)$ & $3$ & $(3,5)$ & (1,2) & $+$ \\
${\bf III_{3}}$ & $\frac{44}{7}$ & $(\frac47, \frac57)$ &  $88$  & $(11,44)$ & (1,1) & $-$ \\ 
${\bf III_{4}}$ & $\frac{36}{5}$ & $(\frac35, \frac45)$ & $144$ & $(12,45)$ & (1,2) & $-$ \\
${\bf III_{5}}$ & $\frac{52}{7}$ & $(\frac47, \frac67)$ &  $156$  & $(13,78)$ & (1,1) & $-$ \\
\sout{${\bf III_{6}}$} & $\frac{17}{2}$ & $(\frac{1}{16}, \frac32)$ & $255$ & $(17,221)$ & $(\frac{1}{256},1)$ & $\cdots$  \\
${\bf III_{7}}$ & $\frac{60}{7}$ & $(\frac37, \frac87)$ &  $210$  & $(10, 285)$ & (1,1) & $-$ \\ 
${\bf III_{8}}$ & $\frac{44}{5}$ & $(\frac{2}{5}, \frac65)$ & $220$ & $(11,275)$ & (1,2) & $-$  \\
\sout{${\bf III_{9}}$} & $\frac{44}{5}$ & $(\frac15, \frac75)$ &  $253$  & $(11, 242)$ & $(\frac{1}{50},1)$ & $\cdots$ \\
\sout{${\bf III_{10}}$} & $9$ & $(\frac{1}{8}, \frac32)$ & $261$ & $(9,456)$ & $(\frac{1}{32},1)$ & $\cdots$ \\
\sout{${\bf III_{11}}$} & $\frac{19}{2}$ & $(\frac{3}{16}, \frac32)$ &  $266$  & $(19, 703)$ & $(\frac{1}{64},1)$ & $\cdots$ \\
${\bf III_{12}}$ & $\frac{68}{7}$ & $(\frac{3}{7}, \frac97)$ & $221$ & $(17,782)$ & (1,1) & $-$  \\
\sout{${\bf III_{13}}$} & $10$ & $(\frac{1}{4}, \frac32)$ &  $270$  & $(5, 960)$ & $(\frac{1}{2},1)$ & $\cdots$ \\
\sout{${\bf III_{14}}$} & $\frac{21}{2}$ & $(\frac{5}{16}, \frac32)$ & $273$ & $(21, 1225)$ & $(\frac{1}{16},1)$ & $\cdots$  \\
\sout{${\bf III_{15}}$} & $11$ & $(\frac{3}{8}, \frac32)$ &  $275$  & $(11, 1496)$ & $(\frac{1}{2},1)$ & $\cdots$ \\
\sout{${\bf III_{16}}$} & $\frac{23}{2}$ & $(\frac{7}{16}, \frac32)$ & $276$ & $(23, 1771)$ & $(\frac{1}{4},1)$ & $\cdots$  \\
${\bf III_{17}}$ & $12$ & $(\frac{3}{5}, \frac75)$ &  $222$  & $(25, 1275)$ & $(2,2)$ & $\cdots$ \\
${\bf V_{18}}$ & $12$ & $(\frac{1}{3}, \frac53)$ & $318$ & $(9, 4374)$ & $(1,1)$ & $\cdots$  \\
${\bf III_{19}}$ & $\frac{25}{2}$ & $(\frac{9}{16}, \frac32)$ &  $275$  & $(25, 2325)$ & (1,1) & $+$ \\ 
${\bf III_{20}}$ & $13$ & $(\frac{5}{8}, \frac32)$ & $273$ & $(26, 2600)$ & (2,1) & $+$  \\
${\bf III_{21}}$ & $\frac{27}{2}$ & $(\frac{11}{16}, \frac32)$ &  $270$  & $(54, 2871)$ & (1,1) & $+$ \\
${\bf III_{22}}$ & $\frac{68}{5}$ & $(\frac{4}{5}, \frac75)$ & $136$ & $(119, 1700)$ & (1,2) & $+$   \\
\sout{${\bf III_{23}}$} & $\frac{68}{5}$ & $(\frac{2}{5}, \frac95)$ &  $374$  & $(119, 12138)$ & $(\frac{1}{50},1)$ & $\cdots$  \\
${\bf III_{24}}$ & $\frac{100}{7}$ & $(\frac{5}{7}, \frac{11}{7})$ & $325$ & $(55, 2925)$ & (1,1) & $-$ \\
${\bf III_{25}}$ & $\frac{100}{7}$ & $(\frac{4}{7}, \frac{12}{7})$ &  $380$  & $(55, 11495)$ & (1,1) & $-$ \\
${\bf III_{26}}$ & $\frac{29}{2}$ & $(\frac{13}{16}, \frac{3}{2})$ & $261$ & $(116, 3393)$ & (1,1) & $+$  \\
${\bf III_{27}}$ & $\frac{76}{5}$ & $(\frac{4}{5}, \frac{8}{5})$ &  $380$  & $(57, 3249)$ & (2,1) & $-$ \\
${\bf III_{28}}$ & $\frac{76}{5}$ & $(\frac{3}{5}, \frac{9}{5})$ & $437$ & $(57, 11875)$ & $(1,2)$ & $-$  \\
${\bf III_{29}}$ & $\frac{108}{7}$ & $(\frac{6}{7}, \frac{11}{7})$ &  $378$  & $(117, 3510)$ & (1,1) & $-$ \\ 
${\bf III_{30}}$ & $\frac{108}{7}$ & $(\frac{4}{7}, \frac{13}{7})$ & $456$ & $(39, 20424)$ & (1,1) & $-$  \\
 ${\bf III_{31}}$ & $\frac{33}{2}$ & $(\frac{17}{16}, \frac{3}{2})$ &  $231$  & $(528, 4301)$ & (1,1) & $+$ \\
${\bf III_{32}}$ & $\frac{116}{7}$ & $(\frac{8}{7}, \frac{10}{7})$ & $348$ & $(725, 1972)$ & (1,1) & $-$ \\
${\bf III_{33}}$ & $\frac{84}{5}$ & $(\frac{6}{5}, \frac{7}{5})$ &  $336$  & $(770, 1452)$ & (2,1) & $-$  \\
\sout{${\bf III_{34}}$} & $\frac{84}{5}$ & $(\frac{1}{5}, \frac{12}{5})$ & $534$ & $(33, 55924)$ & $(\frac{2}{625},1)$ & $\cdots$ \\
${\bf III_{35}}$ & $\frac{124}{7}$ & $(\frac{9}{7}, \frac{10}{7})$ &  $248$  & $(2108, 2108)$ & (1,1) & $-$ \\
\sout{${\bf III_{36}}$} & $18$ & $(\frac{1}{4}, \frac{5}{2})$ & $598$ & $(25, 221\cdot 2^{10})$ & $(\frac{1}{32},1)$ & $\cdots$  \\
${\bf III_{37}}$ & $\frac{92}{5}$ & $(\frac{6}{5}, \frac{8}{5})$ &  $92$  & $(1196, 7475)$ & (1,2) & $+$ \\
\sout{${\bf III_{38}}$} & $\frac{92}{5}$ & $(\frac{3}{5}, \frac{11}{5})$ & $690$ & $(299, 178802)$ & $(\frac{2}{25},1)$ & $\cdots$  \\
${\bf V_{39}}$ & $20$ & $(\frac{4}{3}, \frac{5}{3})$ &  $80$  & $(2430, 17496)$ & $(1,1)$ & $\cdots$ \\
${\bf V_{40}}$ & $20$ & $(\frac{1}{3}, \frac{8}{3})$ & $728$ & $(12, 2\cdot 3^{12})$ & $(1,1)$ & $\cdots$  \\
${\bf V_{41}}$ & $20$ & $(\frac{2}{3}, \frac{7}{3})$ &  $890$  & $(135, 10\cdot 2\cdot 3^9)$ & $(1,1)$ & $\cdots$ \\
${\bf III_{42}}$ & $\frac{108}{5}$ & $(\frac{7}{5}, \frac{9}{5})$ & $27$ & $(2295, 42483)$ & (2,1) & $+$ \\
\sout{${\bf III_{43}}$} & $\frac{108}{5}$ & $(\frac{2}{5}, \frac{14}{5})$ &  $860$  & $(833, 3015426)$ & $(\frac{1}{1250},1)$ & $\cdots$ \\
${\bf III_{44}}$ & $\frac{108}{5}$ & $(\frac{4}{5}, \frac{12}{5})$ & $1404$ & $(459, 153\cdot 5^5)$ & $(1,2)$ & $+$  \\
${\bf III_{45}}$ & $22$ & $(\frac{3}{2}, \frac{7}{4})$ &  $66$  & $(77\cdot 2^6, 11\cdot 2^{11})$ & $(1,2)$ & $+$ \\
${\bf III_{46}}$ & $22$ & $(\frac{3}{4}, \frac{5}{2})$ & $1298$ & $(154, 847\cdot 2^{10})$ & $(2,1)$ & $+$ \\
${\bf III_{47}}$ & $\frac{156}{7}$ & $(\frac{11}{7}, \frac{12}{7})$ &  $78$  & $(5070, 27170)$ & (1,1) & $-$ \\
${\bf III_{48}}$ & $\frac{156}{7}$ & $(\frac{5}{7}, \frac{18}{7})$ & $1248$ & $(130, 799500)$ & (1,1) & $-$  \\
\sout{${\bf III_{49}}$} & $\frac{45}{2}$ & $(\frac{13}{16}, \frac{5}{2})$ &  $1640$  & $(1595, 956449)$ & $(\frac{1}{16},1)$ & $\cdots$ \\
${\bf III_{50}}$ & $23$ & $(\frac{3}{2}, \frac{15}{8})$ & $23$ & $(4600,23\cdot 2^{11})$ & $(1,2)$ & $+$  \\
${\bf III_{51}}$ & $23$ & $(\frac{7}{8}, \frac{5}{2})$ &  $2323$  & $(575, 32683*32)$ & $(2,1)$ & $+$  \\
${\bf III_{52}}$ & $\frac{116}{5}$ & $(\frac{8}{5}, \frac{9}{5})$ & $58$ & $(4959, 27550)$ & $(1,2)$  & $-$ \\
\sout{${\bf III_{53}}$} & $\frac{116}{5}$ & $(\frac{4}{5}, \frac{13}{5})$ &  $1711$  & $(1653, 910803)$ & $(\frac{1}{50},1)$ & $\cdots$ \\
${\bf III_{54}}$ & $\frac{164}{7}$ & $(\frac{11}{7}, \frac{13}{7})$ & $41$ & $(4797, 50922)$ & $(1,1)$ & $-$  \\
\sout{${\bf III_{55}}$} & $\frac{47}{2}$ & $(\frac{15}{16}, \frac{5}{2})$ &  $4371$  & $(4371, 1135003)$ & $(\frac{1}{4},1)$ & $\cdots$ \\
\sout{${\bf III_{56}}$} & $26$ & $(\frac{1}{4}, \frac{7}{2})$ & $1118$ & $(117, 3315\cdot 2^{14})$ & $(\frac{1}{512},1)$ & $\cdots$ \\
\sout{${\bf III_{57}}$} & $\frac{132}{5}$ & $(\frac{3}{5}, \frac{16}{5})$ &  $1536$  & $(2392, 47018049)$ & $(\frac{2}{625},1)$ & $\cdots$ \\
${\bf V_{58}}$ & $28$ & $(\frac{2}{3}, \frac{10}{3})$ & $1948$ & $(225, 11\cdot 2\cdot 3^{14})$ & $(1,1)$ & $\cdots$  \\
\sout{${\bf III_{59}}$} & $30$ & $(\frac{3}{4}, \frac{7}{2})$ &  $2778$  & $(539, 14421\cdot 2^{14})$ & $(\frac{1}{2},1)$ & $\cdots$ \\
\sout{${\bf III_{60}}$} & $\frac{61}{2}$ & $(\frac{13}{16}, \frac{7}{2})$ & $3599$ & $(47763, 264580485)$ & $(\frac{1}{2^{12}},1)$ & $\cdots$  \\
\sout{${\bf III_{61}}$} & $31$ & $(\frac{7}{8}, \frac{7}{2})$ &  $5239$  & $(9269, 2295147\cdot 2^7)$ & $(\frac{1}{32},1)$ & $\cdots$ \\
\sout{${\bf III_{62}}$} & $\frac{156}{5}$ & $(\frac{4}{5}, \frac{18}{5})$ & $3612$ & $(14877, 250774426)$ & $(\frac{1}{1250},1)$ & $\cdots$  \\
${\bf V_{63}}$ & $36$ & $(\frac{2}{3}, \frac{13}{3})$ &  $3384$  & $(324, 8\cdot 3^{20})$ & $(1,1)$ & $\cdots$ \\
\sout{${\bf V_{64}}$} & $44$ & $(\frac{1}{3}, \frac{17}{3})$ & $3146$ & $(13, 19\cdot 3^{25})$ & $(\frac{9}{4},1)$ & $\cdots$  \\
\sout{${\bf III_{65}}$} & $\frac{276}{5}$ & $(\frac{4}{5}, \frac{33}{5})$ &  $13110$  & $(12971091, 4897835680923668)$ & $(\frac{2}{244140625},1)$ & $\cdots$ \\
\hline
\hline
\rowcolor{Mywhite}\caption{Previously uncharacterized admissible character solutions to the  $(3,0)$ MLDE. The ones of type \sout{${\bf III}$} and \sout{${\bf V}$} have $Y_i$ that are fractional and cannot be made integer by rescaling the degeneracies.}
\label{T0}
\end{longtable}

\setlength\LTleft{110pt}
\setlength\LTright{0pt}
\rowcolors{2}{Mygrey}{Mywhite}
\begin{longtable}{c||cccc}
\hline
\hline
\rowcolor{Mywhite}\# & $c$ & $(h_1,h_2)$ & $m_1$ & $(D_1,D_2)$    \\
\hline
\hline
${\bf III^{'}}$ & $8$ & $(\frac12, 1)$ &  $ \mathbb{N}\setminus\{248\}$  & $(1, 1)$   \\ 
${\bf III^{''}}$ & $16$ & $(1, \frac32)$ & $ \mathbb{N}\setminus\{496\}$  & $(1, 1)$  \\
\hline
\hline
\rowcolor{Mywhite}\caption{Previously uncharacterized infinite familes of admissible character solutions to the  $(3,0)$ MLDE with $c = 8, 16$}
\label{T00}
\end{longtable}

As mentioned above, the tables in the following Sections will include the \sout{${\bf III}$} and \sout{${\bf V}$} entries in Table \ref{T0} even though they are already ruled out from being CFTs. For completeness, our tables will also include some already characterised theories from \cite{Das:2021uvd}, as their bilinear pairings are interesting and could be useful for subsequent work. 

One of the intriguing features that will come up is that Virasoro minimal models with $c<1$ appear in the coset pairings, thus making it clear that the coset construction is more general than pairings of theories with Kac-Moody symmetry. This feature was already foreseen in the mathematics literature in \cite{frenkel1992vertex,Dong:1994} and a few examples have appeared in the physics literature in \cite{Hampapura:2016mmz,Das:2022slz,Mukhi:2022bte}.


\Comment{

\section{Methodology}
We summarise procedure below.
\begin{itemize}
\item In the tables, $\mathcal{C}$ $\leftrightarrow\widetilde{\mathcal{C}}$
\item Consider the following,
\begin{align}
    \mathcal{Z} = \left|\chi_0\right|^2 + \sum\limits_{i=1}^2 \, Y_i\left|\chi_i\right|^2 \label{partn_fn}    
\end{align}
where for a given admissible character-like solution, if $Y_1$ and $Y_2$ are fractional then these characters do not from a modular invariant partition function and hence these characters cannot correspond to a genuine RCFT. Such category ${\bf III}$ solutions are colored in \textcolor{red}{red} in the tabulated results. 
\item $\mathcal{N}=a^\mathcal{H}_1$ where $\chi^\mathcal{H}(q)=q^{\alpha_0^\mathcal{H}}(1 + a^\mathcal{H}_1 \, q + a^\mathcal{H}_2 \, q^2 + \cdots)$
\item \textcolor{red}{\bf III}: Category III solutions which have fractional $Y_1$ or $Y_2$ value.
\item \textcolor{blue}{\bf V}: Category V solutions which were presented in \cite{Das:2021uvd}.
\item $B_{r,1}\leftrightarrow B_{c^\mathcal{H}-1-r,1} \, \forall \, r\geq 1$ with $(d_1,d_2)=(1,1)$. This happens whenever $n_1=1$. 
\item $D_{r,1}\leftrightarrow D_{c^\mathcal{H}-r,1} \, \forall \, r\geq 1$ with $(d_1,d_2)=(1,2)$. This happens whenever $n_1=1$.
\item At times we have identified, $A_{1,2}\cong B_{1,1}$, $C_{2,1}\cong B_{2,1}$, $A_{1,1}^{\otimes 2}\cong D_{2,1}$ and $A_{3,1}\cong D_{3,1}$. These pairs are equivalent and exactly match at the level of their characters.
\item $d_1$ or $d_2$ is fractional $\Rightarrow$ either $\mathcal{C}$ or $\tilde{\mathcal{C}}$ has fractional $Y_1$ or $Y_2$ value. However, the converse is not true (e.g. see entry 10 in Table 1 where the dual has fractional $Y_1$ or $Y_2$ value).
\item It is observed that, for $B_{r,1}$ with $(h_1,h_2)=\left(\frac{1}{2},\frac{2r+1}{16}\right)$ we have $(D_1,D_2) = (2r+1,2^r)$.
\item It is observed that, for $D_{r,1}$ with $(h_1,h_2)=\left(\frac{1}{2},\frac{r}{8}\right)$ we have $(D_1,D_2) = (2r,2^{r-1})$.
\item In the tables, for all duals of non-unitary Virasoro minimal models, we have considered $c$ obtained from their MLDEs and not $c\to c-24h$. We shall consider the $c\to c-24h$ separately later.
\item Whenever, $n_1\neq n_2$ we get one theory dual to atmost two theories.
\item In the tables, we haven't considered duals of the two infinite set of theories at $c=8$ and $c=16$.
\item In the tables, we have first matched GHM dual pairs at the level of their $(c,h_1,h_2)\leftrightarrow (\Tilde{c},\Tilde{h}_1,\Tilde{h}_2)$ values. At this point following cases arise:
\begin{enumerate}
    \item One member among a pair is not an admissible solution: Known$\leftrightarrow$NA, IV$\leftrightarrow$NA, FM$\leftrightarrow$NA, New$\leftrightarrow$NA.
    \item $\mathcal{C}$ or(and) $\tilde{\mathcal{C}}$ are known RCFTs: Known$\leftrightarrow$Known, IV$\leftrightarrow$Known, FM$\leftrightarrow$Known and New$\leftrightarrow$Known
    \item $\mathcal{C}$ or(and) $\tilde{\mathcal{C}}$ are IV solutions: Known$\leftrightarrow$IV, IV$\leftrightarrow$IV, FM$\leftrightarrow$IV and New$\leftrightarrow$IV
    \item $\mathcal{C}$ or(and) $\tilde{\mathcal{C}}$ are FM solutions: Known$\leftrightarrow$FM, IV$\leftrightarrow$FM, FM$\leftrightarrow$FM and New$\leftrightarrow$FM
    \item $\mathcal{C}$ or(and) $\tilde{\mathcal{C}}$ are New solutions: Known$\leftrightarrow$New, IV$\leftrightarrow$New, FM$\leftrightarrow$New and New$\leftrightarrow$New
\end{enumerate}
It is observed that the number of pairs of the first kind in the above list goes on increasing as we increase $c^{\mathcal{H}}$. This is perhaps due to the reason that the number of admissible solutions keep on decreasing as we increase the central charge and so we have less and less solutions with higher central charges to pair up with solutions having lower central charges.
\item After matching pairs at the level of their $(c,h_1,h_2)\leftrightarrow (\Tilde{c},\Tilde{h}_1,\Tilde{h}_2)$ values, we move on to compute their bilinear relations upto orders $\mathcal{O}(q^{1000})$ in their q-expansions. A successful bilinear relation is the one where,
\begin{enumerate}
\item $d_1,d_2\in\mathbb{N}$
\item $\chi^\mathcal{H}$ of the meromorphic theory is an admissible solution.
\item \textcolor{brown}{``Curiosities above $c=24$"} claims $\forall \, c^\mathcal{H}\geq 24, \, \mathcal{N}\leq \text{dim}\left(D_{c^\mathcal{H},1}\right)$. Otherwise $c^\mathcal{H}$ would increase with increasing $\mathcal{N}$ (due to Sugwara construction) else, the dual coxeter number would be very high (which it cannot be as each Kac-Moody factor in a theory has to have the same $\frac{g}{k}$ with $g$ being the dual coxeter number and $k$ being the level (see next point)).
\end{enumerate}
\item \textcolor{brown}{Schellekens}: In a given theory with affine subparts, all its Kac-Moody subalgebras have the same value of the ratio: $\frac{g}{k}$ where $g$ is the dual coxeter number and $k$ is the level. So, if $\mathcal{C}\leftrightarrow \tilde{\mathcal{C}}$ for a duality with a given $c^\mathcal{H}$ and $n_1=n_2\neq 1$, then $\frac{g}{k}-\frac{\Tilde{g}}{\Tilde{k}}=0$. Also in this case, $\frac{g_i}{k_i}=\frac{m_1^\mathcal{H}}{c^\mathcal{H}}-1$ (see Eq.(6.27) of MMS (Modular Geometry on Torus)). Using this we can perhaps predict the affine subparts, $\Tilde{k}$ and $\Tilde{g}$ for $\tilde{\mathcal{C}}$ when $\mathcal{C}$ is a known RCFT.
\item For a given duality at $c^\mathcal{H}$ with $n_1 \neq n_2$, if $B_{r,1}\leftrightarrow B_{c^\mathcal{H}-1-r,1}$ and $D_{r,1}\leftrightarrow D_{c^\mathcal{H}-r,1}$, then,
\begin{enumerate}
    \item Let $g_1$ be the dual coxeter number of $B_{r,1}$ and $g_2$ be the dual coxeter number of $B_{c^\mathcal{H}-1-r,1}$. Also both these theories are at level $1$. Now consider $\frac{g_1}{k_1}+\frac{g_2}{k_2}=2r-1+2c^\mathcal{H}-2-2r-1=2c^\mathcal{H}-4$.
    \item Let $g_3$ be the dual coxeter number of $D_{r,1}$ and $g_4$ be the dual coxeter number of $D_{c^\mathcal{H}-r,1}$. Also both these theories are at level $1$. Now consider $\frac{g_3}{k_3}+\frac{g_4}{k_4}=2r-2+2c^\mathcal{H}-2r-2=2c^\mathcal{H}-4$.
\end{enumerate}

\end{itemize}

}

\subsection{Embeddings of Lie algebras}
In this section, we gather facts from Lie algebras, affine Lie algebras etc that we will need to understand coset relations. Typically, the CFTs of ${\cal H}$ and ${\cal C}$ \eqref{chc} have chiral algebras which contain affine Lie-subalgebras, whose Lie algebras are such that the Lie algebra associated with ${\cal C}$ is a subalgebra of that of ${\cal H}$. Denote by $\mathfrak{h} \hookrightarrow \mathfrak{g}$ the corresponding embedding. Here both the subalgebra $\mathfrak{h}$ as well as the embedding map are crucial data. The same subalgebra can be embedded in multiple ways and can potentially result in different cosets; we will see examples of this phenomenon in the next section. 

First we study maximal embeddings; when there is no Lie-subalgebra of $\mathfrak{g}$ that properly contains $\mathfrak{h}$. There are two kinds of maximal embeddings: regular (R) and special (S). The rank of $\mathfrak{h}$ is equal to that of $\mathfrak{g}$ in a regular embedding and is smaller in a special embedding. One can obtain the regular and special embeddings of simple Lie algebras readily from the literature; we use the LieArt 2.0 package (see \cite{feger:2012}) in {\it Mathematica} to obtain all possible maximal embeddings of a given Lie algebra.
For example $E_8$ has five regular maximal embeddings, namely $D_8$, $A_4 \oplus A_4$, $E_6 \oplus A_2$, $E_7 \oplus A_1$ and $A_8$ and six special embeddings, namely $G_2 \oplus F_4$, $A_1 \oplus A_2$, $C_2$, $A_1$, $A_1$ and $A_1$. The last three correspond to $A_1$ embedded into $E_8$ in three different ways; one way to characterize this difference is via the embedding index, which we discuss below. After having understood maximal embeddings, one studies non-maximal embeddings as follows. Let $\mathfrak{l} \hookrightarrow \mathfrak{h}$ and $\mathfrak{h} \hookrightarrow \mathfrak{g}$ be maximal embeddings. By composing the two embedding maps, one obtains a non-maximal embedding  $\mathfrak{l} \hookrightarrow \mathfrak{g}$ and all non-maximal embeddings are obtained in this manner, in steps of maximal embeddings. 

Now given an embedding $\mathfrak{h} \hookrightarrow \mathfrak{g}$, maximal or non-maximal, there exists an important quantity called the embedding index $x_e\in\mathbb{N}$ which can be computed as follows. Pick any non-trivial irrep of $\mathfrak{g}$ say $\Lambda^g$ and consider its branching
\begin{align}
\Lambda^g = \oplus_i ~\Lambda^h_i, \label{branch0} 
\end{align}
where $\Lambda^h_i$s are irreps of $\mathfrak{h}$. The embedding index is then computed using the formula: \begin{align}
x_e = \frac{\sum\limits_{i} \mathscr{L}\left(\Lambda^h_i\right)}{\mathscr{L}\left(\Lambda^g\right)}, \label{xe}
\end{align}
where $\mathscr{L}\left(\Lambda^h_i\right)$ denotes the Dynkin index of the irrep $\Lambda^h_i$. Note that even if $x_e$ is
computed in \eqref{xe} using a particular irrep and its branching \eqref{branch0}, one obtains the same answer for any finite-dimensional irrep. For example, the embedding indices of the various subalgebras (occuring in the maximal embeddings) of $E_8$ are given below in superscript. For regular embeddings we have $D_8^{(1)}$, $A_4^{(1)} \oplus A_4^{(1)}$, $E_6^{(1)} \oplus A_2^{(1)}$, $E_7^{(1)} \oplus A_1^{(1)}$ and $A_8^{(1)}$ and for special embeddings we have
$G_2^{(1)} \oplus F_4^{(1)}$, $A_1^{(16)} \oplus A_2^{(6)}$, $C_2^{(12)}$, $A_1^{(520)}$, $A_1^{(760)}$ and $A_1^{(1240)}$. Computations of branching rules, Dynkin indices, embedding indices etc are performed  using LieArt 2.0 (\cite{feger:2012}).

The relationship between the affine Lie algebras associated with the CFTs of ${\cal H}$ and ${\cal C}$ in \eqref{chc} can now be made explicit. For affine embeddings of the form, $\mathfrak{h}_{\Tilde{k}}\hookrightarrow \mathfrak{g}_{k}$, the levels follow the rule (see section 14.7 of \cite{DiFrancesco:1997nk}):
\begin{align}
    \Tilde{k} = k \, x_e. \label{affine_xe}
\end{align}
Thus, for example, when ${\cal H} = E_{8,1}$, some possibilities for ${\cal C}$ are $D_{8,1}$, $A_{4,1}$, $E_{6,1}$, $E_{7,1}$, $A_{1,1}$, $A_{2,1}$, $G_{2,1}$, $F_{4,1}$.\\

\noindent{\bf Convention:} Throughout this paper, we think of the Ising CFT $\mathcal{M}(4,3)$ as $B_{0,1}$, $A_{1,2}$ as $B_{1,1}$, $C_{2,1}$ as $B_{2,1}$, $U(1)$ (with the appropriate radius) as $D_{1,1}$, $A_{1,1}^{\otimes 2}$ as $D_{2,1}$ and $A_{3,1}$ as $D_{3,1}$.

\subsection{Extension of a chiral algebra}

Consider an affine theory based on a (not necessarily simple) Kac-Moody algebra $\mathfrak{g}_{k}$. Its $n$-character extension, denoted by $\mathcal{E}_n[\mathfrak{g}_{k}]$, is a new theory where the chiral algebra has been extended by adding new generators. The theories based on $\mathfrak{g}_{k}$ and $\mathcal{E}_n[\mathfrak{g}_{k}]$ have the same $c$. The characters of the extension are linear combinations of characters of the original theory that differ in dimension by integers and as a result the extension 
will have fewer characters than the original affine theory. It also has a different Wronskian index in general. Note that a given affine theory may have more than one extension.

One can also consider extensions of more general chiral algebras. For example, a direct product of Kac-Moody and $c<1$ Virasoro minimal models can be extended in the same way. If there is a single minimal model module of central charge $c$, we will denote the extension by $\cE_n[\mfg \otimes L(c)]$ and similarly for the more general case. Such extensions have arisen in \cite{Mukhi:2022bte, Das:2022slz} and will also arise in the cases we consider.

\subsection{More about coset relations}
\label{s3}

As we saw above, coset relations between a pair of CFTs ($\mathcal{C}$ and $\widetilde{\mathcal{C}}$) or admissible characters ($\mathcal{W}$ and $\widetilde{\mathcal{W}}$) are bilinear relations between characters of the form:
\begin{align}
    \chi^\mathcal{H} = \chi_0\Tilde{\chi}_0 + \sum\limits_{i=1}^2 \, d_i \, \chi_i\Tilde{\chi}_i. 
    \label{bilin0}
 \end{align}
Here, $\chi_0, \chi_1, \chi_2$ are the characters of $\mathcal{W}$ and $\tilde{\chi}_0, \tilde{\chi}_1, \tilde{\chi}_2$ are the characters of $\widetilde{\mathcal{W}}$. $(d_1, d_2)$ are positive integers. $\chi_0^{\mathcal{H}}$ is the character of a meromorphic CFT. Sometimes we have the situation of ``self-cosets” when the same CFT/admissible character solution is both $\mathcal{W}$ as well as $\widetilde{\mathcal{W}}$. Also sometimes (as we will see this happens when $D_{4n,1}$ are involved) there may be more than one pairing of the same sets of characters: one with $\chi_0 = \tilde{\chi}_0, \chi_1 = \tilde{\chi}_1,  \chi_2 = \tilde{\chi}_2,$ which results in a standard bilinear relation as in \eqref{bilin0} with a pair of positive integers $(d_1, d_2)$, and one or more distinct ones when the characters are paired differently as described below.  

In Eq.(\ref{bilin0}), the characters $\chi_i$ and $\Tilde{\chi}_i$ are understood to be properly normalised with integral ground-state degeneracies and multiplicities that have been determined. Let the multiplicities of $\chi_i,\tchi_i$ be $Y_i,\Tilde{Y}_i$. Since the standard coset pairing is a pairing at the level of primaries these two multiplicities must be the same for each $i$. Moreover, by modular invariance it follows that the integers $d_i$ in the bilinear relation are also equal to these multiplicities, thus $d_i=Y_i={\Tilde Y}_i$. Hence from now on, we use $d_i$ to denote both $Y_i$ and $\Tilde{Y}_i$ whenever the pairing is of standard type. We will comment on the non-standard pairings as and when they arise.

We now describe in detail three infinite families of coset pairs of CFTs and compute their $(d_1, d_2)$ values. Members of these families will occur often amongst the many coset relations between $(3,0)$ admissible characters that we compute and tabulate in the next section. In one of the families  below the non-standard pairings will also be illustrated.

\textbf{Example 1 :} We start with the case where the meromorphic CFT is the $E_{8,1}$ CFT and the coset pairs are $B_{3,1}$ and $B_{4,1}$. $E_8$ contains a regular maximal sub-algebra $D_8$ which contains a special maximal sub-algebra $B_3 \oplus B_4$. Thus $B_3 \oplus B_4 \hookrightarrow E_8$ constitutes a non-maximal embedding. One finds that the commutant of $B_3$ in $E_8$ is $B_4$ and vice versa. This then means that if $B_3$ is taken to be the Lie algebra associated to the denominator theory ${\cal C}$ in \eqref{chc}, then the Lie algebra associated to the coset theory ${\tilde{C}}$ would be $B_4$. After computing embedding indices and levels, this means that the coset of $E_{8,1}$ by $B_{3,1}$ is $B_{4,1}$ CFT, and vice versa.

The characters of $B_{3,1}$ and $B_{4,1}$ satisfy a bilinear relation with $\chi^{\cal H} = j^{\frac13}$:
\begin{align}
j^{\frac13} = \chi_0 \tilde{\chi}_0 + d_1\, \chi_{\frac12} \tilde{\chi}_{\frac12} + d_2 \, \chi_{\frac{7}{16}} \tilde{\chi}_{\frac{9}{16}}
\label{b3b4}
\end{align}
We can compute the $(d_1, d_2)$ values for this relation using Lie algebra representation theory. For this, let us count spin-1 currents on both sides. On the LHS we have the 248 currents of $E_8$ spanning the adjoint representation. This representation decomposes as follows into irreducible representations of $B_3\oplus B_4$:
\begin{align}
\mathbf{248} = (\mathbf{21}, \mathbf{1}) \oplus (\mathbf{1}, \mathbf{36}) \oplus (\mathbf{7}, \mathbf{9}) \oplus (\mathbf{8}, \mathbf{16})
\label{248b3b4}
\end{align}
This means that on the RHS of \eref{b3b4}, the 248 currents come from: (i) 21 spin-1 currents of $B_{3,1}$ combined with the identity from $B_{4,1}$, (ii) 36 spin-1 currents of $B_{4,1}$ combined with the identity from $B_{3,1}$, (iii) the product of primaries in the $\underline 7$ and $\underline 9$ representations of $B_3$ and $B_4$, (iv) the product of primaries in the $\underline 8$ and $\underline{16}$ spinor representations of $B_3$ and $B_4$ respectively. Of  these, (i) and (ii) can be found in the first term of \eref{b3b4}, (iii) in the second term and (iv) in the third term. Since there are no multiplicities in the above decomposition, it follows that $d_1=d_2=1$.

This example is a special case of a more general phenomenon where the meromorphic CFT is the one-character extension $\cE_1[D_{r,1}]$ for $r = 8, 16, 24, 32, 40 \ldots$ of which $E_{8,1}$ is the $c=8$ case. The single character of each of these CFTs is the modular invariant obtained by combining the identity character $\chi_0$ (which at level-1 contains the adjoint representation $\mathbf{2r^2 - r}$ of $D_r$) and the character $\chi_{\frac{r}{8}}$ for the spinor representation $\mathbf{2^{r-1}}$ of $D_r$. We will find several coset pairs of admissible characters that correspond to the CFTs $\cC = B_{r_1,1}$ and $\tcC = B_{r_2,1}$ for $r_1+r_2+1\equiv r$ where $r$ is a multiple of 8, that satisfy the following bilinear relation to the above meromorphic extension of $D_{r,1}$:
\be 
\chi_0^{{\cal E}_1[D_{r,1}]} = \chi_0 \tilde{\chi}_0 + d_1\, \chi_{\frac12} \tilde{\chi}_{\frac12} + d_2 \, \chi_{\frac{2r_1+1}{16}}\, \tilde{\chi}_{\frac{2r_2+1}{16}}
\label{br1br2a}
 \ee
 The relevant Lie algebra representation content of each of its terms comes from the following two relations:
 \be
 \begin{split}
 \mathbf{2r^2 - r}&= \mathbf{(2r_1^2+r_1,1) \oplus (1,2r_2^2+r_2) \oplus (2r_1+1,2r_2+1)}\\ \mathbf{2^{r-1}} &= \mathbf{(2^{r_1},2^{r_2})}
  \end{split}
\label{br1br2b}
\ee
Now the first two terms on the right hand side of the first line of \eqref{br1br2b} give rise to the spin-1 contributions in the product of identity characters (first term of \eref{br1br2a}), while the third term gives rise to the spin-1 contributions in the product of the characters in the fundamental representations (second term of \eref{br1br2a}). Meanwhile the spinor of $D_r$ decomposes into the product of spinor reprsentations of $B_{r_1}, B_{r_2}$ (second line of \eref{br1br2b}) and this corresponds to the last term in \eref{br1br2a} (note that this contribution in general has spin $\frac{r}{8}$ rather than 1). It follows that $d_1 = 1, d_2 = 1$. We also learn that the dimensions of the spinors of the coset pair add up to $\frac{r}{8}$ rather than 1, and this corresponds to the integer $n_2$ defined in \eref{peq3reln}. The commutant of $B_{r_1}$ inside $D_{r_1 + r_2+1}$ is $B_{r_2}$ (because there is a special maximal embedding $B_{r_1}^{(1)}\oplus B_{r_2}^{(1)} \hookrightarrow D_{r_1 + r_2 +1}$) so we can identify $B_{r_1,1}$ with the denominator theory ${\cal C}$ with ${\cal E}_1[D_{r_1+r_2+1,1}]$ as the numerator theory ${\cal H}$  and $B_{r_2,1}$ as the coset theory $\tilde{{\cal C}}$. Of course one can also exchange the roles of $B_{r_1}$ and $B_{r_2}$. 

\textbf{Example 2 : }Another infinite family of coset pairs is $D_{r_1,1}$  and $D_{r_2,1}$ pairing up in a bilinear relation with a meromorphic extension $\cE_1[D_{r,1}]$ where where $r=r_1+r_2$ is a multiple of 8. The affine theory $D_{r,1}$ has three characters: the identity character $\chi_0$, the vector character $\chi_{\frac12}$ with conformal dimension $\frac12$ and the spinor and conjugate spinor (two representations with the same character)  $\chi_{\frac{r}{8}}$ with conformal dimension $\frac{r}{8}$. The bilinear relation for the coset pair of $D_{r_1,1}$  and $D_{r_2,1}$  is:
\begin{align} 
\chi_0^{{\cal E}_1[D_{r,1}]} = \chi_0 \tilde{\chi}_0 + d_1\, \chi_{\frac12} \tilde{\chi}_{\frac12} + d_2 \, \chi_{\frac{r_1}{8}}\, \tilde{\chi}_{\frac{r_2}{8}}\label{dr1dr2a}
\end{align}
and the Lie algebra representations decompose as:
\be
\begin{split}
\mathbf{2r^2 - r}  &=\mathbf{(2r_1^2-r_1,1) + (1,2r_2^2-r_2) + (2r_1,2r_2)}  \\
\mathbf{2^{r-1}}&=\mathbf{(\overline{2^{{r}_1-1}},2^{r_2-1})+(2^{r_1-1}, \overline{2^{{r}_2-1}})}
\label{dr1dr2b}
\end{split}
\ee
The first two terms on the right hand side of the first line in \eqref{dr1dr2b} are associated with the product of the identity characters while the third term corresponds to the product of the characters in the fundamental representations (hence $d_1 = 1$), and these terms are associated to spin-1 generators on both sides. The two terms on the right hand side of the second line in \eqref{dr1dr2b} correspond to the product of the characters in the spinor representations and since there are two terms we find $d_2 = 2$. These are associated to spin-$\frac{r}{8}$ generators. Finally we note that the commutant of $D_{r_1}$ inside $D_{r_1 + r_2}$ is $D_{r_2}$ (because there is a regular maximal embedding $D_{r_1}^{(1)}\oplus D_{r_2}^{(1)} \hookrightarrow D_{r_1 + r_2}$) which means we can choose $D_{r_1,1}$ for the denominator theory ${\cal C}$ with ${\cal E}_1[D_{r_1+r_2,1}]$ as the numerator theory ${\cal H}$  and get $D_{r_2,1}$ for the coset theory $\tilde{{\cal C}}$; again the roles of $D_{r_1},D_{r_2}$ can be exchanged.

Interestingly, when $r_1,r_2$ are both multiples of 4 there is another way to pair them up to a meromorphic theory that is not $D_{r_1+r_2,1}$. As an example, consider the pair $D_{12,1},D_{12,1}$ (both members of this pair are the same, but that is irrelevant to the discussion). The non-trivial conformal dimensions for each factor are $\half,\frac32$. Of these, the latter -- the spinor representation -- occurs twice because of chirality. Thus $(Y_1,Y_2)=(1,2)$. We find that they have a bilinear pairing to the meromorphic theory $D_{24,1}$ as discussed in Subsection \ref{label:cosets.24}. This is consistent with the fact that:
\be
\frac{\cE[D_{24,1}]}{D_{12,1}}=D_{12,1}
\ee
In this pairing, the vector primaries with $h=\half$ of each $D_{12,1}$ pair up to make $(24)^2=576$ spin-1 fields that, together with the 276 generators of each $D_{12}$, make up the 1128 generators of $D_{24}$. This is a special case of the counting above. This pairing relies on the existence of a modular-invariant extension of $D_{24}$ which is a general phenomenon for all $D_{8n}$. We may therefore consider this a ``standard'' or ``default'' pairing.

However we also find another coset pairing in which the vector representation with $h=\half$ for each $D_{12}$ combines with one of the spinors with $h=\frac32$ of the other as shown in Table \ref{T7}. We see that this time new spin-2 generators arise, but no new spin-1 generators are created. As a result the meromorphic theory formed by this pairing still has Kac-Moody algebra $(D_{12,1})^{\otimes 2}$. The pairing is:
\be
\chi_0\tchi_0+\chi_\half\tchi_\frac32+\chi_\frac32\tchi_\half=\chi^{\cE_1[(D_{12,1})^{\otimes 2}]}=j(\tau)-192
\label{newpairing}
\ee
and corresponds to the coset:
\be
\frac{\cE[(D_{12,1})^{\otimes 2}]}{D_{12,1}}=D_{12,1}
\ee
It exists because of the special modular invariant $\cE_1[(D_{12,1})^{\otimes 2}]$ which is entry 66 of \cite{Schellekens:1992db}. Notice that in \eref{newpairing} not all primaries are used, since each spinor occurs only once rather than twice as in the affine theory $D_{12,1}$. 
Comparing with \eref{bilin0} it seems that we  effectively have $(d_1,d_2)=(1,1)$, and therefore $(d_1,d_2)\ne(Y_1,Y_2)$, but a better way to think of it is that for such special pairings, $(d_1,d_2)$ are not associated to multiplicities of primaries at all. 

This point becomes clearer if we consider two copies of $D_{16,1}$ which pair up in two different ways to a meromorphic $c=32$ theory, corresponding to the distinct cosets:
\be
\frac{\cE_1[D_{32,1}]}{D_{16,1}}=D_{16,1},\quad
\frac{\cE_1[(D_{16,1})^{\otimes 2}]}{D_{16,1}}=D_{16,1}
\ee
Now $D_{16,1}$ has $(h_1,h_2)=(\half,2)$ and hence there seems to be only one bilinear pairing involving the vector  representation having $h_1=\half$, where it pairs with itself. The result is easily seen to be $\cE_1[D_{32,1}]$. One may then wonder what is the other pairing leading to the second coset. The resolution is that in the other pairing we skip the vector representation entirely and take the modular-invariant combination (up to a phase) $\chi_0+\chi_2$ as the single character of each factor, then multiply them. The resulting bilinear relation is:
\be
(\chi_0+\chi_2)(\tchi_0+\tchi_2)=\chi_0\tchi_0+\chi_0\tchi_2+\chi_2\tchi_0+\chi_2\tchi_2
= \chi^{(\cE_1[D_{16,1}])^2}=j(\tau)+248
\ee
Thus in this example the meromorphic extension of the square is actually the square of a meromorphic extension of each factor -- and the corresponding 32-dimensional lattice is the direct sum of two independent 16d lattices (this was not true for the two ways of pairing $D_{12,1}$ however, where the resulting extension is not the product of extensions). In this situation we again see that the numbers $(d_1,d_2)$ are not meaningful per se and should not be compared to $(Y_1,Y_2)$. Fortunately, as emphasised above, this issue arises only for coset pairs involving affine theories of type $D_{4r,1}$.

\textbf{Example 3 : }The third and last example of an infinite family of coset relations is based on the maximal special embedding $B_{r-1}^{(1)} \hookrightarrow D_{r}$. The commutant of $B_{r-1}$ inside $D_r$ is trivial; one can see this from the fact that the branching rule for the adjoint representation of $D_r$ contains no singlets. This means that when ${\cal E}_1[D_r,1]$ is taken to be the numerator theory ${\cal H}$ and $B_{r-1,1}$ as the denominator theory ${\cal C}$, then the coset theory $\tilde{{\cal C}}$ is a CFT with a chiral algebra containing no Kac-Moody currents. Comparing central charges, we see that this CFT has $c=\half$. Since it is unitary, it has to be the Ising CFT, equivalently the ${\cal M}(4,3)$ Virasoro minimal model. We thus have the coset pair, $B_{r-1,1}$ and ${\cal M}(4,3)$;  its bilinear relation is:
\begin{align}
\chi_0^{{\cal E}_1[D_r,1]} = \chi_0 \tilde{\chi}_0 + d_1 \, \chi_{\frac12} \tilde{\chi}_1 +  d_2 \, \chi_{\frac{2r-1}{16}} \tilde{\chi}_2 \label{brisinga}
\end{align}
where $\tilde{\chi}_0, \tilde{\chi}_1, \tilde{\chi_2}$ are the characters of the Ising model. The Lie algebra representation content is:
\be
\mathbf{r(2r-1)} = \mathbf{(r-1)(2r-1)} \oplus \mathbf{2r-1}
\label{brisingb}
\ee
Additionally the spinor representation $\mathbf{2^{r-1}}$ of $D_{r}$ goes directly into the spinor of the same dimension for $B_{r-1}$.
Matching the dimensions of the representations in \eqref{brisingb} and comparing with \eqref{brisinga} we conclude that $d_1 = 1, d_2 = 1$. The coset pair relations amongst $(3,0)$ admissible characters feature this example for $r = 8, 16, 24, 32, 40$. This family of examples can be subsumed under Example 1 above if we denote the Ising CFT by $B_{0,1}$. Following the standard formulae for $B_{r,1}$ we see that $B_{0,1}$ should have $c=\half, h_1=\half, h_2=\frac{1}{16}$ which is precisely the Ising model. 

\subsection{Intermediate Vertex Operator Algebras (IVOA)}\label{ivoa}

There is an intriguing class of characters whose existence was first noted in \cite{Mathur:1988na, Mathur:1988gt} and a few of which were subsequently identified as ``Intermediate Vertex Operator Algebras'' (IVOA) in \cite{Kawasetsu:2014}. For these, some of the fusion rules derived from the modular $S$-matrices via the Verlinde formula \cite{Verlinde:1988sn} turned out to be negative integers. In general these cannot be identified with unitary CFT, though in a few special cases one can exchange characters to find a non-unitary -- but otherwise genuine -- CFT \cite{Mathur:1988gt}.

Such characters do share a number of good properties with RCFT and are of some mathematical interest. Hence we include them in our classification \footnote{IVOA-type characters have also been included in the work of \cite{Kaidi:2021ent, Bae:2021mej}.}. Whether these can be precisely said to be IVOA is beyond the scope of the present work, so we will simply identify them as ``potentially of IVOA-type'' and put them in separate tables.

It is important to realise that having negative fusion rules is quite distinct from non-unitarity. In fact  IVOA's have positive central charges with (some) negative fusion rules, while the non-unitary $c<1$ minimal models have negative central charges but positive fusion rules. Exchanging the choice of identity characters sometimes (but not always) converts an IVOA to a consistent but non-unitary CFT. 
We will find several admissible characters of IVOA type that pair up via bilinear relations into a modular invariant \footnote{For two characters, a bilinear pairing between IVOA-type characters of $c=\frac{2}{5}$ and $c=\frac{118}{5}$ is easily seen from Appendix B.2 of \cite{Chandra:2018pjq}.}.
Our policy when encountering such characters will be to list them separately in tables. They are listed in our conclusions but do not appear in our final list of CFTs, Table \ref{complete30}. Determining whether they are consistent IVOA's within the definitions of \cite{Kawasetsu:2014} is left for future work.

\section{Coset pairs and identification of CFTs}

\label{cosetpairs}

In this section, we tabulate the bilinear relations that exist between pairs of admissible character-like solutions and then discuss what this tells us about possible identification of the solutions with CFTs. To begin with, we list all pairs $\mathcal{W}\leftrightarrow \widetilde{\mathcal{W}}$ which satisfy $c+\tilde{c} = 8N$ and $h_i+\Tilde{h}_i = n_i\in\mathbb{Z} \, \, \forall \, i\in\{1,2\}$. Such a list a priori includes some pairs which when paired up in a bilinear way lead to rational, rather than integral, $d_i$. We then rule out such pairs as inconsistent since they do not satisfy a valid bilinear relation even at the level of characters. 

That leaves us with pairs ($\mathcal{W}\leftrightarrow \widetilde{\mathcal{W}}$) that satisfy the bilinear relations with integral  $d_i$. Then we perform a case-by-case analysis and explain the bilinear relations from the point of view of Lie algebra embeddings. If, for a coset pair, such an embedding exists, then we can readily find the affine subalgebra of the new theory and show that its extension leads to the new theory by computing its characters as linear combinations of the affine characters. Then we can declare it to be a genuine RCFT.

For the remaining relations, in some cases we are able to show that there does not exist an embedding, in which case the bilinear relation only holds at the level of characters but does not lead to a CFT interpretation for the members of the pair. In the remaining cases we must leave the existence of the coset theory unresolved at this stage. One of the tantalising classes in every set corresponds to Intermediate Vertex Operator Algebras (IVOA) as discussed above.

\subsection{Cosets of $c^{\mathcal{H}}=8$}

We first consider coset bilinear relations between $(3,0)$ admissible character solutions with the $c^{\cal H} = 8$ meromorphic CFT viz. the $E_{8,1}$ CFT with character $j^{\frac13}$. This would correspond to $N = 1$ and $n_1 = 1, n_2 = 1$ in \eqref{peq3reln}. Any admissible character that is potentially part of such a coset relation has to have a central charge less than $8$. Hence we consider all admissible characters from \cite{Das:2021uvd} with $c < 8$. For any of them, call it $\mathcal{W}$ with central charge and conformal dimensions $(c, h_1, h_2)$, we ask if there is another admissible character  $\widetilde{\mathcal{W}}$ with central charge and conformal dimensions $( 8 - c, 1 - h_1, 1 - h_2 )$.  For each such pair $\mathcal{W}, \widetilde{\mathcal{W}}$, we then ask if their characters satisfy a coset bilinear relation \eqref{bilin0} and if they do, we would have computed the values of $(d_1, d_2)$ defined in these equations. We collect the details of these coset bilinear relations in two tables, \ref{T1}, \ref{T3}. It is remarkable that every $(3,0)$ admissible character with $c < 8$ is part of a coset relation and is featured somewhere in these tables; this is not necessarily the case for $c>8$. 

\subsection*{\bf Comments on Table \ref{T1}}

This table contains 10 bilinear pairings. Each of these is consistent, as we will recount below -- in other words both members of every pair are genuine CFTs.
Row $1$ is a special case of Example 3 of section \ref{s3}, namely \eqref{brisinga}, \eqref{brisingb} for $r = 8$ (note that ${\cal E}_1[D_{8,1}] = E_{8,1}$). On general grounds, we know that (i) since the  $B_7$ of the denominator theory has a trivial commutant in $E_8$, the coset  must have no Kac-Moody symmetries, and (ii) the central charge of the coset must be $\frac12$. Unitarity then implies that the coset theory is indeed the Ising CFT, as we also explicitly verify. We will see more examples of this phenomenon later -- that the coset $\cH/\cC$, where both $\cH$ and $\cC$ have Kac-Moody symmetries, results in a CFT with no Kac-Moody symmetries, in this case a minimal model. Because of the way it naturally arises as a special case of the $B_{r,1}$ Kac-Moody algebras, we will often denote the Ising model by $B_{0,1}$ in what follows. 

Rows $2$, $4$, $7$ are special cases of Example 2 of section \ref{s3}, namely \eqref{dr1dr2a}, \eqref{dr1dr2b}  for $r = 8$. We thus have $\frac{E_{8,1}}{D_{3,1}} \cong D_{5,1}$ or $\frac{E_{8,1}}{D_{5,1}} \cong D_{3,1}$.  The $(d_1, d_2)$ values  follow the predictions from decomposing the representation as explained in Section \ref{s3}.

Rows $3$, $6$, $8$ are special cases of Example 1 of Section \ref{s3}, hence \eqref{br1br2a}, \eqref{br1br2b}, for $r = 8$. Note that either of the pair of CFTs can serve as the denominator while the other would be the coset, we thus have $\frac{E_{8,1}}{B_{6,1}} \cong B_{1,1}$ and $\frac{E_{8,1}}{B_{1,1}} \cong B_{6,1}$ (this is a very general phenomenon, though if only one member of the pair is known as a CFT then it is more useful to treat that one as the denominator). The $(d_1, d_2)$ values follow the predictions from the decomposition of representations explained in section \ref{s3}. 

Rows 9 and 10 are coset relations between two identical CFTs, namely self-cosets. Row $9$ is a self-coset relation with $d_1=d_2 =2$ and is explained by the regular maximal embedding: $A_4^{(1)} \oplus A_4^{(1)} \hookrightarrow E_8$ so that  the commutant of each $A_4$ is the other one. This gives us the coset $\frac{E_{8,1}}{A_{4,1}} \cong A_{4,1}$. The  computed value $d_1+d_2 = 4$ can be explained from the decomposition: $\mathbf{248} = (\mathbf{24}, \mathbf{1}) \oplus (\mathbf{1}, \mathbf{24})\oplus (\mathbf{5}, \mathbf{10}) \oplus (\overline{\mathbf{5}}, \overline{\mathbf{10}}) \oplus (\overline{\mathbf{10}}, \mathbf{5}) \oplus (\mathbf{10}, \overline{\mathbf{5}})$. The first two terms correspond to the $\chi_0^2$ term of \eqref{bilinrel} while the last four terms correspond to the $\chi_i \tilde{\chi_i}$ term thus giving $d_1= d_2 = 2$. 
Row $10$ is again a self-coset relation. The embedding behind this coset relation is obtained in two steps, each of which is a regular maximal embedding: $A_2 \oplus A_2 \oplus A_2 \hookrightarrow E_6$ and $E_6 \oplus A_2 \hookrightarrow E_8$. Computing the embedding indices we get $A_2^{(1)} \oplus A_2^{(1)} \oplus A_2^{(1)} \oplus A_2^{(1)} \hookrightarrow E_8$. The commutant of one of the $A_2 \oplus A_2$ is the other $A_2 \oplus A_2$. This gives us the coset $\frac{E_{8,1}}{A_{2,1}^{\otimes 2}} \cong A_{2,1}^{\otimes 2}$.  The computed value of $d_1,d_2$ can be explained from the decomposition: $\mathbf{248} = (\mathbf{8},\mathbf{1},\mathbf{1}, \mathbf{1}) \oplus (\mathbf{1},\mathbf{8},\mathbf{1}, \mathbf{1}) \oplus (\mathbf{1},\mathbf{1},\mathbf{8}, \mathbf{1}) \oplus (\mathbf{1},\mathbf{1},\mathbf{1}, \mathbf{8}) \oplus (\mathbf{3},\mathbf{1},\mathbf{3}, \mathbf{3}) \oplus (\mathbf{1},\mathbf{3},\overline{\mathbf{3}}, \mathbf{3}) \oplus (\overline{\mathbf{3}},\overline{\mathbf{3}},\mathbf{1}, \mathbf{3}) \oplus (\overline{\mathbf{3}},\mathbf{1},\overline{\mathbf{3}}, \overline{\mathbf{3}}) \oplus (\mathbf{1},\overline{\mathbf{3}},\mathbf{3}, \overline{\mathbf{3}}) \oplus (\mathbf{3},\mathbf{3},\mathbf{1}, \overline{\mathbf{3}}) \oplus (\overline{\mathbf{3}},\mathbf{3},\mathbf{3}, \mathbf{1}) \oplus (\mathbf{3},\overline{\mathbf{3}}, \overline{\mathbf{3}}, \mathbf{1})$. The first four terms correspond to the $\chi_0\tchi_0$ term of \eqref{bilinrel} while the last eight terms correspond to the $\chi_i \tchi_i$ terms thus giving $d_1=d_2 = 4$. 

All the coset relations described so far (in rows 1--10 except row 5) were between $(3,0)$ admissible characters corresponding to well-known CFTs namely WZW CFTs and Virasoro minimal models. In row $5$ we encounter for the first time a coset relation between a WZW CFT namely $G_{2,1} \otimes G_{2,1}$ and ${\bf III_2}$, an admissible character (see Table \ref{T1}) which has not yet been characterised as a CFT. Coset relations and the general theory of meromorphic cosets will enable us to characterise ${\bf III_2}$ as follows. We first seek a Lie algebra embedding for $E_8$ which contains $G_2 \oplus G_2$. We find it in two steps of maximal embeddings: $G_2 \oplus A_1 \hookrightarrow F_4$ and $G_2 \oplus F_4 \hookrightarrow E_8$ giving $G_2 \oplus G_2 \oplus A_1 \hookrightarrow E_8$. This means that the commutant of $G_2 \oplus G_2$ in $E_8$ is $A_1$. Further, computing the embedding indices, we have $G_2^{(1)} \oplus G_2^{(1)} \oplus A_1^{(8)} \hookrightarrow E_8$, which gives the affine Lie algebra embedding $G_{2,1} \otimes G_{2,1} \otimes A_{1,8} \hookrightarrow E_{8,1}$ (see appendix \ref{c1} -- Example 1,2). This implies that the the coset CFT is $A_{1,8}$. The central charge of $A_{1,8}$ is $\frac{12}{5}$ and  $m_1 = 3$ which matches with that of ${\bf III_2}$. But $A_{1,8}$ is a nine-character CFT and cannot as such be in a coset relation with the three-character $G_{2,1}\otimes G_{2,1}$. Instead, we are dealing with a three-character extension $\mathcal{E}_3[A_{1,8}]$. 

Let us construct this extension explicitly. Denote the three characters of ${\bf III_2}$ by $\{\Tilde{\chi}_{0},\Tilde{\chi}_{\frac{1}{5}},\Tilde{\chi}_{\frac{3}{5}}\}$ and the nine Kac-Weyl characters of $A_{1,8}$ by $\{\chi^{\text{K}}_0,\chi^{\text{K}}_{\frac{3}{40}},\chi^{\text{K}}_{\frac{1}{5}},\chi^{\text{K}}_{\frac{3}{8}},\chi^{\text{K}}_{\frac{3}{5}},\chi^{\text{K}}_{\frac{7}{8}},\chi^{\text{K}}_{\frac{6}{5}},\chi^{\text{K}}_{\frac{63}{40}},\chi^{\text{K}}_2\}$. Then ${\cal E}_3[A_{1,8}]$ is given by:
\begin{align}
    \Tilde{\chi}_0 = \chi^{\text{K}}_0 + \chi^{\text{K}}_2, \qquad    
    \Tilde{\chi}_{\frac{1}{5}} = \chi^{\text{K}}_{\frac{1}{5}} + \chi^{\text{K}}_{\frac{6}{5}}, \qquad \Tilde{\chi}_{\frac{3}{5}} = \chi^{\text{K}}_{\frac{3}{5}}. \label{FMG2_00}
\end{align}
The explicit forms of the left hand sides of \eqref{FMG2_00} are available from the solutions of the $(3,0)$ MLDE  \cite{Das:2021uvd}. The explicit forms of the right hand sides of \eqref{FMG2_00} are also available, from say chapter 14 of \cite{DiFrancesco:1997nk}. This allows us to derive the relevant coefficients in \eqref{FMG2_00}. Further evidence towards the fact that ${\bf III_2}$ is the above extension is provided by the following derivation of the $(d_1, d_2)$ values: $\mathbf{248} = (\mathbf{14}, \mathbf{1}, \mathbf{1}) \oplus (\mathbf{1}, \mathbf{14}, \mathbf{1}) \oplus (\mathbf{1}, \mathbf{1}, \mathbf{3}) \oplus (\mathbf{7}, \mathbf{7}, \mathbf{3}) \oplus (\mathbf{7}, \mathbf{1}, \mathbf{5}) \oplus (\mathbf{1}, \mathbf{7}, \mathbf{5})$. The first three representations are associated with the $\chi_0 \Tilde{\chi}_0$ term in the coset relation, the fourth representation is associated with $d_1 \chi_1 \Tilde{\chi}_1$ and the last two representations are associated with  $d_2 \chi_2 \Tilde{\chi}_2$ thus giving $d_1 = 1, d_2 = 2$. Thus using the coset relation in row no. $10$ we have completed the identification of ${\bf III_2}$ as the three-character extension ${\cal E}_3[A_{1,8}]$ in \eqref{FMG2_00}. 

Note that the modular invariant partition function one can construct from Eq.(\ref{FMG2_00}) is the following (see Table 1 of \cite{Cappelli:2009xj}),
\be
\begin{split}
    \mathcal{Z} &= \left|\chi^{\text{K}}_0 + \chi^{\text{K}}_2\right|^2 + \left|\chi^{\text{K}}_{\frac{1}{5}} + \chi^{\text{K}}_{\frac{6}{5}}\right|^2 + 2\left|\chi^{\text{K}}_{\frac{3}{5}}\right|^2 \\
    &= \left|\Tilde{\chi}_0\right|^2 + \left|\Tilde{\chi}_{\frac{1}{5}}\right|^2 + 2\left|\Tilde{\chi}_{\frac{3}{5}}\right|^2 \label{Z_A18}
\end{split}
\ee
which shows that $(d_1,d_2) = (1,2)$. Thus, $\mathcal{E}_3[A_{1,8}]$ is a $3$-character and $4$-primary extension of $A_{1,8}$. This is the $D$-type non-diagonal invariant of \cite{Gepner:1986wi, Cappelli:2009xj}.

\Comment{
{\color{red} Since $G_{2,1}^{\otimes 2}$ is a $4$ primary theory, this had to be true since in terms of MTCs, $G_{2,1}^{\otimes 2}\in\text{Fibonacci}\times\text{Fibonacci}$ and $\mathcal{E}_3[A_{1,8}]\in\overline{\text{Fibonacci}}\times\overline{\text{Fibonacci}}$ (see appendix Sec. \ref{g2MTC_8} for more details).}
}

\setlength\LTleft{15pt}
\setlength\LTright{0pt}
\rowcolors{2}{Mygrey}{Mywhite}
\begin{longtable}{l||ccccc||ccccc||c}
\hline
\hline
\rowcolor{Mywhite}\# & $c$ & $(h_1,h_2)$ & $m_1$ & \tiny{$(D_1,D_2)$} & $\mathcal{C}$ & $\Tilde{c}$ & $(\Tilde{h}_1,\Tilde{h}_2)$ & $\tilde{m}_1$ & \tiny{$(\Tilde{D}_1,\Tilde{D}_2)$} & $\tilde{\mathcal{C}}$ & \tiny{$(d_1,d_2)$} \\
\hline
\hline
1. & $\frac{1}{2}$ & $\left(\frac{1}{2},\frac{1}{16}\right)$ & $0$ & \tiny{(1,1)} & $B_{0,1}$ & $\frac{15}{2}$ & $\left(\frac{1}{2},\frac{15}{16}\right)$ & $105$ &  \tiny{(15,$2^{7}$)} & $B_{7,1}$  & \tiny{$\left(1,1\right)$} \\
2. & $1$ & $\left(\frac{1}{2},\frac{1}{8}\right)$ & $1$ & \tiny{(2,1)} & $D_{1,1}$ & $7$ & $\left(\frac{1}{2},\frac{7}{8}\right)$ & $98$ & \tiny{(14,64)} & $D_{7,1}$ & \tiny{$\left(1,2\right)$} \\
3. & $\frac{3}{2}$ & $\left(\frac{3}{16},\frac{1}{2}\right)$ & $3$ &  \tiny{(2,3)} & $B_{1,1}$ & $\frac{13}{2}$ & $\left(\frac{13}{16},\frac{1}{2}\right)$ & $78$ & \tiny{(64,13)} & $B_{6,1}$ & \tiny{$\left(1,1\right)$} \\
4. & $2$ & $\left(\frac{1}{2},\frac{1}{4}\right)$ & $6$ & \tiny{(4,2)} & $D_{2,1}$ & $6$ & $\left(\frac{1}{2},\frac{3}{4}\right)$ & $66$ & \tiny{(32,12)} & $D_{6,1}$ & \tiny{$\left(1,2\right)$} \\
5. & $\frac{12}{5}$ & $\left(\frac{1}{5},\frac{3}{5}\right)$ & $3$ & \tiny{(3,5)} & ${\bf III_{2}}$ & $\frac{28}{5}$ & $\left(\frac{4}{5},\frac{2}{5}\right)$ & $28$ &  \tiny{(49,7)} & $G_{2,1}^{\otimes 2}$ & \tiny{$\left(1,2\right)$} \\
6. & $\frac{5}{2}$ & $\left(\frac{5}{16},\frac{1}{2}\right)$ & $10$ &  \tiny{(4,5)} & $B_{2,1}$ & $\frac{11}{2}$ & $\left(\frac{11}{16},\frac{1}{2}\right)$ & $55$ &  \tiny{(32,11)} & $B_{5,1}$ & \tiny{$\left(1,1\right)$} \\
7. & $3$ & $\left(\frac{1}{2},\frac{3}{8}\right)$ & $15$ & \tiny{(6,4)} & $D_{3,1}$  & $5$ & $\left(\frac{1}{2},\frac{5}{8}\right)$ & $45$ & \tiny{(10,16)} & $D_{5,1}$  & \tiny{$\left(1,2\right)$} \\
8. & $\frac{7}{2}$ & $\left(\frac{7}{16},\frac{1}{2}\right)$ & $21$ & \tiny{(8,7)} & $B_{3,1}$ & $\frac{9}{2}$ & $\left(\frac{9}{16},\frac{1}{2}\right)$  & $36$ & \tiny{(16,9)} & $B_{4,1}$& \tiny{$\left(1,1\right)$} \\
9. & $4$ & $\left(\frac{2}{5},\frac{3}{5}\right)$ & $24$ & \tiny{(5,10)} & $A_{4,1}$ & $4$ & $\left(\frac{3}{5},\frac{2}{5}\right)$ & $24$ & \tiny{(10,5)} & $A_{4,1}$ & \tiny{$(2,2)$} \\
10. & $4$ & $\left(\frac{1}{3},\frac{2}{3}\right)$ & $16$ & \tiny{(3,9)} & $A_{2,1}^{\otimes 2}$ & $4$ & $\left(\frac{2}{3},\frac{1}{3}\right)$ &  $16$ & \tiny{(9,3)} & $A_{2,1}^{\otimes 2}$ & \tiny{$(4,4)$} \\
\hline
\hline
\rowcolor{Mywhite}\caption{CFT pairings, $c^\mathcal{H}=8$ with $(n_1,n_2)=(1,1)$. The meromorphic theory $\cH$ to which the solutions pair up is $E_{8,1}$.}
\label{T1}
\end{longtable}

\noindent {\bf Conclusion:} From table \ref{T1} we conclude that ${\bf III_2}$ is identified as a genuine CFT which is $\mathcal{E}_3[A_{1,8}]$.

\subsection*{Comments on Table \ref{T3}}

The bilinear pairings in Table \ref{T3} are pairs of admissible character solutions with central charges $(\frac{4}{7},\frac{52}{7}), (\frac{4}{5},\frac{36}{5})$ and $(\frac{12}{7},\frac{44}{7})$. The fusion rules in all these cases are of IVOA type, that is atleast one of the fusion coefficients is negative. In the first two cases, one of the two members of the pair is a known IVOA -- obtained by reordering the characters of the non-unitary minimal model $\cM(7,2)$  in one case and the product of non-unitary minimal models $\cM(5,2)^{\otimes 2}$ in the other. Here, the notation $\mathcal{I}[\mathcal{W}]$ denotes the ``unitary presentation" of $\mathcal{W}$. It is quite remarkable that these pair up to give the $E_{8,1}$ character though we cannot obtain this result via Lie algebra embeddings. Note how the dimension 248 is realised by the sum $1+156$ (spin-1 currents of the pair) added to $78+13$ (coming from the products of primaries of the two factors and having degeneracies $78,13$ due to the second factor). 
Based on this we would like to claim that $\mathbf{III_5}$ and $\mathbf{III_4}$ are also IVOAs.

The last row contains the pair ${\bf III_1}$ and ${\bf III_3}$, neither of which has previously been characterised. As noted above, these are of IVOA type. 

\setlength\LTleft{0pt}
\setlength\LTright{0pt}
\rowcolors{2}{Mygrey}{Mywhite}
\begin{longtable}{l||ccccc||ccccc||c}
\hline
\hline
\rowcolor{Mywhite}\# & $c$ & $(h_1,h_2)$ & $m_1$ & \tiny{$(D_1,D_2)$} & $\mathcal{W}$ & $\Tilde{c}$ & $(\Tilde{h}_1,\Tilde{h}_2)$ & $\tilde{m}_1$ & \tiny{$(\Tilde{D}_1,\Tilde{D}_2)$} & $\widetilde{\mathcal{W}}$ & \tiny{$(d_1,d_2)$} \\
\hline
\hline
1. & $\frac{4}{7}$ & $\left(\frac{1}{7},\frac{3}{7}\right)$ & $1$ & \tiny{(1,1)} & ${\mathcal I}[\mathcal{M}(7,2)]$ & $\frac{52}{7}$ & $\left(\frac{6}{7},\frac{4}{7}\right)$ & $156$ & \tiny{(78,13)} & ${\bf III_{5}}$ & \tiny{$\left(1,1\right)$} \\
2. & $\frac{4}{5}$ & $\left(\frac{1}{5},\frac{2}{5}\right)$ & $2$ & \tiny{(2,1)} & ${\mathcal I}[\mathcal{M}(5,2)^{\otimes 2}]$ & $\frac{36}{5}$ & $\left(\frac{4}{5},\frac{3}{5}\right)$ & $144$ & \tiny{(45,12)} & ${\bf III_{4}}$ & \tiny{$\left(2,1\right)$} \\
3. & $\frac{12}{7}$ & $\left(\frac{2}{7},\frac{3}{7}\right)$ & $6$ & \tiny{(3,2)} & ${\bf III_{1}}$ & $\frac{44}{7}$ & $\left(\frac{5}{7},\frac{4}{7}\right)$ & $88$ & \tiny{(44,11)} & ${\bf III_{3}}$ & \tiny{$\left(1,1\right)$} \\
\hline
\hline
\rowcolor{Mywhite}\caption{IVOA-type pairings, $c^\mathcal{H}=8$ with $(n_1,n_2)=(1,1)$.}
\label{T3}
\end{longtable}

\noindent {\bf Conclusion:} From Table \ref{T3} we conclude that ${\bf III_1}$, ${\bf III_3}$, ${\bf III_4}$ and ${\bf III_5}$ belong to the IVOA-type class as they have negative fusion rules, and that they are paired as in the table.

\subsection{Cosets of $c^{\mathcal{H}}=16$}

We consider coset bilinear relations between $(3,0)$ admissible character solutions with $c^{\cal H} = 16$ meromorphic character $j^\frac23$.  With reference to \eqref{peq3reln} this would correspond to $N = 2$ and to either $n_1 = 1, n_2 = 2$ or $n_1 = 2, n_2 = 1$. Any admissible character solution that is potentially part of such a coset relation has to have a central charge less than $16$. Hence we consider all admissible character solutions from \cite{Das:2021uvd} with $c < 16$. For any one of them, say $\mathcal{W}$ with central charge and conformal dimensions $(c, h_1, h_2)$, we ask if there is an admissible character solution $\widetilde{\mathcal{W}}$ with central charge and conformal dimensions either $(16-c, 1-h_1, 2 - h_2)$ or $(16-c, 2 - h_1, 1 - h_2)$.  For every such pair $(\mathcal{W}, \widetilde{\mathcal{W}})$, we then ask if their characters satisfy a bilinear relation \eqref{bilin0} and compute $(d_1, d_2)$. The resulting pairs of VVMF are listed in tables \ref{T4}, \ref{T5} and \ref{T6}. The tables provide the details first of $\mathcal{W}$, then of $\widetilde{\mathcal{W}}$, followed by $(d_1, d_2)$. 
\subsubsection*{Comments on Table \ref{T4}}

Table \ref{T4} contains $23$ bilinear relations; 22 of them are such that each member of every pair is an affine theory. There is one bilinear relation (row 8) in which one of the pair ($\mathbf{III_2}$) has been characterised in the previous subsection and the other $\mathbf{III_{22}}$ is to be characterised. The solutions of each row  each pair up to a known meromorphic theory at $c=16$, for which there are two choices of the theory $\cH$, namely $E_{8,1} \otimes E_{8,1}$ and ${\cal E}_1[D_{16,1}]$. For short, we refer to these two cases in the last column of the Table as E and D respectively.

Consider rows 1 and 2. These are both coset relations that involve the Ising CFT ${\cal M}(4,3)=B_{0,1}$. Starting from the central charge and conformal dimensions of the Ising CFT $(c = \frac12, h_1 = \frac{1}{16}, h_2 = \frac12)$, one can obtain two potential coset relation partners, one with $n_1 = 1, n_2 = 2$ which gives $B_{15,1}$ and the other with $n_1 = 2, n_2 = 1$ which gives $E_{8,2}$. Row 1 is a special case of Example 3 of section \ref{s3}, with $r = 16$ and hence the meromorphic CFT for this coset relation is ${\cal E}_1[D_{16,1}]$. Row 2 follows from the well-known coset $\frac{E_{8,1} \otimes E_{8,1}}{E_{8,2}} \cong B_{0,1}$ where the denominator is diagonally embedded.

The coset relations in rows $3$, $7$, $10$, $15$, $18$ and $20$ are all special cases of Example 2 of Section \ref{s3}, corresponding to $(r_1, r_2)$ values $(1,15)$, $(2,14)$, $(3,13)$, $(5,11)$, $(6,10)$ and $(7,9)$ respectively. All these rows thus have $d_1 = 1, d_2 = 2$ and D (standing for $\cE_1[D_{16,1}]$) as the entry in the last column. For row 7, notice that $A_{1,1}^{\otimes 2}$ is identical to $D_{2,1}$. Note that all possible $(r_1, r_2)$ pairs with $r_1 + r_2 = 16$ are realised. 

Next we consider row $4$. In fact the bilinear relations in rows $3$ and $4$ involve the same $D_{1,1}$ factor, but the Lie algebra embedding is different. In the former case, $D_1$ is  embedded via the regular maximal embedding: $D_{1} \hookrightarrow D_{1} \otimes D_{15} \hookrightarrow D_{16}$ while in the latter case it is embedded via a different regular maximal embedding: $D_{1} \hookrightarrow D_{1} \otimes A_{15} \hookrightarrow D_{16}$ (see appendix \ref{c1} -- Example 4). This suggests a coset relation (after considering embedding indices) between $D_{1,1}$ and $A_{15,1}$; but  since the latter is a nine-character theory one should expect the coset relation to involve a three-character (and four-primary) extension of it, ${\cal E}_3[A_{15,1}]$.  There is a $c^{{\cal H}} = 24$ meromorphic CFT, the Schellekens CFT  $\# 63$ whose affine sub-algebra is $D_{9,1}\,A_{15,1}$, indicating a coset relation between the three-character $D_{9,1}$ CFT and a three-character extension of $A_{15,1}$, which is in row  $4$ here. This extension was first found, in precisely this way, in \cite{Gaberdiel:2016zke} and hence we denote this here by ${\cal E}_3[A_{15,1}]=\text{GHM}_{255}$ \footnote{We remind the reader that ``GHM'' indicates that the coset was discovered in \cite{Gaberdiel:2016zke}, and the subscript is the dimension of the algebra listed there. ${\mathbf {III}_{xx}},{\mathbf V}_{xx}$ indicates that the pair is taken from \cite{Das:2021uvd} and it is labelled following the conventions used there.}.

The coset relations in rows $5$, $9$, $11$, $14$, $16$, $19$ and $21$ are all special cases of Example 1 of section \ref{s3}, corresponding to $r_1, r_2$ values $(1,14)$, $(2,13)$, $(3,12)$, $(4,11)$, $(5,10)$, $(6,9)$ and $(7,8)$ respectively. All these rows thus have $d_1 = 1, d_2 = 1$ and D as the entry in the last column. Notice that in row 5, $A_{1,2}$ is identical to $B_{1,1}$ and in row 9, $C_{2,1}$ is identical to $B_{2,1}$. Also note that all possible $(r_1, r_2)$ pairs with $r_1 + r_2 = 15$ are realised.

We will study rows $6$, $12$, $17$, $23$ together. In row $6$, we have the folllowing identification: $D_{2,1}\cong A_{1,1}^{\otimes 2}$. Now each of the three-character  bilinear relations in these rows is derived from two-character bilinear relations involving the pairs $(A_{1,1}, E_{7,1})$, $(A_{2,1}, E_{6,1})$, $(G_{2,1}, F_{4,1})$ and $(D_{4,1}, D_{4,1})$ which form coset pairs with $E_{8,1}$ with $d = 1, 2, 1, 3$ respectively \cite{Das:2021uvd}. The last one is a self-coset relation. Denote any of these pairs by $(\mathbf{g}_1, \widetilde{{\bf g}_1})$ with central charge and conformal dimensions $(c, h)$ and  $(\widetilde{c}, \widetilde{h})$, related by $c + \widetilde{c} = 8,~ h + \widetilde{h} = 1$. Now consider the pair of three-character CFTs, $( {\bf g}_1 \otimes {\bf g}_1, \widetilde{{\bf g}}_1 \otimes \widetilde{{\bf g}}_1)$ whose central charges and conformal dimensions are given by $(2 c, h, 2 h)$, $(2 \widetilde{c}, \widetilde{h}, 2 \widetilde{h})$. We have $2c + 2\widetilde{c} = 16$ and $h + \widetilde{h} = 1, 2h + 2\widetilde{h} = 2$, corresponding to the pairings in this table. If we denote the characters of ${\bf g}_1$ by $\chi_0, \chi_1$ and those of $\widetilde{{\bf g}_1}$ by $\widetilde{\chi}_0, \widetilde{\chi}_1$ and the two-character coset relation by $\chi_0 \widetilde{\chi}_0 + d \,\chi_1 \widetilde{\chi}_1 = j^{\frac13}$ then the characters of ${\bf g}_1\otimes {\bf g}_1$ are $\chi_0^2, \chi_0 \chi_1, \chi_1^2$ and those of $\widetilde{{\bf g}}_1 \otimes \widetilde{{\bf g}}_1$ are $\widetilde{\chi}_0^2, \widetilde{\chi}_0 \widetilde{\chi}_1, \widetilde{\chi}_1^2$. A three-character coset relation is obtained by simply squaring the two-character coset relation: $\chi_0^2\, \widetilde{\chi}_0^2 +  2 d \, \chi_0 \chi_1 \widetilde{\chi}_0 \widetilde{\chi}_1 + d^2 \, \chi_1^2 \widetilde{\chi}_1^2 = j^{\frac23}$. We can read off the $(d_1, d_2)$ values for the three-character relation to be $d_1 = 2 d, d_2 = d^2$. Finally, we identify the meromorphic CFT in the three-character coset relation to be the $E_{8,1} \otimes E_{8,1}$ CFT. In terms of Lie algebra embeddings, each factor of ${\bf g} \oplus {\bf g}$ is embedded into a corresponding factor of $E_8 \oplus E_8$. The commutant of ${\bf g} \otimes {\bf g}$ inside  $E_{8} \otimes E_{8}$ is the direct sum of two copies of the commutant of ${\bf g}$ in $E_8$, i.e. $\widetilde{{\bf g}} \oplus \widetilde{{\bf g}}$. All aspects of the coset relations in rows $6$, $12$, $17$ and $23$ are thus explained from two-character coset relations.

One may ask what happens if we embed ${\bf g} \oplus {\bf g}$ into $E_8 \oplus E_8$ with both copies embedded into the same copy of $E_8$  It turns out that such embeddings, when they are possible, are relations between CFTS with $\ell=0$ and $\ell=6$ (recall that $\ell$ is the Wronskian index). When ${\bf g}_1 = D_{4,1}$, we  do not get anything because $D_{4,1}\otimes D_{4,1}$ has a central charge of $8$ and its commutant is trivial. For ${\bf g}_1 = A_{1,1}$, after recognizing that $A_{1,1}\otimes A_{1,1} \cong D_{2,1}$, from the coset relation in row 3 of table \ref{T1}, we can conclude that the coset  would be $D_{6,1} \otimes E_{8,1}$ which is a three-character CFT whose characters are $j^{\frac13}$ times the characters of $D_{6,1}$. This then means that it is an $\ell=6$ CFT. This is one example of the more general rule that, for $n$ characters, the tensor product of an $\ell=0$ CFT with $E_{8,1}$ is an $\ell=2n$ CFT. For  ${\bf g}_1 = A_{2,1}$, we invoke the coset relation in row  8 of table \ref{T1} and obtain the coset to be $A_{2,1}^{\otimes 2} \otimes E_{8,1}$, another $(3,6)$ CFT. Finally for ${\bf g}_1 = G_{2,1}$, we invoke the coset relation in row 9 of table \ref{T1} to conclude that the coset CFT is ${\cal E}_3[A_{1,8}] \otimes E_{8,1}$, an $\ell=6$ CFT. We have thus anticipated three coset relations between $\ell=0$ and $\ell=6$ CFTs, which would be part of a more thorough study of all such coset relations \cite{DGJM:upcoming}.

In row 8 we find a pairing between ${\bf III_{2}}$ and ${\bf III_{22}}$. The first of these characters was identified from Table \ref{T1} above to be $\cE_3[A_{1,8}]$. We find there is an embedding $A_1\hookrightarrow D_{16}$ with embedding index 8, whose commutant is $C_8$ (see appendix \ref{c1} -- Example 5). It follows that ${\bf III_{22}}$ is the three-character extension $\cE_3[C_{8,1}]$. Below we will find independent confirmation of this fact from another embedding.

In row 13, we pair $D_{4,1}$ (two characters, four primaries) with $D_{12,1}$ (three characters, four primaries). This is a slightly unusual example where the two elements of the pair do not have the same number of characters. They do, however, have the same number of primaries and the coset relation is straightforward if we just pair the primaries with unit coefficient for each term. The three non-trivial primaries of $D_{4,1}$ all have $h=\half$, while one of the non-trivial primaries of $D_{12,1}$ has $\tilh=\half$ and the other two have $\tilh=\frac32$. Thus the bilinear relation is:
\be
\begin{split}
\chi_\cH=j(\tau)^\frac23 &=\chi_0\tchi_0+\chi_\half \tchi_\half + \chi_\half \tchi_\frac32 + \chi_\half \tchi_\frac32\\
&= \chi_0\tchi_0+\chi_\half (\tchi_\half + 2 \tchi_\frac32)
\end{split}
\ee

Row 22 is another self-coset relation. It is a special case of Example $2$ of section \ref{s3} with $r_1 = r_2 = 8$ and $r = 16$. The meromorphic CFT is thus ${\cal E}_1[D_{16,1}]$ CFT which is reflected in the last column. The $d_1 = 1, d_2 = 2$ values are also thereby explained. 

\setlength\LTleft{-10pt}
\setlength\LTright{0pt}
\rowcolors{2}{Mygrey}{Mywhite}
\begin{longtable}{l||ccccc||ccccc||cc}
\hline
\hline
\rowcolor{Mywhite}\# & $c$ & $(h_1,h_2)$ & $m_1$ & \tiny{$(D_1,D_2)$} & $\mathcal{W}$ & $\Tilde{c}$ & $(\Tilde{h}_1,\Tilde{h}_2)$ & $\tilde{m}_1$ & \tiny{$(\Tilde{D}_1,\Tilde{D}_2)$} & $\widetilde{\mathcal{W}}$ & \tiny{$(d_1,d_2)$} & $\mathcal{H}$  \\
\hline
\hline
1. & $\frac{1}{2}$ & $\left(\frac{1}{2},\frac{1}{16}\right)$ & $0$ & \tiny{(1,1)} & $B_{0,1}$ & $\frac{31}{2}$ & $\left(\frac{1}{2},\frac{31}{16}\right)$ & $465$ & \tiny{(31,$2^{15}$)} & $B_{15,1}$ & \tiny{$\left(1,1\right)$} & D \\
2. & $\frac{1}{2}$ & $\left(\frac{1}{16},\frac{1}{2}\right)$ & $0$ & \tiny{(1,1)} & $B_{0,1}$ & $\frac{31}{2}$ & $\left(\frac{15}{16},\frac{3}{2}\right)$ & $248$ & \tiny{(248,3875)} & $E_{8,2}$ &  \tiny{$\left(1,1\right)$} & E \\
3. & $1$ & $\left(\frac{1}{2},\frac{1}{8}\right)$ & $1$ & \tiny{(2,1)} & $D_{1,1}$ & $15$ & $\left(\frac{1}{2},\frac{15}{8}\right)$ & $435$ & \tiny{(30,$2^{14}$)} & $D_{15,1}$ & \tiny{$\left(1,2\right)$} & D \\
4. & $1$ & $\left(\frac{1}{8},\frac{1}{2}\right)$ & $1$ & \tiny{(1,2)} & $D_{1,1}$ & $15$ & $\left(\frac{7}{8},\frac{3}{2}\right)$ & $255$ & \tiny{(120,3640)} & $\text{GHM}_{255}$ & \tiny{$\left(2,1\right)$} & D \\
5. & $\frac{3}{2}$ & $\left(\frac{1}{2},\frac{3}{16}\right)$ & $3$ & \tiny{(3,2)} & $B_{1,1}$ & $\frac{29}{2}$ & $\left(\frac{1}{2},\frac{29}{16}\right)$ & $406$ & \tiny{(29,$2^{14}$)} & $B_{14,1}$ & \tiny{$\left(1,1\right)$} & D \\
$6$. & $2$ & $\left(\frac{1}{4},\frac{1}{2}\right)$ & $6$ & \tiny{(2,4)} & $D_{2,1}$ & $14$ & $\left(\frac{3}{4},\frac{3}{2}\right)$ & $266$ & \tiny{(56,$56^{2}$)}  & $E_{7,1}^{\otimes 2}$ & \tiny{$\left(2,1\right)$} & E \\
$7$. & $2$ & $\left(\frac{1}{2},\frac{1}{4}\right)$  & $6$ &  \tiny{(4,2)} & $D_{2,1}$ & $14$ & $\left(\frac{1}{2},\frac{7}{4}\right)$ & $378$ & \tiny{(28,$2^{13}$)} & $D_{14,1}$ & \tiny{$\left(1,2\right)$} & D \\
8. & $\frac{12}{5}$ & $\left(\frac{1}{5},\frac{3}{5}\right)$ & $3$ & \tiny{(3,5)} & ${\bf III_{2}}$ & $\frac{68}{5}$ & $\left(\frac{4}{5},\frac{7}{5}\right)$ & $136$ & \tiny{$(119,68\cdot 25)$} & ${\bf III_{22}}$ & \tiny{$\left(1,2\right)$} & D \\
$9$. & $\frac{5}{2}$ & $\left(\frac{1}{2},\frac{5}{16}\right)$ & $10$ & \tiny{(5,4)} & $B_{2,1}$ & $\frac{27}{2}$ & $\left(\frac{1}{2},\frac{27}{16}\right)$ & $351$ & \tiny{(27,$2^{13}$)} & $B_{13,1}$ & \tiny{$\left(1,1\right)$} & D \\
$10$. & $3$ & $\left(\frac{1}{2},\frac{3}{8}\right)$ & $15$ & \tiny{(6,4)} & $D_{3,1}$ & 13 & $\left(\frac{1}{2},\frac{13}{8}\right)$ & $325$ & \tiny{(26,$2^{12}$)} & $D_{13,1}$ & \tiny{$\left(1,2\right)$} & D \\
$11$. & $\frac{7}{2}$ & $\left(\frac{1}{2},\frac{7}{16}\right)$ & $21$ & \tiny{(7,8)} & $B_{3,1}$ & $\frac{25}{2}$ & $\left(\frac{1}{2},\frac{25}{16}\right)$ & $300$ & \tiny{(25,$2^{12}$)} & $B_{12,1}$ & \tiny{$\left(1,1\right)$} & D \\
$12$. & 4 & $\left(\frac{1}{3},\frac{2}{3}\right)$ & $16$ & \tiny{(3,9)} & $A_{2,1}^{\otimes 2}$ & 12 & $\left(\frac{2}{3},\frac{4}{3}\right)$ & $156$ & \tiny{(27,$27^2$)} & $E_{6,1}^{\otimes 2}$ & \tiny{$\left(4,4\right)$} & E \\
13. & 4 & $\left(\frac{1}{2},\frac{1}{2}\right)$ & $28$ & \tiny{$(8,2^{3})$} & $D_{4,1}$ & 12 & $\left(\frac{1}{2},\frac{3}{2}\right)$ & $276$ & \tiny{(24,$2^{11}$)} & $D_{12,1}$ & \tiny{$(1,2)$} & D \\
$14$. & $\frac{9}{2}$ & $\left(\frac{1}{2},\frac{9}{16}\right)$  & $36$ & \tiny{(9,16)} & $B_{4,1}$ & $\frac{23}{2}$ & $\left(\frac{1}{2},\frac{23}{16}\right)$ & $253$ & \tiny{(23,$2^{11}$)} & $B_{11,1}$ & \tiny{$\left(1,1\right)$} & D \\
$15$. & 5 & $\left(\frac{1}{2},\frac{5}{8}\right)$ & $45$ & \tiny{(10,16)} & $D_{5,1}$ & 11 & $\left(\frac{1}{2},\frac{11}{8}\right)$ & $231$ & \tiny{(22,1024)} & $D_{11,1}$ & \tiny{$\left(1,2\right)$} & D \\
$16$. & $\frac{11}{2}$ & $\left(\frac{1}{2},\frac{11}{16}\right)$ & $55$ & \tiny{(11,32)} & $B_{5,1}$ & $\frac{21}{2}$ & $\left(\frac{1}{2},\frac{21}{16}\right)$ & $210$ & \tiny{(21,1024)} & $B_{10,1}$ & \tiny{$\left(1,1\right)$} & D \\
$17$. & $\frac{28}{5}$ & $\left(\frac{2}{5},\frac{4}{5}\right)$ & $28$ & \tiny{(7,49)} & $G_{2,1}^{\otimes 2}$ & $\frac{52}{5}$ & $\left(\frac{3}{5},\frac{6}{5}\right)$ & $104$ & \tiny{(26,$26^2$)} & $F_{4,1}^{\otimes 2}$ & \tiny{$\left(2,1\right)$} & E \\
$18$. & 6 & $\left(\frac{1}{2},\frac{3}{4}\right)$ & $66$ & \tiny{(12,32)} & $D_{6,1}$ & 10 & $\left(\frac{1}{2},\frac{5}{4}\right)$ & $190$ & \tiny{(20,512)} & $D_{10,1}$ & \tiny{$\left(1,2\right)$} & D \\
$19$. & $\frac{13}{2}$ & $\left(\frac{1}{2},\frac{13}{16}\right)$ & $78$ & \tiny{(13,64)} & $B_{6,1}$ & $\frac{19}{2}$ & $\left(\frac{1}{2},\frac{19}{16}\right)$ & $171$ & \tiny{(19,512)} & $B_{9,1}$ & \tiny{$\left(1,1\right)$} & D \\
$20$. & 7 & $\left(\frac{1}{2},\frac{7}{8}\right)$ & $91$ & \tiny{(14,64)} & $D_{7,1}$ & 9 & $\left(\frac{1}{2},\frac{9}{8}\right)$ & $153$ & \tiny{(18,256)} & $D_{9,1}$ & \tiny{$\left(1,2\right)$} & D \\
$21$. & $\frac{15}{2}$ & $\left(\frac{1}{2},\frac{15}{16}\right)$ & $105$ & \tiny{(15,128)} & $B_{7,1}$ & $\frac{17}{2}$ & $\left(\frac{1}{2},\frac{17}{16}\right)$ & $136$ & \tiny{(17,256)} & $B_{8,1}$ & \tiny{$\left(1,1\right)$} & D \\
22. & 8 & $\left(\frac{1}{2},1\right)$ & $120$ & \tiny{$(16,2^{7})$} & $D_{8,1}$ & 8 & $\left(\frac{1}{2},1\right)$ & $120$ & \tiny{(16,$2^{7}$)} & $D_{8,1}$ & \tiny{$(1,2)$} & D \\
23. & 8 & $\left(\frac{1}{2},1\right)$ & $56$ & \tiny{$(8,2^{6})$} & $D_{4,1}^{\otimes 2}$ & 8 & $\left(\frac{1}{2},1\right)$ & $56$ & \tiny{$(8,2^{6})$} & $D_{4,1}^{\otimes 2}$ & \tiny{$\left(6,9\right)$} & E \\
\hline
\hline
\rowcolor{Mywhite}\caption{CFT pairings, $c^\mathcal{H}=16$ with $(n_1,n_2)=(1,2)$. The meromorphic theory $\mathcal{H}$ in the last column is $E_{8,1}\otimes E_{8,1}$, denoted E, or $\mathcal{E}_1[D_{16,1}]$, denoted D. }
\label{T4}
\end{longtable}

\noindent {\bf Conclusion:} From Table \ref{T4} we have found that the character ${\bf III_{22}}$ should be identified with $\cE_3[C_{8,1}]$. The remaining entries in the table correspond to known CFTs.

\subsubsection*{Comments on Table \ref{T5}}

This table contains 9 pairs that are all of IVOA type, by which we mean some of their fusion rules as computed from the modular $S$-matrix are negative. The third row of Table \ref{T5} displays a dual pair of IVOAs. This pair is inherited from the simpler pair with two characters that combine  to give $E_{8,1}$. Rows $1,2,4$ contain bilinear pairs that combine to the character $j^\frac23$ and one of which in each case is a known IVOA. We would therefore claim that the duals, $\mathbf{III_{29},III_{30},III_{28}}$ are also IVOAs. However the remaining rows $5-9$ contain pairs where neither member is a known CFT or IVOA. In terms of fusion rules (deduced from the modular $S$-matrix) these are all of IVOA type, but we cannot say more about them. In some of these cases, one member of the pair already appeared in Table \ref{T3}, so if one is able to characterise that one using the $c^\cH=8$ duality then it would provide evidence for existence of its partner as an IVOA.

\setlength\LTleft{-5pt}
\setlength\LTright{0pt}
\rowcolors{2}{Mygrey}{Mywhite}
\begin{longtable}{l||ccccc||ccccc||cc}
\hline
\hline
\rowcolor{Mywhite}\# & $c$ & $(h_1,h_2)$ & $m_1$ & \tiny{$(D_1,D_2)$} & $\mathcal{W}$ & $\Tilde{c}$ & $(\Tilde{h}_1,\Tilde{h}_2)$ & $\tilde{m}_1$ & \tiny{$(\Tilde{D}_1,\Tilde{D}_2)$} & $\widetilde{\mathcal{W}}$ & \tiny{$(d_1,d_2)$} \\
\hline
\hline
1. & $\frac{4}{7}$ & $\left(\frac{1}{7},\frac{3}{7}\right)$ & $1$ & \tiny{(1,1)} & \scriptsize{$\cI(\mathcal{M}(7,2))$} & $\frac{108}{7}$ & $\left(\frac{6}{7},\frac{11}{7}\right)$ & $378$ & \tiny{(117,3510)} & ${\bf III_{29}}$ & \tiny{$\left(1,1\right)$} \\
2. & $\frac{4}{7}$ & $\left(\frac{3}{7},\frac{1}{7}\right)$ & $1$ & \tiny{(1,1)} & \scriptsize{$\cI(\mathcal{M}(7,2))$} & $\frac{108}{7}$ & $\left(\frac{4}{7},\frac{13}{7}\right)$ & $456$ & \tiny{(39,20424)} & ${\bf III_{30}}$ & \tiny{$\left(1,1\right)$} \\
3. & $\frac{4}{5}$ & $\left(\frac{1}{5},\frac{2}{5}\right)$ & 2 & \tiny{(1,1)} & \scriptsize{$\cI(\mathcal{M}(5,2)^{\otimes 2})$} & $\frac{76}{5}$ & $\left(\frac{4}{5},\frac{8}{5}\right)$ & $380$ & \tiny{(57,3249)} & $E_{7.5}^{\otimes 2}$ & \tiny{$\left(2,1\right)$} \\
4. & $\frac{4}{5}$ & $\left(\frac{2}{5},\frac{1}{5}\right)$ & $2$ & \tiny{(1,1)} & \scriptsize{$\cI(\mathcal{M}(5,2)^{\otimes 2})$} & $\frac{76}{5}$ & $\left(\frac{3}{5},\frac{9}{5}\right)$ & $437$ & \tiny{$(57,19\cdot 625)$} & ${\bf III_{28}}$ & \tiny{$\left(1,2\right)$} \\
5. & $\frac{12}{7}$ & $\left(\frac{2}{7},\frac{3}{7}\right)$ & $6$ & \tiny{(3,2)} & ${\bf III_{1}}$ & $\frac{100}{7}$ & $\left(\frac{5}{7},\frac{11}{7}\right)$ & $325$ &  \tiny{(55,2925)} & ${\bf III_{24}}$ & \tiny{$\left(1,1\right)$} \\
6. & $\frac{12}{7}$ & $\left(\frac{3}{7},\frac{2}{7}\right)$ & $6$ & \tiny{(2,3)} & ${\bf III_{1}}$ & $\frac{100}{7}$ & $\left(\frac{4}{7},\frac{12}{7}\right)$ & $380$ & \tiny{(55,11495)} & ${\bf III_{25}}$ & \tiny{$\left(1,1\right)$} \\
7. & $\frac{44}{7}$ & $\left(\frac{4}{7},\frac{5}{7}\right)$ & $88$ &  \tiny{(11,44)} & ${\bf III_{3}}$ & $\frac{68}{7}$ & $\left(\frac{3}{7},\frac{9}{7}\right)$ & $221$ & \tiny{(17,782)} & ${\bf III_{12}}$ & \tiny{$\left(1,1\right)$} \\
8. & $\frac{36}{5}$ & $\left(\frac{3}{5},\frac{4}{5}\right)$ & $144$ &  \tiny{(12,45)} & ${\bf III_{4}}$ & $\frac{44}{5}$ & $\left(\frac{2}{5},\frac{6}{5}\right)$ & $220$ & \tiny{$(11,11\cdot 25)$} & ${\bf III_{8}}$ & \tiny{$\left(1,2\right)$} \\
9. & $\frac{52}{7}$ & $\left(\frac{4}{7},\frac{6}{7}\right)$ & 156 & \tiny{(13,78)} & ${\bf III_{5}}$ & $\frac{60}{7}$ & $\left(\frac{3}{7},\frac{8}{7}\right)$ & 210 & \tiny{(10,285)} & ${\bf III_{7}}$ & \tiny{$\left(1,1\right)$} \\
\hline
\hline
\rowcolor{Mywhite}\caption{IVOA-type pairings, $c^\mathcal{H}=16$ with $(n_1,n_2)=(1,2)$. The two sets of characters pair up to $j^\frac23$.}
\label{T5}
\end{longtable}

\noindent {\bf Conclusion:} From table \ref{T5} we conclude that ${\bf III_7}$, ${\bf III_8}$, ${\bf III_{12}}$, ${\bf III_{24}}$, ${\bf III_{25}}$, ${\bf III_{28}}$, ${\bf III_{29}}$ and ${\bf III_{30}}$ are of IVOA-type as these have negative fusion rules and that they are paired as in the table.

\subsubsection*{Comments on Table \ref{T6}}

As for the two previous cases, the two sets of characters in each line of Table \ref{T6} satisfy a bilinear pairing to the character $j^{\frac23}$. We now argue that all the previously uncharacterised solutions that appear in this table are inconsistent as CFTs. For short, we refer to these as ``inconsistent pairings''. This means that, though the VVMFs do pair up into a modular invariant, these are not coset pairs of CFTs.

In rows 1, 2, 4--7 we find known CFT in the left column paired with the characters $\mathbf{III_{26},III_{21},III_{20},III_{19},V_{18}, III_{17}}$ in the right column. In the first five of these cases, the CFTs in the first column also appear in a coset pair in Table \ref{T4}, in lines 5, 9--12 respectively, where they are paired with known CFTs. However here these theories are paired differently and their partners are previously uncharacterised admissible characters. For the sixth case, $A_{4,1}$ does not appear in Table \ref{T4} but only in Table \ref{T6}. The details of the bilinear relation in row 7 suggests that for $\mathbf{III_{17}}$ to be a CFT, it must be based on a Lie subalgebra, $\mfh$, of $D_{16}$\footnote{An embedding of $E_8 \times E_8$ that contains a $A_4$ factor will result in a CFT with Wronskian index $6$.} which has dimension $222$ and that there must exist a ($246$ dimensional) embedding  $A_4 \times \mfh \hookrightarrow D_{16}$. We listed embeddings of $D_{16}$ in decreasing order of dimensions ($496, 384, 380 \ldots$) till a little beyond $246$ and we did not find any with a $A_4$ factor (there is a $256$ dimensional embedding $A_4 \times D_{11}\times D_1$.) We thus conclude that the character $\mathbf{III_{17}}$ does not correspond to a CFT. We will independently confirm this in a slightly simpler way when we come to $c^\cH=24$, in table \ref{T6}. This story for row 7 repeats for each of rows 1, 2, 4, 5 and 6. 

There is another way to rule out solution ${\bf V_{18}}$ in row 6 of table \ref{T6}.  $A_{2,1}^{\otimes 2}$ is known to have nine primaries and three characters; one primary corresponding to the identity character and each of the other two characters correspond to four primaries each. Thus the multiplicities in the partition function are $Y_1 = 4, ~ Y_2 = 4.$ Any CFT which forms a coset relation with $A_{2,1}^{\otimes 2}$ is also expected to have the same partition function multiplicities $\tilde{Y}_1  = 4, ~ \tilde{Y}_2 = 4$ and  the multiplicities in the bilinear identity are expected to be $d_1 = 4, d_2 = 4$. The MLDE analysis \cite{Das:2021uvd} for the admissible character $\mathbf{V_{18}}$ gives the degeneracies $\tilde{D_1} = 1, ~ \tilde{D_2} = 1$ which results in partition function multiplicities $\tilde{Y}_1  = 3^4, ~ \tilde{Y}_2 = 2^2 \cdot 3^{14}$. A reassignment of degeneracies and multiplicities is allowed as long as $\tilde{Y_i}\,\tilde{D_i}^2$ is kept fixed. There is no reassignment with $\tilde{Y_1} = 4$ simply because  $\tilde{Y_1}\tilde{D_1}^2 = 81$ does not have $4$ as a factor. Another inconsistency comes from the details of the bilinear identity given in row no. 6. A reassignment of degeneracies and multiplicities ($d_1, d_2$) is allowed as long as $d_i\tilde{D_i}$ is kept fixed. But there is no reassignment with $d_1 = 4$ simply because  $d_1\,\tilde{D_1} = 18$ does not have $4$ as a factor. Due to all these details, we conclude that the admissible character $\mathbf{V_{18}}$ does not correspond to a CFT. This agrees with the conclusion based on embeddings.

Rows 3 and rows 8-15 of Table \ref{T6} are inconsistent since every $\widetilde{\mathcal{W}}$ in these rows has fractional $Y_1,Y_2$ values. These are the entries of type \sout{${\bf III}$}, \sout{${\bf V}$}.

Row 16 is interesting because both members of the pair are known affine theories. However this is not a consistent bilinear pairing since the coefficient $d_1$ in the bilinear relation is fractional. This enables us to rule it out without even computing $d_2$. There is an important consistency test that explains why this pairing failed. Had it succeeded, there would have been a meromorphic theory at $c=16$ involving an extension of $D_{4,1}^{\otimes 2}D_{8,1}$ with a total of $120+56+128=304$ Kac-Moody generators. Such an extension is known not to exist (since there are just two $c=16$ meromorphic theories, both having 496 Kac-Moody generators) which is why the pairing also should not exist.

\setlength\LTleft{0pt}
\setlength\LTright{0pt}
\rowcolors{2}{Mygrey}{Mywhite}
\begin{longtable}{l||ccccc||ccccc||cc}
\hline
\hline
\rowcolor{Mywhite}\# & $c$ & $(h_1,h_2)$ & $m_1$ & \tiny{$(D_1,D_2)$} & $\mathcal{W}$ & $\Tilde{c}$ & $(\Tilde{h}_1,\Tilde{h}_2)$ & $\tilde{m}_1$ & \tiny{$(\Tilde{D}_1,\Tilde{D}_2)$} & $\widetilde{\mathcal{W}}$ & \tiny{$(d_1,d_2)$}  \\
\hline
\hline
1. & $\frac{3}{2}$ & $\left(\frac{3}{16},\frac{1}{2}\right)$ & $3$ & \tiny{(2,3)} & $B_{1,1}$ & $\frac{29}{2}$ & $\left(\frac{13}{16},\frac{3}{2}\right)$ & $261$ & \tiny{$(29\cdot 4,3393)$} & ${\bf III_{26}}$ & \tiny{$\left(1,1\right)$}  \\
2. & $\frac{5}{2}$ & $\left(\frac{5}{16},\frac{1}{2}\right)$ & $10$ & \tiny{(4,5)} & $B_{2,1}$ & $\frac{27}{2}$ & $\left(\frac{11}{16},\frac{3}{2}\right)$ & $270$ & \tiny{$(27\cdot 2,2871)$} & ${\bf III_{21}}$ & \tiny{$\left(1,1\right)$}  \\
3. & $\frac{12}{5}$ & $\left(\frac{3}{5},\frac{1}{5}\right)$ & 3 & \tiny{(5,3)} & ${\bf III_{2}}$ & $\frac{68}{5}$ & $\left(\frac{2}{5},\frac{9}{5}\right)$ & 374 & \tiny{(119,12138)} & \sout{${\bf III_{23}}$} & \tiny{$\left(\frac{1}{5},1\right)$}  \\
4. & $3$ & $\left(\frac{3}{8},\frac{1}{2}\right)$ & $15$ & \tiny{(4,6)} & $D_{3,1}$ & 13 & $\left(\frac{5}{8},\frac{3}{2}\right)$ & $273$ & \tiny{$(13\cdot 2,325\cdot 8)$} & ${\bf III_{20}}$ & \tiny{$\left(2,1\right)$}  \\
5. & $\frac{7}{2}$ & $\left(\frac{7}{16},\frac{1}{2}\right)$ & $21$ &  \tiny{(8,7)} & $B_{3,1}$ & $\frac{25}{2}$ & $\left(\frac{9}{16},\frac{3}{2}\right)$ & $275$ & \tiny{(25,2325)} & ${\bf III_{19}}$ & \tiny{$\left(1,1\right)$}  \\
6. & 4 & $\left(\frac{2}{3},\frac{1}{3}\right)$ & $16$ & \tiny{(9,3)} & $A_{2,1}^{\otimes 2}$ & 12 & $\left(\frac{1}{3},\frac{5}{3}\right)$ & $318$ & \tiny{(9,4374)} & ${\bf V_{18}}$ & \tiny{$\left(2,2\right)$}  \\
7. & 4 & $\left(\frac{2}{5},\frac{3}{5}\right)$ & $24$ &  \tiny{(5,10)} & $A_{4,1}$ & 12 & $\left(\frac{3}{5},\frac{7}{5}\right)$ & $222$ & \tiny{$(1\cdot 25,51\cdot 25)$} & ${\bf III_{17}}$ & \tiny{$\left(2,2\right)$}  \\
8. & $\frac{9}{2}$ & $\left(\frac{9}{16},\frac{1}{2}\right)$ & 36 & \tiny{(16,9)} & $B_{4,1}$ & $\frac{23}{2}$ & $\left(\frac{7}{16},\frac{3}{2}\right)$ & 276 & \tiny{(23,1771)} & \sout{${\bf III_{16}}$} & \tiny{$\left(\frac{1}{2},1\right)$}  \\
9. & 5 & $\left(\frac{5}{8},\frac{1}{2}\right)$ & 45 & \tiny{(16,10)} & $D_{5,1}$ & 11 & $\left(\frac{3}{8},\frac{3}{2}\right)$ & 275 & \tiny{(11,1496)} & \sout{${\bf III_{15}}$} & \tiny{$\left(1,1\right)$}  \\
10. & $\frac{11}{2}$ & $\left(\frac{11}{16},\frac{1}{2}\right)$ & 55 & \tiny{(32,11)} & $B_{5,1}$ & $\frac{21}{2}$ & $\left(\frac{5}{16},\frac{3}{2}\right)$ & 273 & \tiny{(21,1225)} & \sout{${\bf III_{14}}$} & \tiny{$\left(\frac{1}{4},1\right)$}  \\
11. & 6 & $\left(\frac{3}{4},\frac{1}{2}\right)$ & 66 & \tiny{(32,12)} & $D_{6,1}$ & $10$ & $\left(\frac{1}{4},\frac{3}{2}\right)$ & 270 & \tiny{(5,960)} & \sout{${\bf III_{13}}$} & \tiny{$\left(1,1\right)$}  \\
12. & $\frac{13}{2}$ & $\left(\frac{13}{16},\frac{1}{2}\right)$ & 78 & \tiny{(64,13)} & $B_{6,1}$ & $\frac{19}{2}$ & $\left(\frac{3}{16},\frac{3}{2}\right)$ & 266 & \tiny{(19,703)} & \sout{${\bf III_{11}}$} & \tiny{$\left(\frac{1}{8},1\right)$}  \\
13. & 7 & $\left(\frac{7}{8},\frac{1}{2}\right)$ & 91 & \tiny{(64,14)} & $D_{7,1}$ & 9 & $\left(\frac{1}{8},\frac{3}{2}\right)$ & 261 & \tiny{(9,456)} & \sout{${\bf III_{10}}$} & \tiny{$\left(\frac{1}{4},1\right)$}  \\
14. & $\frac{36}{5}$ & $\left(\frac{4}{5},\frac{3}{5}\right)$ & 144 & \tiny{(45,12)} & ${\bf III_{4}}$ & $\frac{44}{5}$ & $\left(\frac{1}{5},\frac{7}{5}\right)$ & 253 & \tiny{(11,242)} & \sout{${\bf III_{9}}$} & \tiny{$\left(\frac{1}{5},1\right)$}  \\
15. & $\frac{15}{2}$ & $\left(\frac{15}{16},\frac{1}{2}\right)$ & 105 & \tiny{(128,15)} & $B_{7,1}$ & $\frac{17}{2}$ & $\left(\frac{1}{16},\frac{3}{2}\right)$ & 255 & \tiny{(17,221)} & \sout{${\bf III_{6}}$} & \tiny{$\left(\frac{1}{16},1\right)$}  \\
16. & $8$ & $\left(\frac{1}{2},1\right)$ & 56 & \tiny{(8,64)} & $D_{4,1}^{\otimes 2}$ & $8$ & $\left(\frac{1}{2},1\right)$ & 120 & \tiny{(16,$2^7$)} & $D_{8,1}$ & \tiny{$\left(\frac{5}{2},\frac{9}{2}\right)$}  \\
\hline
\hline
\rowcolor{Mywhite}\caption{Inconsistent pairings, $c^\mathcal{H}=16$ with $(n_1,n_2)=(1,2)$. The two sets of characters pair up to $j^\frac23$.}
\label{T6}
\end{longtable}


\noindent {\bf Conclusion:} From table \ref{T6}  we conclude that ${\bf III_{17}}$, ${\bf V_{18}}$, ${\bf III_{19}}$, ${\bf III_{20}}$, ${\bf III_{21}}$ and ${\bf III_{26}}$ are not valid CFTs \footnote{In \cite{Duan:2022ltz} it is claimed that ${\bf III_{17}}$ is a CFT, however we disagree with this.}. 

\subsection{Cosets of $c^{\mathcal{H}}=24$}

\label{label:cosets.24}

With $c=24$, and considering that we are working throughout with Wronskian index $\ell=0$, \eref{datareln} gives us the constraint $n_1+n_2=4$. This can be satisfied in two ways, with $(n_1,n_2)=(2,2)$ or $(1,3)$. Each choice leads to a distinct set of bilinear pairings. We address each class in turn. 

The character of the meromorphic theory to which the two entries in each row pair up, can be written $\chi(\tau)=j(\tau)-744+\cN$. In this way of writing it, $\cN$ is the dimension of the Kac-Moody algebra of the meromophic theory, if any. Below, wherever relevant we provide the serial number(s) in the list of \cite{Schellekens:1992db} which specifies the meromorphic CFT(s) with that $\cN$.

\subsubsection*{\underline{$(n_1,n_2)=(2,2)$}}

This set comprises Tables \ref{T7}, \ref{T8} and \ref{T9}. We discuss each one in turn. There is some overlap between this section and the papers \cite{Bae:2021mej, Duan:2022ltz}. The main focus of the former is fermionic CFT and of the latter,  Hecke relations, and both references present some bilinear pairs of admissible three-character VVMFs. However these references mostly restrict to pairings with total central charge $c^\cH=24$, and moreover the sub-case $(n_1,n_2)=(2,2)$ that we consider in this subsection. In some of these cases the bilinear pairing was used to identify admissible characters as CFTs. Thus there is some overlap between the results of these references and our Table \ref{T7}, which we will point out below.

\subsubsection*{Comments on Table \ref{T7}}

In this table we will go into considerable detail to illustrate the way to correctly choose the degeneracies $D_i$ for the type ${\bf III}$ and ${\bf V}$ characters which, since they were discovered via MLDE, did not
automatically come with a fixed normalisation. We will not be so detailed about this point in the remaining tables.

All cosets in this table are of the form explained in the discussion below \eref{sumbound}, where the coset simply ``deletes''  simple factors (at most two) from a Schellekens theory and leaves behind the remaining simple factors. These cases are labelled as follows: ``GHM'' indicates that the coset was discovered in \cite{Gaberdiel:2016zke}, and the subscript is the dimension of the algebra listed there. ${\mathbf {III}_{xx}},{\mathbf V}_{xx}$ indicates that the pair is taken from \cite{Das:2021uvd} and it is labelled following the conventions used there and reviewed here in Section \ref{background} and in Table \ref{T0}. Rows 22, 23, 18, 20 were for some reason missed in both these references. Interestingly the first two are ``self-cosets'' where $\cC,\tcC$ are the same theory. This implies that $\tcC$ is actually an affine theory rather than an extension of one. 
The table provides the correct degeneracies for both the non-trivial primaries of $\tilde{W}$.

In table \ref{T7} (and in other tables of this paper) we have arranged the coset relations in an increasing order of central charge for the admissible character solution in the left column, so that, naturally the solution on the right has a decreasing central charge, and the self-cosets (if any) are at  the bottom of the table. But it makes sense to discuss the coset relations in a slightly different order. We discuss first the batch of rows $1, 3, 5, 7, 10, 12, 15, 18$, all of which have a $B_{r,1}$ CFT in the left column. Then we discuss the batch of rows $2, 4, 6, 11, 14, 16, 19, 20$ all of which have a $D_{r,1}$ CFT in the left column. Then we discuss row 9 which is a sporadic case. After that we take up the batch of rows $8, 13, 17, 21, 23$ where the CFT in the left column is a tensor product CFT.  This then leaves us with row $22$ which is a self-coset relation. 

The case of row 1 is different from the others: here $\mfh,\mfg$, and consequently also $\tmfh$, are empty. This is the coset pairing of the Ising model, here denoted $\cM(4,3)$, with the Baby Monster CFT \cite{Hoehn:Baby8}. This bilinear pairing was previously studied in \cite{Hampapura:2016mmz}. The latter character was obtained as an admissible character in \cite{Das:2021uvd} with degeneracies $\tilde{D_1} = 4371, ~\tilde{D_2} = 47$ which then results in the multiplicities in the partition function as $\tilde{Y_1} = 1, ~\tilde{Y_2} = 2^{22}$. Requiring that $\tilde{Y_i} \tilde{D_i}^2$ is unchanged we can redefine : $\tilde{D_1} = 4371, ~\tilde{D_2} = 47 \cdot 2^{11}$ and $\tilde{Y_1} = 1, ~\tilde{Y_2} = 1$. These new degeneracies then enter into the computation of the bilinear identity to give the multiplicities there as $d_1 = 1, ~ d_2 = 1$. We thus have a consistent coset relation between to three-primary CFTs. 

In row 3, we have four pairs of coset relations. Each of the theories $\tilde{\mathcal{C}}$ have a common set of characters which were obtained by solving the MLDE in \cite{Das:2021uvd}. The degeneracies of the characters as obtained from the MLDE, for conformal dimensions $\frac32$ ($\tilde{\chi}_{\frac32}$) and $\frac{29}{16}$ ($\tilde{\chi}_{\frac{29}{16}}$) are $\tilde{D_1} = 4785$ and $\tilde{D_2} = 45$ respectively. The multiplicities in the partition function were then computed to be $\tilde{Y_1} = 1, ~\tilde{Y_2} = 2^{20}$. With these degeneracies the bilinear identity then gives multiplicities of $d_1 = 1$ and $d_2 = 1024$ respectively. If we redefine our degeneracies to be $\tilde{D_1} = 4785$ and $\tilde{D_2} = 45 \times 2^{10}$, then the multiplicities would be $d_1 = 1$ and $d_2 = 1$ respectively (which is what we display in the table). With this assignment of degeneracies and multiplicities, we have the interpretation for the coset relation as between two three-primary CFTs. We can justify the above redefinition, for the first of the four theories of row 3, where it can be realised as a three-character extension of $A_{1,2}^{\otimes 15}$. Let us denote the characters of $A_{1,2}$ to be $\chi_0$, $\chi_{\frac12}$ and $\chi_{\frac{3}{16}}$ and note that they have degeneracies of $3$ and $2$ respectively. It turns out that the leading term of $\tilde{\chi}_{\frac32}$ is given by $35 \chi_0^{12} \chi_{\frac12}^3 + 15 \chi_0^7 \chi_{\frac{3}{16}}^8$ from which it follows that the degeneracy is $35 \times 3^3 + 15 \times 2^8 = 4785$. Similarly the leading term of $\tilde{\chi}_{\frac{29}{16}}$ is given by $120 \chi_0^7 \chi_{\frac{3}{16}}^7 \chi_{\frac12}$ which gives it a degeneracy of $120 \times 2^7 \times 3$ which is also equal to $2^{10} \times 45$. Thus, at least for one of the theories of row 3, we have derived the degeneracies that will make the multiplicities to be each equal to $1$. We expect this to hold for the other theories in row 3 as well. Furthermore, the new degeneracies implies that the multiplicities in the partition function are now $\tilde{Y_1} = 1, ~\tilde{Y_2} = 1$, which is consistent with $d_1 = 1, d_2 = 1$.

For the three coset relations in row 5 the degeneracies of the characters obtained from the MLDE are $\tilde{D_1} = 5031$ and $\tilde{D_2} = 43$ respectively. The multiplicities in the partition function was then computed to be $\tilde{Y_1} = 1, ~\tilde{Y_2} = 2^{18}$. With these degeneracies the bilinear identity then gives multiplicities of $d_1 = 1$ and $d_2 = 512$ respectively.  If we redefine our degeneracies to be $\tilde{D_1} = 5031$ and $\tilde{D_2} = 43 \times 2^9$, then the multiplicities would be $d_1 = 1$ and $d_2 = 1$ respectively. Furthermore, these new degeneracies change the multiplicities in the partition function to $\tilde{Y_1} = 1,~ \tilde{Y_2} = 1$ With this assignment of degeneracies and multiplicities, we have the interpretation for the coset relations in row no. 5 as between two three-primary CFTs. This same phenomenon repeats itself in rows nos. 7, 10, 12 and 15. We need to multiply the degeneracy obtained by solving the MLDE, for the character paired with the spinor character, by $2^8$, $2^7$, $2^6$ and $2^5$ respectively. We would then have multiplicities of $1$ and $1$ in each case and consequently the correct interpretation between two three-primary CFTs.  

In row 18, we have a coset relation between two three-primary CFTs both of which are WZW CFTs. The bilinear gives rise to  meromorphic theory $\#62$ in the list of \cite{Schellekens:1992db} which is a non-lattice theory. This case was in fact the basis for the prediction in \cite{Das:2022slz} of an infinite series of non-lattice meromorphic theories at increasing central charges, and is the $m=0$ case of entry $\#15$ in Table 3 of that reference. Similarly, the bilinear relations in rows 1, 3, 5, 7, 10, 12 and 15 were the basis for the prediction in \cite{Das:2022slz} of $14$ infinite series of non-lattice meromorphic theories at increasing central charges, corresponding to entries $\#1 - \#14$ in Table 3 there.

For the coset relation in row 2, the MLDE computations for the degeneracies are $\tilde{D_1} = 575, \tilde{D_2} = 23$ which gives the degeneracies in the partition function to be $\tilde{Y_1} = 64, ~\tilde{Y_2} = 2^{23}$. With these degeneracies the bilinear identity then gives multiplicities of  $d_1 = 8, d_2 = 4096.$  The MLDE and the bilinear identity are also consistent with the following new assignment viz. $\tilde{D_1} = 575 \times 8 , \tilde{D_2} = 23 \times 2^{11}$ and $d_1 = 1, d_2 = 2$. This new assignment of the degeneracies results in the following partition function multiplicities: $\tilde{Y_1} = 1, \tilde{Y_2} = 2$. Row 2 is thus a coset relation between two four-primary CFTs.) More significantly, we can now conclude that the admissible character solution $\mathbf{III_{50}}$ corresponds to a genuine CFT,  a three-character extension of $D_{1,1}^{\otimes 23}$. Thus the coset relation in row $2$ has resulted in the discovery of a new CFT.

In row 4, we have a coset bilinear relation between $D_{2,1}$ and $\mathbf{III_{45}}$. We are able to redefine the degeneracies to obtain partition function multiplicities to be $\tilde{Y_1} = 1,~ \tilde{Y_2} = 2$ and the parameters in the bilinear identity to be $d_1 = 1, ~ d_2 = 2$, indicating a pairing between two four-primary CFTs. There are six meromorphic theories, $\#15$ - $\#20$,  of \cite{Schellekens:1992db} with $D_{2,1} = A_{1,1}^{\otimes2}$ as a factor of the affine part of their chiral algebras, which means that  $\mathbf{III_{45}}$  corresponds to six different CFTs. Each of these are three-character extensions of the remaining factors of the affine part of the chiral algebras, viz. $A_{1,1}^{\otimes 22}$, $A_{3,2}^{\otimes 4}A_{1,1}^{\otimes 2}$, $A_{5,3}D_{4,3}A_{1,1}$, $A_{7,4}A_{1,1}$, $D_{5,4}C_{3,2}$ and $D_{6,5}^{\otimes 22}$ respectively.  Thus, the coset relation in row  4 has resulted in the discovery of six new CFTs. In rows 6, 11, 14, 16, 19, we have bilinear relations between $D_{3,1}, D_{5,1}, D_{6,1}, D_{7,1}, D_{9,1}$ on the left  with CFTs already discovered in \cite{Gaberdiel:2016zke}.  What we are able to do new here is give exact details of the characters: the degeneracies of the non-identity characters that lead to partition function multiplicities $\tilde{Y_1} = 1,~ \tilde{Y_2} = 2$ and the multiplicities in the bilinear identity to be  $d_1 = 1, d_2 = 2$. Thus each of these rows describe pairings between four-primary CFTs.

In row  20, we have a  coset relation between two  four-primary CFTs both of which are WZW CFTs. The bilinear gives rise to meromorphic theory $\#64$ in the list of \cite{Schellekens:1992db}. This case was in fact the basis for the prediction in \cite{Das:2022slz} of an infinite series of meromorphic theories at increasing central charge, and is the $m=0$ case of  entry $\#33$ in Table 3 of that reference. Similarly, the bilinear relations in rows 2, 4, 6, 11, 14, 16, and 19 were the basis for the prediction in \cite{Das:2022slz} of $17$ infinite series of meromorphic theories at increasing central charges, corresponding to entries $\#16-\#32$ in Table 3 there.

Row 9 is a bilinear relation between $A_{4,1}$ and a CFT already discovered in \cite{Gaberdiel:2016zke}. Again what we do new here is to give exact details of the characters: the degeneracies $\tilde{D_1}, \tilde{D_2}$ that lead to partition function multiplicities $\tilde{Y_1} = 2, ~\tilde{Y_2} = 2$ and the multiplicities in the bilinear identity to be $d_1 = 2, ~ d_2 = 2$. This establishes a pairing between two five-primary CFTs. 

We now study bilinear relations where one of the solutions is the three-character CFT obtained from a tensor product of two copies of two-character CFTs. There are $7$ such CFTs viz. $A_{1,1}^{\otimes2}$, $A_{2,1}^{\otimes2}$, $G_{2,1}^{\otimes2}$, $D_{4,1}^{\otimes2}$, $F_{4,1}^{\otimes2}$, $E_{6,1}^{\otimes2}$ and $E_{7,1}^{\otimes2}$. The first has been studied in row 4 (as $D_{2,1}$) and the last one  in row 20. The remaining five are in rows 8, 13, 17, 21 and 23 and for some reason these were missed out in \cite{Gaberdiel:2016zke}. 

In row 8, we have a bilinear relation between $A_{2,1}^{\otimes2}$ and $\mathbf{V_{39}}$. The former is a nine-primary theory with multiplicities $Y_1 = 4, ~Y_2 = 4.$  We are able to obtain an assignment of degeneracies for the latter so the partition function multiplicities are $\tilde{Y_1} = 4, ~\tilde{Y_2} = 4$ and the multiplicities in the bilinear identity are $d_1 = 4, ~ d_2 = 4$, so that we have a pairing between two nine-primary CFTs. Furthermore we find three meromorphic CFTs in \cite{Schellekens:1992db} viz. $\#24$, $\#26$ and $\#27$ that contain a factor of $A_{2,1}^{\otimes 2}$, giving rise to  three new CFTs that are the three-character extensions of the remaining factors viz. $A_{2,1}^{\otimes 10}$, $A_{5,2}^{\otimes 2}C_{2,1}$ and $A_{8,3}$ respectively. Thus the coset relation in row 8 has enabled us to characterize the MLDE solution $\mathbf{V_{39}}$ as corresponding to three CFTs. 

In row 13, we have a bilinear relation between $G_{2,1}^{\otimes2}$ and $\mathbf{III_{37}}$. The former is a four-primary theory with multiplicities $Y_1 = 2, ~Y_2 = 1.$  We are able to obtain an assignment of degeneracies for the latter so the partition function multiplicities are $\tilde{Y_1} = 2, ~\tilde{Y_2} = 1$ and the multiplicities in the bilinear identity are $d_1 = 2, ~ d_2 = 1$, so that we have a pairing between two four-primary CFTs. Furthermore we find a meromorphic CFT in \cite{Schellekens:1992db} viz. $\#32$ that contains a factor of $G_{2,1}^{\otimes 2}$, giving rise to  a new CFT that is a three-character extension of the remaining factors, namely $E_{6,3}G_{2,1}$. Thus the coset relation in row 13 has enabled us to characterize the MLDE solution $\mathbf{III_{37}}$ as corresponding to a genuine CFT. 

Row 17 is a bilinear relation between $D_{4,1}^{\otimes2}$ and an admissible character solution that is one of the infinite family of solutions given in table \ref{T00} viz. ${\bf III''}$ with $m_1 = 112$. The former is a sixteen-primary theory with multiplicities $Y_1 = 6, ~Y_2 = 9.$  We are able to obtain an assignment of degeneracies for the latter so the partition function multiplicities are $\tilde{Y_1} = 6, ~\tilde{Y_2} = 9$ and the multiplicities in the bilinear identity are $d_1 = 5, ~ d_2 = 9$, so that we have a pairing between two sixteen-primary CFTs. Furthermore we find a meromorphic CFT in \cite{Schellekens:1992db} namely $\#42$ that contains a factor of $D_{4,1}^{\otimes 2}$, giving rise to  a new CFT that is a three-character extension of the remaining factor  $D_{4,1}^{\otimes 4}$. Thus the coset relation in row 17 has enabled us to characterise the MLDE solution  $\mathbf{III''}$ with $m_1 = 112$ of Table \ref{T00} as  a sixteen-primary CFT, denoted $\mathcal{E}_3[D_{4,1}^{\otimes 4}]$. This is the CFT with the largest number of primaries but just three characters in this paper that is not a tensor product theory (of course  $D_{4,1}^{\otimes2}$ has the same properties, but it is a tensor product). 

In row 21 we find a coset relation involving $F_{4,1}^2$. The unique meromorphic theory with this factor at $c=24$ is $\#52$ of  \cite{Schellekens:1992db} with Kac-Moody algebra $F_{4,1}^2C_{8,1}$. This proves that ${\bf III_{22}}$ is equivalent to $\cE_3[C_{8,1}]$, confirming the result obtained from Table \ref{T4}.

In row 23 we find a self-coset relation for $E_{6,1}^{\otimes 2}$. This is a pairing between nine-primary CFTs.  This comes about because of  the existence of a meromorphic theory in \cite{Schellekens:1992db} namely $\#58$ which is the extension $\cE_1[E_{6,1}^{\otimes 4}]$. 

Row 22 is a self-coset relation involving $D_{12,1}$; the meromorphic theory is $\#66$ of \cite{Schellekens:1992db} which is the extension  $\cE_1[D_{12,1}^{\otimes 2}]$. Even though the $D_{12,1}$ is a four-primary theory, the pairing of characters is not the usual one which gives the bilinear identity parameters $(d_1,d_2)=(1,2)$ or $(2,1)$. Instead we have $(d_1,d_2) = (1,1)$. This has been explained in the discussion around \eref{newpairing}. 
This unusual coset pairing will appear in our future tables between $D_{4k,1}$ theories with odd $k$, at $c^{\cH}>24$ and whenever $n_1,n_2\neq 1$. 

\setlength\LTleft{-30pt}
\setlength\LTright{0pt}
\begin{longtable}{l||ccccc||ccccc||c||c}
\hline
\hline
\rowcolor{Mywhite}\# & $c$ & $(h_1,h_2)$ & $m_1$ & \tiny{$(D_1,D_2)$} & $\mathcal{C}$ & $\Tilde{c}$ & $(\Tilde{h}_1,\Tilde{h}_2)$ & $\tilde{m}_1$ & \tiny{$(\Tilde{D}_1,\Tilde{D}_2)$} & $\Tilde{\mathcal{C}}$ & \tiny{$(d_1, d_2)$} & S\# \\
\hline
\hline
\rowcolor{Mywhite}1. & $\frac{1}{2}$ & $(\frac{1}{2},\frac{1}{16})$ & $0$ & \tiny{$(1,1)$} & $B_{0,1}$ & $\frac{47}{2}$ & $(\frac{3}{2},\frac{31}{16})$ & $0$ & \tiny{$(4371,47\cdot 2^{11})$} & $\text{BM}$ & \tiny{$(1,1)$} & 0 \\
\rowcolor{Mygrey}2. & $1$ & $(\frac{1}{2},\frac{1}{8})$ & $1$ & \tiny{$(2,1)$} & $D_{1,1}$ & $23$ & $(\frac{3}{2},\frac{15}{8})$ & $23$ & \tiny{$(4600,23\cdot 2^{11})$} & ${\bf III_{50}}$ & \tiny{$(1,2)$} & 1  \\
\rowcolor{Mywhite}3. & $\frac{3}{2}$ & $(\frac{1}{2},\frac{3}{16})$ & $3$ & \tiny{$(3,2)$} & $B_{1,1}$ & $\frac{45}{2}$ & $(\frac{3}{2},\frac{29}{16})$ & $45$ & \tiny{$(4785,45\cdot 2^{10})$} & $\text{GHM}_{45}$ & \tiny{$(1,1)$} & 5 \\
\rowcolor{Mywhite} &  &  &  &  &  &  &  &  &  &  &  & 7 \\
\rowcolor{Mywhite} &  &  &  &  &  &  &  &  &  &  &  & 8 \\
\rowcolor{Mywhite} &  &  &  &  &  &  &  &  &  &  &  & 10 \\
\rowcolor{Mygrey}4. & $2$ & $(\frac{1}{2},\frac{1}{4})$ & $6$ & \tiny{$(4,2)$} & $D_{2,1}$ & $22$ & $(\frac{3}{2},\frac{7}{4})$ & $66$ & \tiny{$(77\cdot 2^6, 11\cdot 2^{11})$} & ${\bf III_{45}}$ & \tiny{$(1,2)$} & 15  \\
\rowcolor{Mygrey} &  &  &  &  &  &  &  &  &  &  &  & 16 \\
\rowcolor{Mygrey} &  &  &  &  &  &  &  &  &  &  &  & 17 \\
\rowcolor{Mygrey} &  &  &  &  &  &  &  &  &  &  &  & 18 \\
\rowcolor{Mygrey} &  &  &  &  &  &  &  &  &  &  &  & 19 \\
\rowcolor{Mygrey} &  &  &  &  &  &  &  &  &  &  &  & 20  \\
\rowcolor{Mywhite}5. & $\frac{5}{2}$ & $(\frac{1}{2},\frac{5}{16})$ & $10$ & \tiny{$(5,4)$} & $B_{2,1}$ & $\frac{43}{2}$ & $(\frac{3}{2},\frac{27}{16})$ & $86$ & \tiny{$(5031,43\cdot 2^9)$} & $\text{GHM}_{86}$ & \tiny{$(1,1)$} & 25  \\
\rowcolor{Mywhite} &  &  &  &  &  &  &  &  &  &  &  & 26  \\
\rowcolor{Mywhite} &  &  &  &  &  &  &  &  &  &  &  & 28  \\
\rowcolor{Mygrey}6. & $3$ & $(\frac{1}{2},\frac{3}{8})$ & $15$ & \tiny{$(6,4)$} & $D_{3,1}$ & $21$ & $(\frac{3}{2},\frac{13}{8})$ & $105$ & \tiny{$(5096,21\cdot 2^9)$} & $\text{GHM}_{105}$ & \tiny{$(1,2)$} & 30  \\
\rowcolor{Mygrey} &  &  &  &  &  &  &  &  &  &  &  & 31  \\
\rowcolor{Mygrey} &  &  &  &  &  &  &  &  &  &  &  & 33  \\
\rowcolor{Mygrey} &  &  &  &  &  &  &  &  &  &  &  & 34  \\
\rowcolor{Mygrey} &  &  &  &  &  &  &  &  &  &  &  & 35 \\  
\rowcolor{Mywhite}7. & $\frac{7}{2}$ & $(\frac{1}{2},\frac{7}{16})$ & $21$ & \tiny{$(7,8)$} & $B_{3,1}$ & $\frac{41}{2}$ & $(\frac{3}{2},\frac{25}{16})$ & $123$ & \tiny{$(5125,41\cdot 2^8)$} & $\text{GHM}_{123}$ & \tiny{$(1,1)$} & 39  \\
\rowcolor{Mywhite} &  &  &  &  &  &  &  &  &  &  &  & 40  \\
\rowcolor{Mygrey}8. & $4$ & $(\frac{1}{3},\frac{2}{3})$ & $16$ & \tiny{$(3,9)$}& $A_{2,1}^{\otimes 2}$ & $20$ & $(\frac{5}{3},\frac{4}{3})$ & $80$ & \tiny{$(4\cdot3^7, 5\cdot 3^5)$} & ${\bf V_{39}}$ & \tiny{$(4,4)$} & 24  \\
\rowcolor{Mygrey} &  & & &  &  &  &  &  &  &  &  & 26  \\
\rowcolor{Mygrey} &  & & & &  &  &  &  &  &  &  & 27  \\
\rowcolor{Mywhite}9. & $4$ & $(\frac{2}{5},\frac{3}{5})$ & $24$ & \tiny{$(5, 10)$} & $A_{4,1}$ & $20$ & $(\frac{8}{5},\frac{7}{5})$ & $120$ & \tiny{$(13 \cdot 5^4, 4\cdot 5^4)$} & $\text{GHM}_{120}$  & \tiny{$(2, 2)$} & 37  \\
\rowcolor{Mygrey}10. & $\frac{9}{2}$ & $(\frac{1}{2},\frac{9}{16})$ & $36$ & \tiny{$(9,16)$} & $B_{4,1}$ & $\frac{39}{2}$ & $(\frac{3}{2},\frac{23}{16})$ & $156$ & \tiny{$(5083,39\cdot 2^7)$} & $\text{GHM}_{156}$ & \tiny{$(1,1)$} & 47 \\
\rowcolor{Mygrey} &  &  &  &  &  &  &  &  &  &  &  & 48  \\
\rowcolor{Mywhite}11. & $5$ & $(\frac{1}{2},\frac{5}{8})$ & $45$ & \tiny{$(10,16)$} & $D_{5,1}$ & $19$ & $(\frac{3}{2},\frac{11}{8})$ & $171$  & \tiny{$(5016,19 \cdot 2^7)$} & $\text{GHM}_{171}$ & \tiny{$(1,2)$} & 49  \\

\rowcolor{Mygrey}12. & $\frac{11}{2}$ & $(\frac{1}{2},\frac{11}{16})$ & $55$ & \tiny{$(11,32)$} & $B_{5,1}$ & $\frac{37}{2}$ & $(\frac{3}{2},\frac{21}{16})$ & $185$ & \tiny{$(4921,37\cdot 2^6)$} & $\text{GHM}_{185}$ & \tiny{$(1,1)$} & 53 \\
\rowcolor{Mywhite}13. & $\frac{28}{5}$ & $(\frac{2}{5},\frac{4}{5})$ & $28$ & \tiny{$(7, 49)$}& $G_{2,1}^{\otimes 2}$ & $\frac{92}{5}$ & $(\frac{8}{5},\frac{6}{5})$ & $92$ & \tiny{$(7475, 1196)$} & ${\bf III_{37}}$ & \tiny{$(2,1)$} & 32  \\ 
\rowcolor{Mygrey}14. & $6$ & $(\frac{1}{2},\frac{3}{4})$ & $66$ & \tiny{$(12,32)$} & $D_{6,1}$ & $18$ & $(\frac{3}{2},\frac{5}{4})$ & $198$ & \tiny{$(75 \cdot 2^6,9 \cdot 2^7)$} & $\text{GHM}_{198}$ & \tiny{$(1,2)$} & 54  \\
\rowcolor{Mygrey} &  &  & & &  &  &  &  &  &  &  & 55 \\
\rowcolor{Mywhite}15. & $\frac{13}{2}$ & $(\frac{1}{2},\frac{13}{16})$ & $78$ & \tiny{$(13,64)$} & $B_{6,1}$ & $\frac{35}{2}$ & $(\frac{3}{2},\frac{19}{16})$ & $210$ & \tiny{$(4655,35\cdot 2^5)$} & $\text{GHM}_{210}$ & \tiny{$(1,1)$} & 56 \\
\rowcolor{Mygrey}16. & $7$ & $(\frac{1}{2},\frac{7}{8})$ & $91$ & \tiny{$(14,64)$} & $D_{7,1}$ & $17$ & $(\frac{3}{2},\frac{9}{8})$ & $221$ & \tiny{$(4488, 544)$} & $\text{GHM}_{221}$ & \tiny{$(1,2)$} & 59  \\ 
\rowcolor{Mywhite}17. & $8$ & $\left(\frac{1}{2},1\right)$ & 56 & \tiny{(8,64)} & $D_{4,1}^{\otimes 2}$ & $16$ & $\left(\frac{3}{2},1\right)$ & 112 & \tiny{($2^{11}$,$2^7$)} & $\cE_3[D_{4,1}^{\otimes 4}]$ & \tiny{$\left(6,9\right)$} & 42 \\
\rowcolor{Mygrey}18. & $\frac{17}{2}$ & $(\frac{1}{2},\frac{17}{16})$ & $136$ & \tiny{$(17,256)$} & $B_{8,1}$ & $\frac{31}{2}$ & $(\frac{3}{2},\frac{15}{16})$ & $248$ & \tiny{$(3875,248)$} & $E_{8,2}$ & \tiny{$(1,1)$} & 62  \\
\rowcolor{Mywhite}19. & $9$ & $(\frac{1}{2},\frac{9}{8})$ & $153$ & \tiny{$(18, 256)$} & $D_{9,1}$ & $15$ & $(\frac{3}{2},\frac{7}{8})$ & $255$ & \tiny{$(3640, 120)$} & $\text{GHM}_{255}$ & \tiny{$(1,2)$} & 63  \\ 
\rowcolor{Mygrey}20. & $10$ & $(\frac{1}{2},\frac{5}{4})$ & $190$ &  \tiny{$(20,512)$}& $D_{10,1}$ & $14$ & $(\frac{3}{2},\frac{3}{4})$ & $266$ & \tiny{$(56^2, 56)$}& $E_{7,1}^{\otimes 2}$ & \tiny{$(1,2)$} & 64  \\
\rowcolor{Mywhite}21. & $\frac{52}{5}$ & $(\frac{3}{5},\frac{6}{5})$ & $104$ & \tiny{$(26, 26^2)$} & $F_{4,1}^{\otimes 2}$ & $\frac{68}{5}$ & $(\frac{7}{5},\frac{4}{5})$ & $136$ & \tiny{$(1700, 119)$} & ${\bf III_{22}}$ & \tiny{$(2,1)$} & 52  \\
\rowcolor{Mygrey}22. & $12$ & $(\frac{1}{2},\frac{3}{2})$ & $276$ & \tiny{$(24,2^{11})$} &$D_{12,1}$ & $12$ & $(\frac{3}{2},\frac{1}{2})$ & $276$ & \tiny{$(2^{11},24)$} & $D_{12,1}$ & \tiny{$(1,1)$} & 66  \\
\rowcolor{Mywhite}23. & $12$ & $(\frac{2}{3},\frac{4}{3})$ & $156$ & \tiny{$(27,27^2)$} & $E_{6,1}^{\otimes 2}$ & $12$ & $(\frac{4}{3},\frac{2}{3})$ & $156$ & \tiny{$(27^2,27)$} & $E_{6,1}^{\otimes 2}$ & \tiny{$(4,4)$} & 58  \\
\hline
\hline
\rowcolor{Mywhite}\caption{CFT pairings, $c^\mathcal{H}=24$ with $(n_1,n_2)=(2,2)$. The entry in the last column identifies the meromorphic theory by its row number in the table of \cite{Schellekens:1992db}.}
\label{T7}
\end{longtable}

\noindent {\bf Conclusion:} From Table \ref{T7} we have deduced the following new identifications for type ${\bf III}$ and ${\bf V}$ solutions:
\be
\begin{split}
&{\bf III_{37}}=\mathcal{E}_3[E_{6,3}G_{2,1}]\\
&{\bf V_{39}}=
\cE_3[A_{2,1}^{\otimes 10}],~\cE_3[A_{5,2}^{\otimes 2}C_{2,1}],~\cE_3[A_{8,3}]\\
&{\bf III_{45}}=\mathcal{E}_3[A_{1,1}^{\otimes 22}],~
\mathcal{E}_3[A_{3,2}^{\otimes 4}A_{1,1}^{\otimes 2}],~
\mathcal{E}_3[A_{5,3}D_{4,3}A_{1,1}],~
\mathcal{E}_3[A_{7,4}A_{1,1}],
~\mathcal{E}_3[D_{5,4}C_{3,2}],~
\mathcal{E}_3[D_{6,5}]
\\
&{\bf III_{50}}=\mathcal{E}_3[D_{1,1}^{\otimes 23}] 
\\
&{\bf III''}(m_1 = 112)=\mathcal{E}_3[D_{4,1}^{\otimes 4}]
\label{idents}
\end{split}
\ee
We also confirm the conclusion from Table \ref{T4} that ${\bf III_{22}}$ is identified with $\cE_3[C_{8,1}]$. We note here that the above identifications, with the exception of ${\bf III''}(m_1=112)$, have been made in \cite{Duan:2022ltz}. 

Let us briefly comment on the three-character extension $\cE_3[D_{4,1}^{\otimes 4}]$ at $c=16$. Though we had excluded $c=8,16$ solutions from the classification at the outset, we felt it worth noting the existence of this one at $c=16$, since it is of the ``GHM'' type \cite{Gaberdiel:2016zke}.

\subsubsection*{Comments on Table \ref{T8}}

This table has several bilinear pairs that we have shown to be of IVOA type. However in a number of cases (rows 1,  5--9) the pairing does not lead to a valid meromorphic CFT as it does not correspond to any entry in \cite{Schellekens:1992db}. In three cases, however, the pairing does reproduce a meromorphic theory -- these are rows $2,3,4$. These examples appear more favourable for identification of the pair as some variant of CFTs.


\setlength\LTleft{-40pt}
\setlength\LTright{0pt}
\rowcolors{2}{Mygrey}{Mywhite}
\begin{longtable}{l||ccccc||ccccc||ccc} 
\hline
\hline
\rowcolor{Mywhite}\# & $c$ & $(h_1,h_2)$ & $m_1$ & \tiny{$(D_1,D_2)$} & $\mathcal{W}$ & $\Tilde{c}$ & $(\Tilde{h}_1,\Tilde{h}_2)$ & $\tilde{m}_1$ & \tiny{$(\Tilde{D}_1,\Tilde{D}_2)$} & $\widetilde{\mathcal{W}}$ & \tiny{$(d_1, d_2)$} & $\mathcal{N}$ & S\#   \\
\hline
\hline
1. & $\frac{4}{7}$ & $(\frac{1}{7},\frac{3}{7})$ & $1$ & \tiny{$(1,1)$} &   \scriptsize${\mathcal{I}[\mathcal{M}(7,2)]}$ & $\frac{164}{7}$ & $(\frac{13}{7},\frac{11}{7})$ & $41$ & \tiny{$(50922,4797)$} & ${\bf III_{54}}$ & \tiny{(1,1)} & 42 & ---   \\
2. & $\frac{4}{5}$ & $(\frac{1}{5},\frac{2}{5})$ & $2$ & \tiny{$(2,1)$} &   \scriptsize${\mathcal{I}[\mathcal{M}(5,2)^{\otimes 2}]}$ & $\frac{116}{5}$ & $(\frac{9}{5},\frac{8}{5})$ & $58$ & \tiny{$(27550,4959)$} & ${\bf III_{52}}$ & \tiny{(2,1)} & 60 & 12-14   \\
3. & $\frac{12}{7}$ & $(\frac27,\frac37)$ & $6$ & \tiny{$(3,2)$} &  ${\bf III_{1}}$ & $\frac{156}{7}$ & $(\frac{12}{7},\frac{11}{7})$ & $78$ & \tiny{$(27170,5070)$} & ${\bf III_{47}}$ & \tiny{(1,1)} & 84 & 22-23   \\
4. & $\frac{44}{7}$ & $(\frac{4}{7},\frac{5}{7})$ & $88$ & \tiny{$(11,44)$} & ${\bf III_{3}}$ & $\frac{124}{7}$ & $(\frac{10}{7},\frac{9}{7})$ & $248$ & \tiny{$(2108,2108)$} & ${\bf III_{35}}$ & \tiny{$(1,1)$} & 336 & 60 \\
5. & $\frac{36}{5}$ & $(\frac{3}{5},\frac{4}{5})$ & $144$ & \tiny{$(12,45)$} &  ${\bf III_{4}}$ & $\frac{84}{5}$ & $(\frac{7}{5},\frac{6}{5})$ & $336$ & \tiny{$(1452,770)$} & ${\bf III_{33}}$ & \tiny{(1,2)} & 480 & ---    \\
6. & $\frac{52}{7}$ & $(\frac{4}{7},\frac{6}{7})$ & $156$ & \tiny{$(13,78)$} & ${\bf III_{5}}$ & $\frac{116}{7}$ & $(\frac{10}{7},\frac{8}{7})$ & $348$ & \tiny{$(1972,725)$} & ${\bf III_{32}}$ & \tiny{(1,1)} & 504 & ---   \\
7. & $\frac{60}{7}$ & $(\frac{3}{7},\frac{8}{7})$ & $210$ & \tiny{$(10,285)$} & ${\bf III_{7}}$ & $\frac{108}{7}$ & $(\frac{11}{7},\frac{6}{7})$ & $378$ & \tiny{$(3510,117)$} & ${\bf III_{29}}$ & \tiny{(1,1)} & 588 & ---   \\
8. & $\frac{44}{5}$ & $(\frac{2}{5},\frac{6}{5})$ & $220$ & \tiny{$(11,275)$} & ${\bf III_{8}}$ & $\frac{76}{5}$ & $(\frac{8}{5},\frac{4}{5})$ & $380$ & \tiny{$(3249,57)$} & ${\bf III_{27}}$ & \tiny{(1,2)} & 600 & ---   \\
9. & $\frac{68}{7}$ & $(\frac{3}{7},\frac{9}{7})$ & $221$ & \tiny{$(17,782)$} & ${\bf III_{12}}$ & $\frac{100}{7}$ & $(\frac{11}{7},\frac{5}{7})$ & $325$ & \tiny{$(2925,55)$} & ${\bf III_{24}}$ & \tiny{(1,1)} & 546 & ---   \\
\hline
\hline
\rowcolor{Mywhite}\caption{IVOA-type pairings, $c^\mathcal{H}=24$ with $(n_1,n_2)=(2,2)$. Wherever present, the entry in the last column identifies
the meromorphic theory by its row number in the table of [40].}
\label{T8}
\end{longtable}

\noindent {\bf Conclusion:} From table \ref{T8} we conclude that ${\bf III_{27}}$, ${\bf III_{32}}$, ${\bf III_{33}}$, ${\bf III_{35}}$, ${\bf III_{47}}$, ${\bf III_{52}}$ and ${\bf III_{54}}$ belong to the IVOA category. These have at least one negative fusion rule, and the above pairings are always between two such solutions.

\subsubsection*{Comments on Table \ref{T9}}

This table lists all the pairs where one can rule out at least one member  being a CFT, or in several cases both members. For rows 1, 4--11, 13--16, the solution in the second column should arise as the commutant of some embedding of the known algebra in the first column in a meromorphic theory. However there is no candidate meromorphic theory for these cases, since the value of the integer denoting the constant term in the meromorphic character $\chi(\tau)=j(\tau)-744+\cN$ does not appear in any entry of the table in \cite{Schellekens:1992db}. This immediately rules out the solution in the right column of every case from being a CFT. 

In some of these cases, namely rows 10, 11, 13--16, the entry in the right column was already ruled out by considerations of non-integral multiplicities $(d_1,d_2)$. That leaves rows 1, 4--9 where we can now rule out the solutions in the right column, namely ${\bf III_{42}}$, ${\bf III_{26}}$, ${\bf III_{21}}$, ${\bf III_{20}}$,
${\bf III_{19}}$, ${\bf V_{18}}$, ${\bf III_{17}}$. The last six of these were already ruled out by Table \ref{T6}, a nice confirmation of the internal consistency of our method. Notice that the reasons for ruling out these six solutions are slightly different in the two tables -- in Table \ref{T6}, the pairings gave a sensible character $j^\frac23$ that actually describes two distinct meromorphic CFT, but there was no possible embedding to justify the coset relation and this ruled out the uncharacterised solution. However in Table \ref{T9}, the same solutions were ruled out more easily because the pairing produced no known theory in the (complete) classification of \cite{Schellekens:1992db}. Meanwhile the solution ${\bf III_{42}}$ is being ruled out for the first time.

Let us move on to the three remaining cases in rows 2, 3 and 12. In row 3 we cannot say anything about ${\bf III_{28}}$ because its partner \sout{${\bf III_9}$} is already ruled out. Thus there are no grounds, from this table, to decide whether ${\bf III_{28}}$ is a CFT or not. Fortunately ${\bf III_{28}}$ has already been identified as being of IVOA-type in Table \ref{T9}. In row 12, although the pairing is formally to an invariant that corresponds to a genuine meromorphic theory from the list of \cite{Schellekens:1992db}, the solution in the right column was already ruled out from the beginning and we get no new information. That leaves row 2 where the pairing gives rise to a modular invariant $j-744+\cN$ with the integer $\cN=336$. This appears in the list of \cite{Schellekens:1992db} and has the Kac-Moody algebra $A_{12,1}^2$. However we have verified that there is no embedding of $B_{7,1}$ in the above algebra that would give rise to the character ${\bf III_{31}}$. It follows that ${\bf III_{31}}$ is not a CFT.

\setlength\LTleft{-40pt}
\setlength\LTright{0pt}
\rowcolors{2}{Mygrey}{Mywhite}
\begin{longtable}{l||ccccc||ccccc||ccc}
\hline
\hline
\rowcolor{Mywhite}\# & $c$ & $(h_1,h_2)$ & $m_1$ & \tiny{$(D_1,D_2)$} & $\mathcal{W}$ & $\Tilde{c}$ & $(\Tilde{h}_1,\Tilde{h}_2)$ & $\tilde{m}_1$ & \tiny{$(\Tilde{D}_1,\Tilde{D}_2)$} & $\widetilde{\mathcal{W}}$ & \tiny{$(d_1, d_2)$} & $\mathcal{N}$ & S\#   \\
\hline
\hline
1. & $\frac{12}{5}$ & $(\frac{1}{5},\frac{3}{5})$ & $3$ & \tiny{$(3,5)$} & ${\bf III_{2}}$ & $\frac{108}{5}$ & $(\frac{9}{5},\frac{7}{5})$ & $27$ & \tiny{$(42483,2295)$} & ${\bf III_{42}}$ & \tiny{(1,2)} & 30 & ---    \\
2. & $\frac{15}{2}$ & $(\frac{1}{2},\frac{15}{16})$ & $105$ & \tiny{$(15,128)$} & $B_{7,1}$ & $\frac{33}{2}$ & $(\frac{3}{2},\frac{17}{16})$ & $231$ & \tiny{$(4301,528)$} & ${\bf III_{31}}$ & \tiny{(1,1)} & 336 & 60   \\
3. & $\frac{44}{5}$ & $(\frac{1}{5},\frac{7}{5})$ & $253$ & \tiny{$(11,242)$} &  \sout{${\bf III_{9}}$} & $\frac{76}{5}$ & $(\frac{9}{5},\frac{3}{5})$ & $437$ & \tiny{$(11875,57)$} & ${\bf III_{28}}$ & \tiny{$(\frac15,1)$} & 690 & ---   \\
4. & $\frac{19}{2}$ & $(\frac{1}{2},\frac{19}{16})$ & $171$ & \tiny{$(19,512)$} & $B_{9,1}$ & $\frac{29}{2}$ & $(\frac{3}{2},\frac{13}{16})$ & $261$ & \tiny{$(3393,116)$} & ${\bf III_{26}}$ & \tiny{(1,1)} & 432 & ---  \\
5. & $\frac{21}{2}$ & $(\frac{1}{2},\frac{21}{16})$ & $210$ & \tiny{$(21,2^{10})$} & $B_{10,1}$ & $\frac{27}{2}$ & $(\frac{3}{2},\frac{11}{16})$ & $270$ & \tiny{$(2871,54)$} & ${\bf III_{21}}$ & \tiny{(1,1)} & 480 & ---   \\
6. & $11$ & $(\frac{1}{2},\frac{11}{8})$ & $231$ & \tiny{$(22,2^{10})$} & $D_{11,1}$ & $13$ & $(\frac{3}{2},\frac{5}{8})$ & $273$ & \tiny{$(2600,26)$} & ${\bf III_{20}}$ & \tiny{$(1,2)$} & 504 & ---   \\
7. & $\frac{23}{2}$ & $(\frac{1}{2},\frac{23}{16})$ & $253$ & \tiny{$(23,2^{11})$} & $B_{11,1}$ & $\frac{25}{2}$ & $(\frac{3}{2},\frac{9}{16})$ & $275$ & \tiny{$(2325,25)$} & ${\bf III_{19}}$ & \tiny{(1,1)} & 528 & ---   \\
8. & $12$ & $(\frac{1}{3},\frac{5}{3})$ & $318$ & \tiny{$(9,4374)$} & ${\bf V_{18}}$ & $12$ & $(\frac{5}{3},\frac{1}{3})$ & $318$ & \tiny{$(4374,9)$} & ${\bf V_{18}}$ & \tiny{$(1,1)$} & 636 & ---   \\
9. & $12$ & $(\frac{3}{5},\frac{7}{5})$ & $222$ & \tiny{$(25,1275)$} & ${\bf III_{17}}$ & $12$ & $(\frac{7}{5},\frac{3}{5})$ & $222$ & \tiny{$(1275,25)$} & ${\bf III_{17}}$ & \tiny{$(2,2)$} & 444 & ---   \\
10. & $\frac{25}{2}$ & $(\frac{1}{2},\frac{25}{16})$ & $300$ & \tiny{$(25,2^{12})$} & $B_{12,1}$ & $\frac{23}{2}$ & $(\frac{3}{2},\frac{7}{16})$ & $276$ & \tiny{$(1771,23)$} & \sout{${\bf III_{16}}$} & \tiny{$(1,\frac12)$} & 576 & ---  \\
11. & $13$ & $(\frac{1}{2},\frac{13}{8})$ & $325$ & \tiny{$(26,2^{12})$} & $D_{13,1}$ & $11$ & $(\frac{3}{2},\frac{3}{8})$ & $275$ & \tiny{$(1496,11)$} & \sout{${\bf III_{15}}$} & \tiny{(1,1)} & 600 & ---   \\
12. & $\frac{27}{2}$ & $(\frac{1}{2},\frac{27}{16})$ & $351$ & \tiny{$(27,2^{13})$} & $B_{13,1}$ & $\frac{21}{2}$ & $(\frac{3}{2},\frac{5}{16})$ & $273$ & \tiny{$(1225,21)$} & \sout{${\bf III_{14}}$} & \tiny{$(1,\frac14)$} & 624 & 67  \\
13. & $14$ & $(\frac{1}{2},\frac{7}{4})$ & $378$ & \tiny{$(28,2^{13})$} & $D_{14,1}$ & $10$ & $(\frac{3}{2},\frac{1}{4})$ & $270$ & \tiny{$(960,5)$} & \sout{${\bf III_{13}}$} & \tiny{(1,1)} & 648 & ---   \\
14. & $\frac{29}{2}$ & $(\frac{1}{2},\frac{29}{16})$ & $406$ & \tiny{$(29,2^{14})$} & $B_{14,1}$ & $\frac{19}{2}$ & $(\frac{3}{2},\frac{3}{16})$ & $266$ & \tiny{$(703,19)$} & \sout{${\bf III_{11}}$} & \tiny{$(1,\frac18)$} & 672 & ---  \\
15. & $15$ & $(\frac{1}{2},\frac{15}{8})$ & $435$ & \tiny{$(30,2^{14})$} & $D_{15,1}$ & $9$ & $(\frac{3}{2},\frac{1}{8})$ & $261$ & \tiny{$(456,9)$} & \sout{${\bf III_{10}}$} & \tiny{$(1,\frac14)$} & 696 & ---   \\
16. & $\frac{31}{2}$ & $(\frac{1}{2},\frac{31}{16})$ & $465$ & \tiny{$(31,2^{15})$} & $B_{15,1}$ & $\frac{17}{2}$ & $(\frac{3}{2},\frac{1}{16})$ & $255$ & \tiny{$(221,17)$} & \sout{${\bf III_{6}}$} & \tiny{$(1,\frac{1}{16})$} & 720 & --- \\
\hline
\hline
\rowcolor{Mywhite}\caption{Inconsistent pairings, $c^\mathcal{H}=24$ with $(n_1,n_2)=(2,2)$. 
Wherever present, the entry in the last column identifies
a candidate meromorphic theory by its row number in the table of [40].}
\label{T9}
\end{longtable}

\noindent {\bf Conclusion:} From table \ref{T9} we concluded that ${\bf III_{31}}$ and ${\bf III_{42}}$ are not valid CFTs, and confirmed that the same holds for ${\bf III_{17}}$, ${\bf V_{18}}$, ${\bf III_{19}}$, ${\bf III_{20}}$,
${\bf III_{21}}$, ${\bf III_{26}}$ which were already ruled out previously.

\subsubsection*{\underline{$(n_1,n_2)=(1,3)$}}

We now turn to bilinear pairs of solutions with $(n_1,n_2)=(1,3)$, a class never previously explored to our knowledge. This set consists of a list of CFT pairings as well as Tables \ref{T119} and \ref{T11}. We do not need a table for the consistent CFT pairings with these values of $n_1,n_2$ as all the pairs are cosets of the meromorphic theory $\cE_1[D_{24,1}]$ that appears in \cite{Schellekens:1992db} as the final entry $\#71$, by $B,D$ type WZW models at level 1. $D_{24}$ has dimension 1128, so the integer $\cN$ in the meromorphic character is $384$ for all these cases. These cosets are obtained through regular embeddings of $B_{r,1}$ or $D_{r,1}$ into $D_{24,1}$ as discussed in Section \ref{s3}. Thus we have pairings of (i) $B_{r_1,1}$ and $B_{r_2,1}$ with  $r_1 + r_2 = 23,~ 0 \leq r_1, r_2 \leq 23$, (ii) $D_{r_1,1}$ and $D_{r_2,1}$ with $r_1 + r_2 = 24, ~ 1 \leq r_1, r_2 \leq 23$. Recall that $B_{0,1}$ is identified with $\cM(4,3)$, the Ising model.

\subsubsection*{Comments on Table \ref{T119}}

In this table we have four pairs that are all of IVOA type. 7 of these 8 solutions have appeared in previous coset pairs where the meromorphic theory had $c=8$ or 16 (Tables \ref{T3} and \ref{T5}). The only new one is ${\bf III_{48}}$ with $c=\frac{156}{7}$. 

\setlength\LTleft{-10pt}
\setlength\LTright{0pt}
\rowcolors{2}{Mygrey}{Mywhite} 
\begin{longtable}{l||ccccc||ccccc||cc}
\hline
\hline
\rowcolor{Mywhite}\# & $c$ & $(h_1,h_2)$ & $m_1$ & \tiny{$(D_1,D_2)$} & $\mathcal{W}$ & $\Tilde{c}$ & $(\Tilde{h}_1,\Tilde{h}_2)$ & $\tilde{m}_1$ & \tiny{$(\Tilde{D}_1,\Tilde{D}_2)$} & $\widetilde{\mathcal{W}}$ & \tiny{$(d_1, d_2)$} & $\mathcal{N}$   \\
\hline
\hline
1. & $\frac{12}{7}$ & $\left(\frac{2}{7},\frac{3}{7}\right)$ & 6 & \tiny{(3,2)} & ${\bf III_{1}}$ & $\frac{156}{7}$ & $\left(\frac{5}{7},\frac{18}{7}\right)$ & 1248 & \tiny{(130,799500)} & ${\bf III_{48}}$ & \tiny{$\left(1,1\right)$} & $1644$ \\
2. & $\frac{60}{7}$ & $\left(\frac{3}{7},\frac{8}{7}\right)$ & $210$ & \tiny{(10,285)} & ${\bf III_{7}}$ & $\frac{108}{7}$ & $\left(\frac{4}{7},\frac{13}{7}\right)$ & $456$ & \tiny{(39,20424)} & ${\bf III_{30}}$ & \tiny{$\left(1,1\right)$} & $1056$ \\
3. & $\frac{44}{5}$ & $\left(\frac{2}{5},\frac{6}{5}\right)$ & $220$ & \tiny{(11,11)} & ${\bf III_{8}}$ & $\frac{76}{5}$ & $\left(\frac{3}{5},\frac{9}{5}\right)$ & $437$ & \tiny{(57,19)} & ${\bf III_{28}}$ & \tiny{$\left(1,1\right)$} & $1056$ \\
4. & $\frac{68}{7}$ & $\left(\frac{3}{7},\frac{9}{7}\right)$ & $221$ & \tiny{(17,782)} & ${\bf III_{12}}$ & $\frac{100}{7}$ & $\left(\frac{4}{7},\frac{12}{7}\right)$ & $380$ & \tiny{(55,11495)} & ${\bf III_{25}}$ & \tiny{$\left(1,1\right)$} & $1536$ \\
\hline
\hline
\rowcolor{Mywhite}\caption{IVOA-type pairings, $c^\mathcal{H}=24$ with $(n_1,n_2)=(1,3)$. The integer $\cN$ in the last column gives the total number of dimension-1 states in the meromorphic character $j-744+\cN$.}
\label{T119}
\end{longtable}

\noindent {\bf Conclusion:} From table \ref{T119}  we conclude that ${\bf III_{48}}$ belong to the IVOA category as this has negative fusion rules and also satisfies the above pairing.

\subsubsection*{Comments on Table \ref{T11}}

In this table, rows 1, 2, 4, 7, 10--12, 14--16, 18, 19 are pairings of solutions of \sout{${\bf III}$} type with consistent CFTs (we used the fact that ${\bf III_2}$ was identified as a CFT in Table \ref{T1}). These pairings mostly give us fractional values of $\cN$ in the meromorphic character, so we do not learn anything from them. In a few cases we get integer values of $\cN$ but these too do not feature in \cite{Schellekens:1992db}.

In row 17 both solutions were previously ruled out. This leaves rows 3, 5, 6, 8, 9, 13, where we can hope to get new information. In all these cases except row 13, the solutions ${\bf III_{51}, III_{46}, III_{44}, V_{41}, V_{40}}$ are paired with known CFTs. However the result of the pairing is not a meromorphic CFT as one readily sees from \cite{Schellekens:1992db}. That means these five solutions are ruled out as corresponding to CFTs. 

\setlength\LTleft{-45pt}
\setlength\LTright{0pt}
\rowcolors{2}{Mygrey}{Mywhite}
\begin{longtable}{l||ccccc||ccccc||cc}
\hline
\hline
\rowcolor{Mywhite}\# & $c$ & $(h_1,h_2)$ & $m_1$ & \tiny{$(D_1,D_2)$} & $\mathcal{W}$ & $\Tilde{c}$ & $(\Tilde{h}_1,\Tilde{h}_2)$ & $\Tilde{m}_1$ & \tiny{$(\Tilde{D}_1,\Tilde{D}_2)$} & $\widetilde{\mathcal{W}}$ & \tiny{$(d_1,d_2)$} & $\mathcal{N}$  \\
\hline
\hline
1. & $\frac{1}{2}$ & $\left(\frac{1}{16},\frac{1}{2}\right)$ & 0 & \tiny{(1,1)} & \scriptsize{$B_{0,1}$} & $\frac{47}{2}$ & $\left(\frac{15}{16},\frac{5}{2}\right)$ & 4371 & \tiny{(4371,1135003)} & \sout{${\bf III_{55}}$} & \tiny{$\left(\frac{1}{2},1\right)$} & $\frac{13113}{2}$ \\
2. & $\frac{4}{5}$ & $\left(\frac{1}{5},\frac{2}{5}\right)$ & 0 & \tiny{(1,1)} & \scriptsize{$\mathcal{I}[\mathcal{M}(5,2)^{\otimes 2}]$} & $\frac{116}{5}$ & $\left(\frac{4}{5},\frac{13}{5}\right)$ & 1711 & \tiny{(1653,910803)} & \sout{${\bf III_{53}}$} & \tiny{$\left(\frac{1}{5},1\right)$} & $\frac{10218}{5}$ \\
3. & $1$ & $\left(\frac{1}{8},\frac{1}{2}\right)$ & 1 & \tiny{(1,2)} & $D_{1,1}$ & 23 & $\left(\frac{7}{8},\frac{5}{2}\right)$ & 2323 & \tiny{$(575,32683\cdot2^5)$} & ${\bf III_{51}}$ & \tiny{$\left(2,1\right)$} & $3474$ \\
4. & $\frac{3}{2}$ & $\left(\frac{3}{16},\frac{1}{2}\right)$ & 3 & \tiny{(2,3)} & $B_{1,1}$ & $\frac{45}{2}$ & $\left(\frac{13}{16},\frac{5}{2}\right)$ & 1640 & \tiny{(1595,956449)} & \sout{${\bf III_{49}}$} & \tiny{$\left(\frac{1}{4},1\right)$} & $\frac{4881}{2}$ \\
5. & 2 & $\left(\frac{1}{4},\frac{1}{2}\right)$ & 6 & \tiny{(2,4)} & $D_{2,1}$ & 22 & $\left(\frac{3}{4},\frac{5}{2}\right)$ & 1298 & \tiny{$(154,847\cdot2^{10})$} & ${\bf III_{46}}$ & \tiny{$\left(2,1\right)$} & $1920$ \\
6. & $\frac{12}{5}$ & $\left(\frac{1}{5},\frac{3}{5}\right)$ & 3 & \tiny{$(3,5)$} & ${\bf III_{2}}$ & $\frac{108}{5}$ & $\left(\frac{4}{5},\frac{12}{5}\right)$ & 1404 & \tiny{$(459,153\cdot5^5)$} & ${\bf III_{44}}$ & \tiny{$\left(1,2\right)$} & $2784$ \\
7. & $\frac{12}{5}$ & $\left(\frac{3}{5},\frac{1}{5}\right)$ & 3 & \tiny{$(5,3)$} & ${\bf III_{2}}$ & $\frac{108}{5}$ & $\left(\frac{2}{5},\frac{14}{5}\right)$ & 860 & \tiny{$(833,3015426)$} & \sout{${\bf III_{43}}$} & \tiny{$\left(\frac{1}{25},1\right)$} & $\frac{5148}{5}$ \\
8. & $4$ & $\left(\frac{1}{3},\frac{2}{3}\right)$ & 16 & \tiny{(3,9)} & $A_{2,1}^{\otimes 2}$ & 20 & $\left(\frac{2}{3},\frac{7}{3}\right)$ & 890 & \tiny{$(135,10\cdot2\cdot3^9)$} & ${\bf V_{41}}$ & \tiny{$\left(2,2\right)$} & $1716$ \\
9. & $4$ & $\left(\frac{2}{3},\frac{1}{3}\right)$ & 16 & \tiny{$(9,3)$} & $A_{2,1}^{\otimes 2}$ & 20 & $\left(\frac{1}{3},\frac{8}{3}\right)$ & 728 & \tiny{$(12,2\cdot3^{12})$} & ${\bf V_{40}}$ & \tiny{$\left(2,2\right)$} & $960$ \\
10. & $\frac{28}{5}$ & $\left(\frac{2}{5},\frac{4}{5}\right)$ & 28 & \tiny{(7,49)} & $G_{2,1}^{\otimes 2}$ & $\frac{92}{5}$ & $\left(\frac{3}{5},\frac{11}{5}\right)$ & 690 & \tiny{(299,178802)} & \sout{${\bf III_{38}}$} & \tiny{$\left(\frac{2}{5},1\right)$} & $\frac{7776}{5}$ \\
11. & 6 & $\left(\frac{3}{4},\frac{1}{2}\right)$ & 66 & \tiny{(32,12)} & $D_{6,1}$ & 18 & $\left(\frac{1}{4},\frac{5}{2}\right)$ & 598 & \tiny{$(25,221\cdot2^{10})$} & \sout{${\bf III_{36}}$} & \tiny{$\left(\frac{1}{4},1\right)$} & $864$ \\
12. & $\frac{36}{5}$ & $\left(\frac{4}{5},\frac{3}{5}\right)$ & 144 & \tiny{$(45,12)$} & ${\bf III_{4}}$ & $\frac{84}{5}$ & $\left(\frac{1}{5},\frac{12}{5}\right)$ & 534 & \tiny{(33,55924)} & \sout{${\bf III_{34}}$} & \tiny{$\left(\frac{2}{25},1\right)$} & $\frac{3984}{5}$ \\
13. & 8 & $\left(\frac{1}{2},1\right)$ & 56 & \tiny{(8,64)} & $D_{4,1}^{\otimes 2}$ & 16 & $\left(\frac{1}{2}, 2\right)$ & 496 & \tiny{(32,$2^{15}$)} & $D_{16,1}$ & \tiny{$\left(\frac{8}{3},\frac{13}{3}\right)$} & $\frac{3704}{3}$ \\
14. & $\frac{17}{2}$ & $\left(\frac{1}{16},\frac{3}{2}\right)$ & $255$ & \tiny{(17,221)} & \sout{${\bf III_{6}}$} & $\frac{31}{2}$ & $\left(\frac{15}{16},\frac{3}{2}\right)$ & $248$ & \tiny{(248,3875)} & $E_{8,2}$ & \tiny{$\left(\frac{1}{16},1\right)$} & $\frac{1533}{2}$ \\
15. & $\frac{44}{5}$ & $\left(\frac{1}{5},\frac{7}{5}\right)$ & $253$ & \tiny{(11,242)} & \sout{${\bf III_{9}}$} & $\frac{76}{5}$ & $\left(\frac{4}{5},\frac{8}{5}\right)$ & $380$ & \tiny{(57,3249)} & $E_{7.5}^{\otimes 2}$ & \tiny{$\left(\frac{1}{5},1\right)$} & $\frac{3792}{5}$ \\
16. & 9 & $\left(\frac{1}{8},\frac{3}{2}\right)$ & $261$ & \tiny{$(9,456)$} & \sout{${\bf III_{10}}$} & 15 & $\left(\frac{7}{8},\frac{3}{2}\right)$ & $255$ & \tiny{$(120,3640)$} & $\mathcal{E}[A_{15,1}]$ & \tiny{$\left(\frac14,1\right)$} & $786$ \\
17. & $\frac{19}{2}$ & $\left(\frac{3}{16},\frac{3}{2}\right)$ & $266$ & \tiny{(19,703)} & \sout{${\bf III_{11}}$} & $\frac{29}{2}$ & $\left(\frac{13}{16},\frac{3}{2}\right)$ & $261$ & \tiny{$(116,3393)$} & ${\bf III_{26}}$ & \tiny{$\left(\frac{1}{8},1\right)$} & $\frac{1605}{2}$ \\
18. & $10$ & $\left(\frac{1}{4},\frac{3}{2}\right)$ & $270$ & \tiny{$(5,960)$} & \sout{${\bf III_{13}}$} & 14 & $\left(\frac{3}{4},\frac{3}{2}\right)$ & $266$ & \tiny{(56,$56^2$)} & $E_{7,1}^{\otimes 2}$ & \tiny{$\left(1,1\right)$} & $816$ \\
19. & $\frac{52}{5}$ & $\left(\frac{3}{5},\frac{6}{5}\right)$ & $104$ & \tiny{(26,$26^2$)} & $F_{4,1}^{\otimes 2}$ & $\frac{68}{5}$ & $\left(\frac{2}{5},\frac{9}{5}\right)$ & $374$ & \tiny{(119,12138)} & \sout{${\bf III_{23}}$} & \tiny{$\left(\frac{1}{5},1\right)$} & $\frac{5484}{5}$ \\
20. & $\frac{21}{2}$ & $\left(\frac{5}{16},\frac{3}{2}\right)$ & $273$ & \tiny{(21,1225)} & \sout{${\bf III_{14}}$} & $\frac{27}{2}$ & $\left(\frac{11}{16},\frac{3}{2}\right)$ & $270$ & \tiny{$(54,2871)$} & ${\bf III_{21}}$ & \tiny{$\left(\frac{1}{4},1\right)$} & $\frac{1653}{2}$ \\
21. & 11 & $\left(\frac{3}{8},\frac{3}{2}\right)$ & $275$ & \tiny{$(11,1496)$} & \sout{${\bf III_{15}}$} & 13 & $\left(\frac{5}{8},\frac{3}{2}\right)$ & $273$ & \tiny{$(26,2600)$} & ${\bf III_{20}}$ & \tiny{$\left(1,1\right)$} & $834$ \\
22. & $\frac{23}{2}$ & $\left(\frac{7}{16},\frac{3}{2}\right)$ & $276$ & \tiny{(23,1771)} & \sout{${\bf III_{16}}$} & $\frac{25}{2}$ & $\left(\frac{9}{16},\frac{3}{2}\right)$ & $275$ & \tiny{(25,2325)} & ${\bf III_{19}}$ & \tiny{$\left(\frac{1}{2},1\right)$} & $\frac{1677}{2}$ \\
23. & 12 & $\left(\frac{2}{3},\frac{4}{3}\right)$ & $156$ & \tiny{(27,$27^2$)} & $E_{6,1}^{\otimes 2}$ & 12 & $\left(\frac{1}{3},\frac{5}{3}\right)$ & $318$ & \tiny{$(9,4374)$} & ${\bf V_{18}}$ & \tiny{$\left(2,2\right)$} & $960$ \\
\hline
\hline
\rowcolor{Mywhite}\caption{Inconsistent pairings, $c^\mathcal{H}=24$ with $(n_1,n_2)=(1,3)$. The integer $\cN$ in the last column gives the total number of dimension-1 states in the meromorphic character $j-744+\cN$.}
\label{T11}
\end{longtable}

\noindent {\bf Conclusion:} From Table \ref{T11} we conclude that ${\bf V_{40}}$, ${\bf V_{41}}$, ${\bf III_{44}}$, ${\bf III_{46}}$ and ${\bf III_{51}}$ are not valid CFTs.

\subsection{Cosets of $c^{\mathcal{H}}=32$}

Now we move on to list coset pairs for $c^{\mathcal H}=32$. The meromorphic character in this case can be written:
\be
\chi(\tau)=j(\tau)^\frac13(j(\tau)-992+\cN)\sim q^{-\frac43}(1+\cN q+\cdots)
\label{charH.32}
\ee
so that $\cN$ is the dimension of its Kac-Moody algebra. 

Since we have $(p,\ell)=(3,0)$, we get $n_1+n_2=5$ from \eref{datareln}. This again implies that  we have two sub-cases: $(n_1,n_2)=(1,4)$ or $(2,3)$ that lead to distinct sets of coset theories. We address each one in turn.

\subsubsection*{\underline{$(n_1,n_2)=(1,4)$}}

Here any admissible character solution that is potentially part of a coset relation has to have a central charge less than $32$. Hence we consider all admissible character solutions from \cite{Das:2021uvd} with $c < 32$. The consistent cosets all turn out to arise through regular embeddings of $B_{r,1}$ or $D_{r,1}$ into $D_{32,1}$ as discussed in Section \ref{s3}.  
Thus we have pairings of (i) $B_{r_1,1}$ and $B_{r_2,1}$ with  $r_1 + r_2 = 31,~ 0 \leq r_1, r_2 \leq 31$, (ii) $D_{r_1,1}$ and $D_{r_2,1}$ with $r_1 + r_2 = 32, ~ 1 \leq r_1, r_2 \leq 31$.
It turns out there are no IVOA-type bilinear pairs with $(n_1,n_2)=(1,4)$ so we go on directly to the table of inconsistent pairings in Table \ref{T14}.

\subsubsection*{Comments on Table \ref{T14}}

All the pairs ($\mathcal{W}\leftrightarrow \widetilde{\mathcal{W}}$) listed in Table \ref{T14} satisfy a bilinear relation to a potential $c=32$ character of the form in \eref{charH.32}. However the relation is problematic in one or more ways. In rows 1--4, 6, 7, 14, 18, 19 we have theories that were found to be inconsistent at the outset, paired with a known CFT. There is nothing left to determine in these cases. Next, in rows 9, 10, 12, 17 both members of the pair are already ruled out. 

Rows 5, 8, 15 seem more promising as the pairings lead to integer values of $\cN$ as seen in the last column of the table. However in these cases $\cN$ is  greater than 2016, which is the dimension of $D_{32}$. It can be shown that the dimension of the Kac-Moody algebra for all meromorphic theories at $c=8N$ is less than or equal to the dimension of $D_{8N}$, we do this in Appendix \ref{upperb}.  For rows 5 and 15 this means the bilinear pairing in these cases does not produce a valid meromorphic theory at $c=32$. In turn, this rules out $\mathbf{V_{58}}$ in row 5 since it is paired with a valid theory. However in row 15 we have already ruled out ${\bf V_{18}}$ so we cannot say anything definite about ${\bf V_{41}}$. Fortunately this was ruled out in Table \ref{T11}. And in row 8 both partners in the pairing are consistent, it is the pairing which is inconsistent as shown by the fractional values of $d_1,d_2$.

This leaves rows 11, 13, 16. Rows 11 and 13 are inconclusive since the solution in the first column is inconsistent. Fortunately, again the  solutions ${\bf III_{51}}$ and ${\bf III_{46}}$ in the second column were already ruled out by Table \ref{T11}. Row 16 is inconclusive for a different reason: we do not know if a meromorphic theory with $\cN=1532$ exists. However again ${\bf V_{40}}$ was also ruled out in Table \ref{T11}.

\setlength\LTleft{-50pt}
\setlength\LTright{0pt}
\rowcolors{2}{Mygrey}{Mywhite}
\begin{longtable}{l||ccccc||ccccc||cc} 
\hline
\hline
\rowcolor{Mywhite}\# & $c$ & $(h_1,h_2)$ & $m_1$ & \tiny{$(D_1,D_2)$} & $\mathcal{W}$ & $\Tilde{c}$ & $(\Tilde{h}_1,\Tilde{h}_2)$ & $\Tilde{m}_1$ & \tiny{$(\Tilde{D}_1,\Tilde{D}_2)$} & $\widetilde{\mathcal{W}}$ & \tiny{$(d_1,d_2)$} & $\mathcal{N}$  \\
\hline
\hline
1. & $\frac{4}{5}$ & $\left(\frac{1}{5},\frac{2}{5}\right)$ & 2 & \tiny{(1,1)} & \scriptsize{$\mathcal{I}[\mathcal{M}(5,2)^{\otimes 2}]$} & $\frac{156}{5}$ & $\left(\frac{4}{5},\frac{18}{5}\right)$ & 3612 & \tiny{($14877,250774426$)} & \sout{${\bf III_{62}}$} & \tiny{$\left(\frac{1}{25},1\right)$} & $\frac{105227}{25}$ \\
2. & $1$ & $\left(\frac{1}{8},\frac{1}{2}\right)$ & 1 & \tiny{(1,2)} & $D_{1,1}$ & 31 & $\left(\frac{7}{8},\frac{7}{2}\right)$ & 5239 & \tiny{$(9269,2295147\cdot2^7)$} & \sout{${\bf III_{61}}$} & \tiny{$\left(\frac{1}{4},1\right)$} & $\frac{30229}{4}$ \\
3. & $\frac{3}{2}$ & $\left(\frac{3}{16},\frac{1}{2}\right)$ & 3 & \tiny{(2,3)} & $B_{1,1}$ & $\frac{61}{2}$ & $\left(\frac{13}{16},\frac{7}{2}\right)$ & 3599 & \tiny{(47763,264580485)} & \sout{${\bf III_{60}}$} & \tiny{$\left(\frac{1}{64},1\right)$} & $\frac{163027}{32}$ \\
4. & 2 & $\left(\frac{1}{4},\frac{1}{2}\right)$ & 6 & \tiny{(2,4)} & $D_{2,1}$ & 30 & $\left(\frac{3}{4},\frac{7}{2}\right)$ & 2778 & \tiny{$(539,14421\cdot2^{14})$} & \sout{${\bf III_{59}}$} & \tiny{$\left(1,1\right)$} & $3862$ \\
5. & $4$ & $\left(\frac{1}{3},\frac{2}{3}\right)$ & 16 & \tiny{(3,9)} & $A_{2,1}^{\otimes 2}$ & 28 & $\left(\frac{2}{3},\frac{10}{3}\right)$ & 1948 & \tiny{$(225,11\cdot2\cdot3^{14})$} & ${\bf V_{58}}$ & \tiny{$\left(2,2\right)$} & $3314$ \\
6. & $\frac{28}{5}$ & $\left(\frac{2}{5},\frac{4}{5}\right)$ & 28 & \tiny{(7,49)} & $G_{2,1}^{\otimes 2}$ & $\frac{132}{5}$ & $\left(\frac{3}{5},\frac{16}{5}\right)$ & 1536 & \tiny{(2392,47018049)} & \sout{${\bf III_{57}}$} & \tiny{$\left(\frac{2}{25},1\right)$} & $\frac{72588}{25}$ \\
7. & 6 & $\left(\frac{3}{4},\frac{1}{2}\right)$ & 66 & \tiny{(32,12)} & $D_{6,1}$ & 26 & $\left(\frac{1}{4},\frac{7}{2}\right)$ & 1118 & \tiny{$(117,3315\cdot2^{14})$} & \sout{${\bf III_{56}}$} & \tiny{$\left(\frac{1}{16},1\right)$} & $1418$ \\
8. & 8 & $\left(\frac{1}{2},1\right)$ & 56 & \tiny{(8,64)} & $D_{4,1}^{\otimes 2}$ & 24 & $\left(\frac{1}{2}, 3\right)$ & 1128 & \tiny{(48,$2^{23}$)} & $D_{24,1}$ & \tiny{$\left(3,\frac{59}{16}\right)$} & $2336$ \\
9. & $\frac{17}{2}$ & $\left(\frac{1}{16},\frac{3}{2}\right)$ & 255 & \tiny{(17,221)} & \sout{${\bf III_{6}}$} & $\frac{47}{2}$ & $\left(\frac{15}{16},\frac{5}{2}\right)$ & 4371 & \tiny{(4371,1135003)} & \sout{${\bf III_{55}}$} & \tiny{$\left(\frac{1}{32},1\right)$} & $\frac{222339}{32}$ \\
10. & $\frac{44}{5}$ & $\left(\frac{1}{5},\frac{7}{5}\right)$ & 253 & \tiny{(11,242)} & \sout{${\bf III_{9}}$} & $\frac{116}{5}$ & $\left(\frac{4}{5},\frac{13}{5}\right)$ & 1711 & \tiny{(1653,910803)} & \sout{${\bf III_{53}}$} & \tiny{$\left(\frac{1}{50},1\right)$} & $\frac{116383}{50}$ \\
11. & 9 & $\left(\frac{1}{8},\frac{3}{2}\right)$ & 261 & \tiny{$(9,456)$} & \sout{${\bf III_{10}}$} & 23 & $\left(\frac{7}{8},\frac{5}{2}\right)$ & 2323 & \tiny{$(575,32683\cdot2^5)$} & ${\bf III_{51}}$ & \tiny{$\left(\frac{1}{4},1\right)$} & $\frac{15511}{4}$ \\
12. & $\frac{19}{2}$ & $\left(\frac{3}{16},\frac{3}{2}\right)$ & 266 & \tiny{(19,703)} & \sout{${\bf III_{11}}$} & $\frac{45}{2}$ & $\left(\frac{13}{16},\frac{5}{2}\right)$ & 1640 & \tiny{(1595,956449)} & \sout{${\bf III_{49}}$} & \tiny{$\left(\frac{1}{32},1\right)$} & $\frac{91297}{32}$ \\
13. & $10$ & $\left(\frac{1}{4},\frac{3}{2}\right)$ & 270 & \tiny{$(5,960)$} & \sout{${\bf III_{13}}$} & 22 & $\left(\frac{3}{4},\frac{5}{2}\right)$ & 1298 & \tiny{$(154,847\cdot2^{10})$} & ${\bf III_{46}}$ & \tiny{$\left(1,1\right)$} & $2338$ \\
14. & $\frac{52}{5}$ & $\left(\frac{3}{5},\frac{6}{5}\right)$ & 104 & \tiny{(26,$26^2$)} & $F_{4,1}^{\otimes 2}$ & $\frac{108}{5}$ & $\left(\frac{2}{5},\frac{14}{5}\right)$ & 860 & \tiny{(833,3015426)} & \sout{${\bf III_{43}}$} & \tiny{$\left(\frac{1}{25},1\right)$} & $\frac{45758}{25}$ \\
15. & 12 & $\left(\frac{1}{3},\frac{5}{3}\right)$ & 318 & \tiny{$(9,1)$} & ${\bf V_{18}}$ & 20 & $\left(\frac{2}{3},\frac{7}{3}\right)$ & 890 & \tiny{$(135,10\cdot2\cdot3^{16})$} & ${\bf V_{41}}$ & \tiny{$\left(1,1\right)$} & $2423$ \\
16. & 12 & $\left(\frac{2}{3},\frac{4}{3}\right)$ & 156 & \tiny{(27,$27^2$)} & $E_{6,1}^{\otimes 2}$ & 20 & $\left(\frac{1}{3},\frac{8}{3}\right)$ & 728 & \tiny{$(12,2\cdot3^12)$} & ${\bf V_{40}}$ & \tiny{$\left(2,2\right)$} & $1532$ \\
17. & $\frac{68}{5}$ & $\left(\frac{2}{5},\frac{9}{5}\right)$ & 374 & \tiny{(119,12138)} & \sout{${\bf III_{23}}$} & $\frac{92}{5}$ & $\left(\frac{3}{5},\frac{11}{5}\right)$ & 690 & \tiny{(299,178802)} & \sout{${\bf III_{38}}$} & \tiny{$\left(\frac{1}{25},1\right)$} & $\frac{62181}{25}$ \\
18. & 14 & $\left(\frac{3}{4},\frac{3}{2}\right)$ & 266 & \tiny{(56,$56^2$)} & $E_{7,1}^{\otimes 2}$ & 18 & $\left(\frac{1}{4},\frac{5}{2}\right)$ & 598 & \tiny{$(25,221\cdot2^{10})$} & \sout{${\bf III_{36}}$} & \tiny{$\left(\frac{1}{4},1\right)$} & $1214$ \\
19. & $\frac{76}{5}$ & $\left(\frac{4}{5},\frac{8}{5}\right)$ & 380 & \tiny{(57,$57^{2}$)} & $E_{7.5}^{\otimes 2}$ & $\frac{84}{5}$ & $\left(\frac{1}{5},\frac{12}{5}\right)$ & 534 & \tiny{(33,55924)} & \sout{${\bf III_{34}}$} & \tiny{$\left(\frac{2}{25},1\right)$} & $\frac{26612}{25}$ \\
\hline
\hline
\rowcolor{Mywhite}\caption{Inconsistent pairings, $c^\mathcal{H}=32$  with $(n_1,n_2)=(1,4)$. The integer $\cN$ in the last column gives the total number of dimension-1 states in the meromorphic character $j^{\frac23}(j-992+\cN)$.}
\label{T14}
\end{longtable}

\noindent {\bf Conclusion:} From table \ref{T14}  we obtain the new information that ${\bf V_{58}}$ is not a valid CFT.


\subsubsection*{$\underline{(n_1,n_2)=(2,3)}$}

We go on to consider bilinear pairings to meromorphic characters of $c=32$ where the integers $(n_1,n_2)=(2,3)$. In this category we find consistent, IVOA-type and inconsistent solutions that are listed in Tables \ref{T15}, \ref{T989} and \ref{T17} respectively. 

\subsection*{Comments on Table \ref{T15}}

This table has 22 bilinear pairs, all of which we will argue to be consistent CFTs. In row 1 the Baby Monster CFT with $c=\frac{47}{2}$ makes its first appearance in which it is not paired with the Ising model $\cM(4,3)$, being paired instead with $B_{8,1}$. This has previously appeared as one in a family of pairings in \cite{Das:2022slz} (entry 1 of Table 2) where it was argued that, since the existence of $B_{r,1}$ as well as the Baby Monster CFT are established, the pairing actually {\em predicts} a non-lattice CFT at $c=32$.

The pairings in rows 3, 5-7, 10--13, 15, 17--19, 21 all involve the pairing of an affine theory with a CFT that was explicitly constructed as a coset in \cite{Gaberdiel:2016zke}. Row 20 is slightly different, being a pairing between two theories from \cite{Gaberdiel:2016zke}, a phenomenon we are seeing for the first time. All these theories already made an appearance in our Table \ref{T7} which is the context in which they were originally discovered in \cite{Gaberdiel:2016zke}. Their re-appearance illustrates a phenomenon that was highlighted in \cite{Das:2022slz}: once a new CFT appears as a coset, it appears repeatedly in distinct coset pairings at higher total central charge.

Rows 9, 16, 22 are pairings between affine theories. Even though these are known theories, the pairings merit some discussion. Row 9 is a case that was analysed in Example 2 of Section \ref{s3}, and involves a pairing of $D_{12,1}$ and $D_{20,1}$ that is distinct from the standard pairing to $D_{32,1}$. In the present case the pairing gives rise to the $c=32$ lattice theory $\cE_1[D_{12,1}D_{20,1}]$ without an enhancement of the Kac-Moody algebra. This is a known Kervaire lattice \cite{kervaire1994unimodular}.
Row 16 pairs $E_{7,1}^{\otimes 2}$ with $D_{18,1}$ to a meromorphic character whose Kac-Moody algebra has dimension 896. From this pairing one would be led to predict the existence of a meromorphic theory at $c=32$ with Kac-Moody algebra $\cE_1[E_{7,1}^2 D_{18,1}]$ of rank 32 and dimension 896. Because this algebra has only simply-laced factors at level 1, and its rank equals the central charge, it must be a lattice theory. And indeed, this is again a known Kervaire lattice \cite{kervaire1994unimodular}. Row 22 pairs $E_{8,2}$ with $B_{16,1}$ and predicts a new meromorphic theory at $c=32$ that must be a non-lattice theory (given that the rank is less than maximal, one factor has a level greater than 1, and one factor is non-simply-laced). This is again part of an infinite family in \cite{Das:2022slz}, corresponding to the $m=1$ case of entry $\#15$ of Table 3 in that reference.

Next we turn to the remaining cases in rows 2,4,8,14. For row 2, the dual of $D_{9,1}$ 
is $\mathbf{III_{50}}$ which was previously identified from Table \ref{T7} as the three-character extension $\cE_3[D_{1,1}^{23}]$. Here we see it paired to give a meromorphic theory at $c=32$ with a total of 176 generators. Of these, $D_{9,1}$ contributes 153 generators and a central charge 9, leaving 23 residual generators and a residual central charge of 23. These two conditions can only be met by $U(1)^{23}$. Thus we predict a lattice theory at $c=32$ with Kac-Moody algebra $D_{9,1}U(1)^{23}$. Comparing with \cite{king2003mass}, we see that there is indeed a lattice with 144 roots (plus 32 Cartan generators) having a $D_{9,1}$ factor.  This verifies the prediction following from our coset pairing and the fact that $\mathbf{III_{50}}$ was previously characterised. Note that this is not a Kervaire lattice, since apart from $D_{9,1}$ the remaining symmetries are all Abelian.

Moving on to row 4, the dual ${\bf III_{45}}$ of $D_{10,1}$ has been identified in \eref{idents} as one of six possible three-character extensions. This means the pairing in the present table predicts six meromorphic theories at $c=32$. Only one of these corresponds to a lattice, with algebra $D_{10,1}A_{1,1}^{22}$ and this indeed exists -- it is a Kervaire lattice \cite{kervaire1994unimodular} with 224 roots. For the remaining five cases one has a prediction for new meromorphic theories at $c=32$, and this is again part of the infinite series of predictions in \cite{Das:2022slz} where they correspond to the $m=1$ case for entries 18-22 in Table 3. 

Row 8 pairs $E_{6,1}^{\otimes 2}$ with ${\bf V_{39}}$ which was identified in \eref{idents} with three distinct three-character extensions.  Thus again we have predictions for three meromorphic theories at $c=32$. One is a lattice theory with algebra $E_{6,1}^2 A_{2,1}^{10}$ that corresponds to a Kervaire lattice \cite{kervaire1994unimodular} and the other two are non-lattice theories that were predicted in entries 2,3 of Table 6 \cite{Das:2022slz}. These theories are part of a finite, rather than infinite, collection. 

Finally in row 14 we have a pairing of ${\bf III_{22}}$ and ${\bf III_{37}}$ which have been identified previously as $\cE_3[C_{8,1}]$ and $\cE_3[E_{6,3}G_{2,1}]$ respectively. This leads to a prediction of a new meromorphic theory at $c=32$ corresponding to $\cE_1[C_{8,1}E_{6,3}G_{2,1}]$. This is entry 4 of Table 6 in \cite{Das:2022slz}.

\setlength\LTleft{-35pt}
\setlength\LTright{0pt}
\rowcolors{2}{Mygrey}{Mywhite}
\begin{longtable}{l||ccccc||ccccc||cc} 
\hline
\hline
\rowcolor{Mywhite}\# & $c$ & $(h_1,h_2)$ & $m_1$ & \tiny{$(D_1,D_2)$} & $\mathcal{W}$ & $\Tilde{c}$ & $(\Tilde{h}_1,\Tilde{h}_2)$ & $\tilde{m}_1$ & \tiny{$(\Tilde{D}_1,\Tilde{D}_2)$} & $\widetilde{\mathcal{W}}$ & \tiny{$(d_1,d_2)$} & $\mathcal{N}$  \\
\hline
\hline
1. & $\frac{17}{2}$ & $\left(\frac{1}{2},\frac{17}{16}\right)$ & $136$ & \tiny{(17,256)} & $B_{8,1}$ & $\frac{47}{2}$ & $\left(\frac{3}{2},\frac{31}{16}\right)$ & $0$ & \tiny{$(4371,47\cdot2^{11})$} & BM & \tiny{$\left(1,1\right)$} & $136$ \\
2. & 9 & $\left(\frac{1}{2},\frac{9}{8}\right)$ & $153$ & \tiny{(18,256)} & $D_{9,1}$ & 23 & $\left(\frac{3}{2},\frac{15}{8}\right)$ & $23$ &  \tiny{$(4600 ,23\cdot2^{11})$} & ${\bf III_{50}}$ & \tiny{$\left(1,2\right)$} & $176$ \\
3. & $\frac{19}{2}$ & $\left(\frac{1}{2},\frac{19}{16}\right)$ & $171$ & \tiny{(19,$2^9$)} & $B_{9,1}$ & $\frac{45}{2}$ & $\left(\frac{3}{2},\frac{29}{16}\right)$ & $45$ & \tiny{$(4785,45\cdot2^{10})$} & $\text{GHM}_{45}$ & \tiny{$\left(1,1\right)$} & $216$ \\
4. & $10$ & $\left(\frac{1}{2},\frac{5}{4}\right)$ & $190$ & \tiny{(20,$2^9$)} & $D_{10,1}$ & 22 & $\left(\frac{3}{2},\frac{7}{4}\right)$ & $66$ & \tiny{$(77\cdot2^6,11\cdot2^{11})$} & ${\bf III_{45}}$ & \tiny{$\left(1,2\right)$} & $256$ \\
5. & $\frac{21}{2}$ & $\left(\frac{1}{2},\frac{21}{16}\right)$ & $210$ & \tiny{(21,$2^{10}$)} & $B_{10,1}$ & $\frac{43}{2}$ & $\left(\frac{3}{2},\frac{27}{16}\right)$ & $86$ &  \tiny{$(5031,43\cdot2^9)$} & $\text{GHM}_{86}$ & \tiny{$\left(1,1\right)$} & $296$ \\
6. & 11 & $\left(\frac{1}{2},\frac{11}{8}\right)$ & $231$ & \tiny{(22,$2^{10}$)} & $D_{11,1}$ & 21 & $\left(\frac{3}{2},\frac{13}{8}\right)$ & $105$ & \tiny{$(5096,21\cdot2^9)$} & $\text{GHM}_{105}$ & \tiny{$\left(1,2\right)$} & $336$ \\
7. & $\frac{23}{2}$ & $\left(\frac{1}{2},\frac{23}{16}\right)$ & $253$ & \tiny{(23,$2^{11}$)} & $B_{11,1}$ & $\frac{41}{2}$ & $\left(\frac{3}{2},\frac{25}{16}\right)$ & $123$ & \tiny{$(5125,41\cdot2^8)$} & $\text{GHM}_{123}$ & \tiny{$\left(1,1\right)$} & $376$ \\
8. & 12 & $\left(\frac{2}{3},\frac{4}{3}\right)$ & $156$ & \tiny{(27,$27^2$)} & $E_{6,1}^{\otimes 2}$ & 20 & $\left(\frac{4}{3},\frac{5}{3}\right)$ & $80$ & \tiny{$(5\cdot3^5,4\cdot3^7)$} & ${\bf V_{39}}$ & \tiny{$\left(4,4\right)$} & $236$ \\
9. & 12 & $\left(\frac{3}{2},\frac{1}{2}\right)$ & $276$ & \tiny{($2^{11}$,24)} & $D_{12,1}$ & 20 & $\left(\frac{1}{2},\frac{5}{2}\right)$ & $780$ & \tiny{(40,$2^{19}$)} & $D_{20,1}$ & \tiny{$(1,1)$} & $1056$ \\
10. & 12 & $\left(\frac{1}{2},\frac{3}{2}\right)$ & 276 & \tiny{(24,$2^{11}$)} & $D_{12,1}$ & 20 & $\left(\frac{3}{2},\frac{3}{2}\right)$ & 140 & \tiny{(5120,5120)} & $\text{GHM}_{140}$ & \tiny{$\left(1,2\right)$} & 416  \\
11. & $\frac{25}{2}$ & $\left(\frac{1}{2},\frac{25}{16}\right)$ & $300$ & \tiny{(25,$2^{12}$)} & $B_{12,1}$ & $\frac{39}{2}$ & $\left(\frac{3}{2},\frac{23}{16}\right)$ & $156$ & \tiny{$(5083,39\cdot2^7)$} & $\text{GHM}_{156}$ & \tiny{$\left(1,1\right)$} & $456$ \\
12. & 13 & $\left(\frac{1}{2},\frac{13}{8}\right)$ & $325$ & \tiny{(26,$2^{12}$)} & $D_{13,1}$ & 19 & $\left(\frac{3}{2},\frac{11}{8}\right)$ & $171$ & \tiny{$(5016,19\cdot2^7)$}  & $\text{GHM}_{171}$ & \tiny{$\left(1,2\right)$} & $496$ \\
13. & $\frac{27}{2}$ & $\left(\frac{1}{2},\frac{27}{16}\right)$ & $351$ & \tiny{(27,$2^{13}$)} & $B_{13,1}$ & $\frac{37}{2}$ & $\left(\frac{3}{2},\frac{21}{16}\right)$ & $185$ & \tiny{(4921,$37\cdot2^6$)} & $\text{GHM}_{185}$ & \tiny{$\left(1,1\right)$} & $536$ \\
14. & $\frac{68}{5}$ & $\left(\frac{4}{5},\frac{7}{5}\right)$ & $136$ & \tiny{(119,1700)} & $\mathbf{III_{22}}$ & $\frac{92}{5}$ & $\left(\frac{6}{5},\frac{8}{5}\right)$ & $92$ & \tiny{(1196,7475)} & $\mathbf{III_{37}}$ & \tiny{$\left(1,2\right)$} & $228$ \\
15. & 14 & $\left(\frac{3}{4},\frac{3}{2}\right)$ & $266$ & \tiny{(56,$56^2$)} & $E_{7,1}^{\otimes 2}$ & 18 & $\left(\frac{5}{4},\frac{3}{2}\right)$ & $198$ & \tiny{$(9\cdot2^7,75\cdot2^6)$} & $\text{GHM}_{198}$ & \tiny{$\left(2,1\right)$} & $464$ \\
16. & 14 & $\left(\frac{3}{2},\frac{3}{4}\right)$ & $266$ & \tiny{($56^2$,56)} & $E_{7,1}^{\otimes 2}$ & 18 & $\left(\frac{1}{2},\frac{9}{4}\right)$ & $630$ & \tiny{(36,$2^{17}$)} & $D_{18,1}$ & \tiny{$\left(1,2\right)$} & $896$ \\
17. & 14 & $\left(\frac{1}{2},\frac{7}{4}\right)$ & $378$ & \tiny{(28,$2^{13}$)} & $D_{14,1}$ & 18 & $\left(\frac{3}{2},\frac{5}{4}\right)$ & $198$ & \tiny{$(75\cdot2^6,9\cdot2^7)$} & $\text{GHM}_{198}$ & \tiny{$\left(1,2\right)$} & $576$ \\
18. & $\frac{29}{2}$ & $\left(\frac{1}{2},\frac{29}{16}\right)$ & $406$ & \tiny{(29,$2^{14}$)} & $B_{14,1}$ & $\frac{35}{2}$ &  $\left(\frac{3}{2},\frac{19}{16}\right)$ & $210$ & \tiny{$(4655,35\cdot2^5)$} & $\text{GHM}_{210}$ & \tiny{$\left(1,1\right)$} & $616$ \\
19. & 15 & $\left(\frac{1}{2},\frac{15}{8}\right)$ & $435$ & \tiny{(30,$2^{14}$)} & $D_{15,1}$ & 17 & $\left(\frac{3}{2},\frac{9}{8}\right)$ & $221$ & \tiny{$(4488,544)$} & $\text{GHM}_{221}$ & \tiny{$\left(1,2\right)$} & $656$ \\
20. & 15 & $\left(\frac{7}{8},\frac{3}{2}\right)$ & $255$ & \tiny{(120,3640)} & $\text{GHM}_{255}$ & 17 & $\left(\frac{9}{8},\frac{3}{2}\right)$ & $221$ & \tiny{(544,4488)} & $\text{GHM}_{221}$ & \tiny{$\left(2,1\right)$} & $476$ \\
21. & 15 & $\left(\frac{3}{2},\frac{7}{8}\right)$ & $255$ & \tiny{$(3640,120)$} & $\text{GHM}_{255}$ & 17 & $\left(\frac{1}{2},\frac{17}{8}\right)$ & $561$ & \tiny{(34,$2^{16}$)} & $D_{17,1}$ & \tiny{$\left(1,2\right)$} & $816$ \\
22. & $\frac{31}{2}$ & $\left(\frac{3}{2},\frac{15}{16}\right)$ & $248$ & \tiny{(3875,248)} & $E_{8,2}$ & $\frac{33}{2}$ & $\left(\frac{1}{2},\frac{33}{16}\right)$ & $528$ & \tiny{(33,$2^{16}$)} & $B_{16,1}$ & \tiny{$\left(1,1\right)$} & $776$ \\
\hline
\hline
\rowcolor{Mywhite}\caption{CFT pairings, $c^\mathcal{H}=32$ with $(n_1,n_2)=(2,3)$. The integer $\cN$ in the last column gives the total number of dimension-1 states in the meromorphic character $j^{\frac23}(j-992+\cN)$.}
\label{T15}
\end{longtable}

\noindent {\bf Conclusion:} From Table \ref{T15} we were not able to characterise any admissible solutions as CFTs or otherwise, but rather started to see several predictions of meromorphic theories at $c=32$. Details of these were presented in \cite{Das:2022slz}.

\subsubsection*{Comments on Table \ref{T989}}

Here all the entries are of IVOA-type and all of these were previously characterised.

\setlength\LTleft{-10pt}
\setlength\LTright{0pt}
\rowcolors{2}{Mygrey}{Mywhite}
\begin{longtable}{l||ccccc||ccccc||cc} 
\hline
\hline
\rowcolor{Mywhite}\# & $c$ & $(h_1,h_2)$ & $m_1$ & \tiny{$(D_1,D_2)$} & $\mathcal{W}$ & $\Tilde{c}$ & $(\Tilde{h}_1,\Tilde{h}_2)$ & $\tilde{m}_1$ & \tiny{$(\Tilde{D}_1,\Tilde{D}_2)$} & $\widetilde{\mathcal{W}}$ & \tiny{$(d_1,d_2)$} & $\mathcal{N}$  \\
\hline
\hline
1. & $\frac{60}{7}$ & $\left(\frac{3}{7},\frac{8}{7}\right)$ & $210$ & \tiny{(10,285)} & ${\bf III_{7}}$ & $\frac{164}{7}$ & $\left(\frac{11}{7},\frac{13}{7}\right)$ & $41$ & \tiny{(4797,50922)} & ${\bf III_{54}}$ & \tiny{$\left(1,1\right)$} & $251$ \\
2. & $\frac{44}{5}$ & $\left(\frac{2}{5},\frac{6}{5}\right)$ & $220$ &  \tiny{(11,275)} & ${\bf III_{8}}$ & $\frac{116}{5}$ & $\left(\frac{8}{5},\frac{9}{5}\right)$ & $58$ & \tiny{(4959,27550)} & ${\bf III_{52}}$ & \tiny{$\left(1,2\right)$} & $278$ \\
3. & $\frac{68}{7}$ & $\left(\frac{3}{7},\frac{9}{7}\right)$ & $221$ & \tiny{(17,782)} & ${\bf III_{12}}$ & $\frac{156}{7}$ & $\left(\frac{11}{7},\frac{12}{7}\right)$ & $78$ & \tiny{(5070,27170)} & ${\bf III_{47}}$ & \tiny{$\left(1,1\right)$} & $299$ \\
4. & $\frac{68}{7}$ & $\left(\frac{9}{7},\frac{3}{7}\right)$ & $221$ &  \tiny{(782,17)} & ${\bf III_{12}}$ & $\frac{156}{7}$ & $\left(\frac{5}{7},\frac{18}{7}\right)$ & $1248$ & \tiny{(130,799500)} & ${\bf III_{48}}$ & \tiny{$\left(1,1\right)$} & $1469$ \\
5. & $\frac{76}{5}$ &  $\left(\frac{4}{5},\frac{8}{5}\right)$ & $380$ & \tiny{(57,$57^{2}$)} & $E_{7.5}^{\otimes 2}$ & $\frac{84}{5}$ & $\left(\frac{6}{5},\frac{7}{5}\right)$ & $336$ & \tiny{$(770,1452)$} & ${\bf III_{33}}$ & \tiny{$\left(2,1\right)$} & $716$ \\
6. & $\frac{76}{5}$ & $\left(\frac{3}{5},\frac{9}{5}\right)$ & $437$ & \tiny{(57,11875)} & ${\bf III_{28}}$ & $\frac{84}{5}$ & $\left(\frac{7}{5},\frac{6}{5}\right)$ & $336$ & \tiny{(1452,770)} & ${\bf III_{33}}$ & \tiny{$\left(1,2\right)$} & $773$ \\
7. & $\frac{100}{7}$ & $\left(\frac{4}{7},\frac{12}{7}\right)$ & $380$ & \tiny{(55,11495)} & ${\bf III_{25}}$ & $\frac{124}{7}$ & $\left(\frac{10}{7},\frac{9}{7}\right)$ & $248$ & \tiny{(2108,2108)} & ${\bf III_{35}}$ & \tiny{$\left(1,1\right)$} & $628$ \\
8. & $\frac{100}{7}$ & $\left(\frac{5}{7},\frac{11}{7}\right)$ & $325$ & \tiny{(55,2925)} & ${\bf III_{24}}$ & $\frac{124}{7}$ & $\left(\frac{9}{7},\frac{10}{7}\right)$ & $248$ & \tiny{(2108,2108)} & ${\bf III_{35}}$ & \tiny{$\left(1,1\right)$} & $573$ \\
9. & $\frac{108}{7}$ & $\left(\frac{6}{7},\frac{11}{7}\right)$ & $378$ & \tiny{(117,3510)} & ${\bf III_{29}}$ & $\frac{116}{7}$ & $\left(\frac{8}{7},\frac{10}{7}\right)$ & $348$ & \tiny{(725,1972)} & ${\bf III_{32}}$ & \tiny{$\left(1,1\right)$} & $726$ \\
10. & $\frac{108}{7}$ & $\left(\frac{4}{7},\frac{13}{7}\right)$ & $456$ & \tiny{(39,20424)} & ${\bf III_{30}}$ & $\frac{116}{7}$ & $\left(\frac{10}{7},\frac{8}{7}\right)$ & $348$ & \tiny{(1972,725)} & ${\bf III_{32}}$ & \tiny{$\left(1,1\right)$} & $804$ \\
\hline
\hline
\rowcolor{Mywhite}\caption{IVOA-type pairings, $c^\mathcal{H}=32$ with $(n_1,n_2)=(2,3)$. The integer $\cN$ in the last column gives the total number of dimension-1 states in the meromorphic character $j^{\frac23}(j-992+\cN)$.}
\label{T989}
\end{longtable}



\subsubsection*{Comments on Table \ref{T17}}

This table consists entirely of inconsistent pairings. In row 1 we see such a pairing between known theories: the value of $d_1$ is fractional. This corresponds to the non-existence of a meromorphic extension $\cE_1[D_{4,1}^2D_{24,1}]$. If such an extension existed it would be a Kervaire lattice, however this does not appear in the list of Kervaire lattices, in agreement with the fact that the pairing is inconsistent. 

In row 6 we have a pairing between $D_{9,1}$ and ${\bf III_{51}}$, however the result has $\cN=2476$ which is greater than the maximum allowed value of 2016 at $c=32$. This means ${\bf III_{51}}$ is not a CFT, consistent with our conclusion from Table \ref{T11}.

In row 14 we have a pairing between $D_{10,1}$ and ${\bf III_{46}}$ with a total $\cN=1488$. However ${\bf III_{46}}$ has been ruled out, and we now argue that this implies the corresponding meromorphic character is not a CFT. This crucially depends on the fact that the pairing has $n_1,n_2>1$. In such pairings, the meromorphic theory -- if any -- has a Kac-Moody algebra that is the direct sum of those of the paired solutions. Thus we can conclude that there is no meromorphic theory at $c^\cH=32$ with $\cN=1488$ {\em and} a $D_{10,1}$ factor. Similar considerations hold for rows 15, 16, 23, 24, 27-32, 36, 37, 39, 40 where in each case we get constraints ruling out specific possibilities for meromorphic theories at $c=32$. The details are a little complicated to work out in some cases, where the valid CFT in the pairing is of GHM type. In these cases one has to look in \cite{Gaberdiel:2016zke} for the definition of the theory in terms of a meromorphic theory of Schellekens type and then read off the answer from \cite{Schellekens:1992db}. The results are summarised below.

In rows 25, 26, both solutions are of type ${\bf V}$. However for row 25 we have ruled out one member, ${\bf V_{18}}$, and characterised ${\bf V_{39}}$ in \eref{idents}, and for row 26 we have already ruled out both members of the pair ${\bf V_{18}}$ (again) and ${\bf V_{40}}$. Note that we do not get a constraint on meromorphic theories in these cases. All remaining rows have an entry of \sout{${\bf III}$} type, from which we typically do not get new information.

\setlength\LTleft{-35pt}
\setlength\LTright{0pt}
\rowcolors{2}{Mygrey}{Mywhite}
\begin{longtable}{l||ccccc||ccccc||cc}
\hline
\hline
\rowcolor{Mywhite}\# & $c$ & $(h_1,h_2)$ & $m_1$ & \tiny{$(D_1,D_2)$} & $\mathcal{W}$ & $\Tilde{c}$ & $(\Tilde{h}_1,\Tilde{h}_2)$ & $\tilde{m}_1$ & \tiny{$(\Tilde{D}_1,\Tilde{D}_2)$} & $\widetilde{\mathcal{W}}$ & \tiny{$(d_1,d_2)$} & $\mathcal{N}$  \\
\hline
\hline
1. & $\frac{17}{2}$ & $\left(\frac{17}{16},\frac{1}{2}\right)$ & $136$ & \tiny{(256,17)} & $B_{8,1}$ & $\frac{47}{2}$ & $\left(\frac{15}{16},\frac{5}{2}\right)$ & $4371$ & \tiny{(4371,1135003)} & \sout{${\bf III_{55}}$} & \tiny{$\left(\frac{1}{2},1\right)$} & $4507$ \\
2. & $\frac{17}{2}$ & $\left(\frac{1}{16},\frac{3}{2}\right)$ & $255$ & \tiny{(17,221)} & \sout{${\bf III_{6}}$} & $\frac{47}{2}$ & $\left(\frac{31}{16},\frac{3}{2}\right)$ & $0$ &  \tiny{$(47\cdot2^{11},4371)$} & BM & \tiny{$\left(\frac{1}{16},1\right)$} & $255$ \\
3. & $\frac{17}{2}$ & $\left(\frac{3}{2},\frac{1}{16}\right)$ & $255$ &  \tiny{(221,17)} & \sout{${\bf III_{6}}$} & $\frac{47}{2}$ & $\left(\frac{1}{2},\frac{47}{16}\right)$ & $1081$ & \tiny{(47,$2^{23}$)} & $B_{23,1}$ & \tiny{$\left(1,\frac{1}{16}\right)$} & $1336$ \\
4. & $\frac{44}{5}$ & $\left(\frac{1}{5},\frac{7}{5}\right)$ & $253$ &  \tiny{(11,242)} & \sout{${\bf III_{9}}$} & $\frac{116}{5}$ & $\left(\frac{9}{5},\frac{8}{5}\right)$ & $58$ & \tiny{$(27550,4959)$} & ${\bf III_{52}}$ & \tiny{$\left(\frac15,1\right)$} & $311$ \\
5. & $\frac{44}{5}$ & $\left(\frac{6}{5},\frac{2}{5}\right)$ & $220$ & \tiny{$(275,11)$} & ${\bf III_{8}}$ & $\frac{116}{5}$ & $\left(\frac{4}{5},\frac{13}{5}\right)$ & $1711$ &  \tiny{(1653,910803)} & \sout{${\bf III_{53}}$} & \tiny{$\left(\frac15,1\right)$} & $1931$ \\
6. & 9 & $\left(\frac{9}{8},\frac{1}{2}\right)$ & $153$ & \tiny{(256,18)} & $D_{9,1}$ & 23 & $\left(\frac{7}{8},\frac{5}{2}\right)$ & $2323$ & \tiny{$(575,32683\cdot2^5)$} & ${\bf III_{51}}$ & \tiny{$\left(2,1\right)$} & $2476$ \\
7. & 9 & $\left(\frac{1}{8},\frac{3}{2}\right)$ & $261$ & \tiny{$(9,456)$} & \sout{${\bf III_{10}}$} & 23 & $\left(\frac{15}{8},\frac{3}{2}\right)$ & $23$ & \tiny{$(23\cdot2^{11},4600)$} & ${\bf III_{50}}$ & \tiny{$\left(\frac14,1\right)$} & 284 \\
8. & 9 & $\left(\frac{3}{2},\frac{1}{8}\right)$ & $261$ & \tiny{$(456,9)$} & \sout{${\bf III_{10}}$} & 23 & $\left(\frac{1}{2},\frac{23}{8}\right)$ & $1035$ & \tiny{(46,$2^{22}$)} & $D_{23,1}$ & \tiny{$\left(1,\frac{1}{4}\right)$} & $1296$ \\
9. & $\frac{19}{2}$ & $\left(\frac{19}{16},\frac{1}{2}\right)$ & $171$ & \tiny{($2^9$,19)} & $B_{9,1}$ & $\frac{45}{2}$ & $\left(\frac{13}{16},\frac{5}{2}\right)$ & $1640$ & \tiny{(1595,956449)} & \sout{${\bf III_{49}}$} & \tiny{$\left(\frac{1}{4},1\right)$} & $1811$ \\
10. & $\frac{19}{2}$ & $\left(\frac{3}{16},\frac{3}{2}\right)$ & $266$ & \tiny{(19,703)} & \sout{${\bf III_{11}}$} & $\frac{45}{2}$ & $\left(\frac{29}{16},\frac{3}{2}\right)$ & $45$ & \tiny{$(45\cdot2^{10},4785)$} & $\text{GHM}_{45}$ & \tiny{$\left(\frac18,1\right)$} & $311$ \\
11. & $\frac{19}{2}$ & $\left(\frac{3}{2},\frac{3}{16}\right)$ & $266$ & \tiny{(703,19)} & \sout{${\bf III_{11}}$} & $\frac{45}{2}$ & $\left(\frac{1}{2},\frac{45}{16}\right)$ & $990$ & \tiny{(45,$2^{22}$)} & $B_{22,1}$ & \tiny{$\left(1,\frac{1}{8}\right)$} & $1256$ \\
12. & $10$ & $\left(\frac{1}{4},\frac{3}{2}\right)$ & $270$ & \tiny{$(5,960)$} & \sout{${\bf III_{13}}$} & 22 & $\left(\frac{7}{4},\frac{3}{2}\right)$ & $66$ & \tiny{$(11\cdot2^{11},77\cdot2^6)$} & ${\bf III_{45}}$ & \tiny{$\left(1,1\right)$} & $336$ \\
13. & $10$ & $\left(\frac{3}{2},\frac{1}{4}\right)$ & $270$ & \tiny{$(960,5)$} & \sout{${\bf III_{13}}$} & 22 & $\left(\frac{1}{2},\frac{11}{4}\right)$ & $946$ & \tiny{(44,$2^{21}$)} & $D_{22,1}$ & \tiny{$\left(1,1\right)$} & $1216$ \\
14. & $10$ & $\left(\frac{5}{4},\frac{1}{2}\right)$ & $190$ & \tiny{($2^9$,20)} & $D_{10,1}$ & 22 & $\left(\frac{3}{4},\frac{5}{2}\right)$ & $1298$ & \tiny{$(154,847\cdot2^{10})$} & ${\bf III_{46}}$ & \tiny{$\left(2,1\right)$} & $1488$ \\
15. & $\frac{52}{5}$ & $\left(\frac{3}{5},\frac{6}{5}\right)$ & $104$ & \tiny{(26,$26^2$)} & $F_{4,1}^{\otimes 2}$ & $\frac{108}{5}$ & $\left(\frac{7}{5},\frac{9}{5}\right)$ & $27$ & \tiny{(2295,42483)} & ${\bf III_{42}}$ & \tiny{$\left(2,1\right)$} & $131$ \\
16. & $\frac{52}{5}$ & $\left(\frac{6}{5},\frac{3}{5}\right)$ & $104$ & \tiny{($26^2$,26)} & $F_{4,1}^{\otimes 2}$ & $\frac{108}{5}$ & $\left(\frac{4}{5},\frac{12}{5}\right)$ & $1404$ & \tiny{$(459,153\cdot5^5)$} & ${\bf III_{44}}$ & \tiny{$\left(1,2\right)$} & $1508$ \\
17. & $\frac{21}{2}$ & $\left(\frac{5}{16},\frac{3}{2}\right)$ & $273$ &  \tiny{(21,1225)} & \sout{${\bf III_{14}}$} & $\frac{43}{2}$ & $\left(\frac{27}{16},\frac{3}{2}\right)$ & $86$ & \tiny{$(43\cdot2^9,5031)$} & $\text{GHM}_{86}$ & \tiny{$\left(\frac14,1\right)$} & $359$ \\
18. & $\frac{21}{2}$ & $\left(\frac{3}{2},\frac{5}{16}\right)$ & $273$ & \tiny{(1225,21)} & \sout{${\bf III_{14}}$} & $\frac{43}{2}$ & $\left(\frac{1}{2},\frac{43}{16}\right)$ & $903$ & \tiny{(43,$2^{21}$)} & $B_{21,1}$ & \tiny{$\left(1,\frac{1}{4}\right)$} & $1176$ \\
19. & 11 & $\left(\frac{3}{8},\frac{3}{2}\right)$ & $275$ & \tiny{$(11,1496)$} & \sout{${\bf III_{15}}$} & 21 & $\left(\frac{13}{8},\frac{3}{2}\right)$ & $105$ & \tiny{$(21\cdot2^9,5096)$} & $\text{GHM}_{105}$ & \tiny{$\left(1,1\right)$} & $380$ \\
20. & 11 & $\left(\frac{3}{2},\frac{3}{8}\right)$ & $275$ & \tiny{$(1496,11)$} & \sout{${\bf III_{15}}$} & 21 & $\left(\frac{1}{2},\frac{21}{8}\right)$ & $861$ & \tiny{(42,$2^{20}$)} & $D_{21,1}$ & \tiny{$\left(1,1\right)$} & $1136$ \\
21. & $\frac{23}{2}$ & $\left(\frac{7}{16},\frac{3}{2}\right)$ & $276$ & \tiny{(23,1771)} & \sout{${\bf III_{16}}$} & $\frac{41}{2}$ & $\left(\frac{25}{16},\frac{3}{2}\right)$ & $123$ & \tiny{$(41\cdot2^8,5125)$} & $\text{GHM}_{123}$ & \tiny{$\left(\frac12,1\right)$} & $399$ \\
22. & $\frac{23}{2}$ & $\left(\frac{3}{2},\frac{7}{16}\right)$ & $276$ & \tiny{(1771,23)} & \sout{${\bf III_{16}}$} & $\frac{41}{2}$ & $\left(\frac{1}{2},\frac{41}{16}\right)$ & $820$ & \tiny{(41,$2^{20}$)} & $B_{20,1}$ & \tiny{$\left(1,\frac{1}{2}\right)$} & $1096$ \\
23. & 12 & $\left(\frac{4}{3},\frac{2}{3}\right)$ & $156$ & \tiny{($27^2$,27)} & $E_{6,1}^{\otimes 2}$ & 20 & $\left(\frac{2}{3},\frac{7}{3}\right)$ & $890$ & \tiny{$(135,20\cdot3^9)$} & ${\bf V_{41}}$ & \tiny{$\left(2,2\right)$} & $1046$ \\
24. & 12 & $\left(\frac{3}{5},\frac{7}{5}\right)$ & $222$ &  \tiny{$(25,1275)$} & ${\bf III_{17}}$ & 20 & $\left(\frac{7}{5},\frac{8}{5}\right)$ & $120$ & \tiny{$(4\cdot5^4,13\cdot5^4)$} & $\text{GHM}_{120}$ & \tiny{$\left(2,2\right)$} & $342$ \\
25. & 12 & $\left(\frac{1}{3},\frac{5}{3}\right)$ & $318$ & \tiny{$(3^2,2\cdot3^7)$} & ${\bf V_{18}}$ & 20 & $\left(\frac{5}{3},\frac{4}{3}\right)$ & $80$ & \tiny{$(2^3\cdot3^7,10\cdot3^5)$} & ${\bf V_{39}}$ & \tiny{$\left(1,1\right)$} & $398$ \\
26. & 12 & $\left(\frac{5}{3},\frac{1}{3}\right)$ & $318$ & \tiny{$(2\cdot3^7,3^2)$} & ${\bf V_{18}}$ & 20 & $\left(\frac{1}{3},\frac{8}{3}\right)$ & $718$ & \tiny{$(12,2\cdot3^{12})$} & ${\bf V_{40}}$ & \tiny{$\left(1, 1\right)$} & $1046$ \\
27. & $\frac{25}{2}$ & $\left(\frac{9}{16},\frac{3}{2}\right)$ & $275$ & \tiny{(25,2325)} & ${\bf III_{19}}$ & $\frac{39}{2}$ & $\left(\frac{23}{16},\frac{3}{2}\right)$ & $156$ & \tiny{$(39\cdot2^7,5083)$} & $\text{GHM}_{156}$ & \tiny{$\left(1,1\right)$} & $431$ \\
28. & $\frac{25}{2}$ & $\left(\frac{3}{2},\frac{9}{16}\right)$ & $275$ &  \tiny{(2325,25)} & ${\bf III_{19}}$ & $\frac{39}{2}$ & $\left(\frac{1}{2},\frac{39}{16}\right)$ & $741$ & \tiny{(39,$2^{19}$)} & $B_{19,1}$ & \tiny{$\left(1,1\right)$} & $1016$ \\
29. & 13 & $\left(\frac{5}{8},\frac{3}{2}\right)$ & $273$ & \tiny{$(26,2600)$} & ${\bf III_{20}}$ & 19 & $\left(\frac{11}{8},\frac{3}{2}\right)$ & $171$ & \tiny{$(19\cdot2^7,5016)$} & $\text{GHM}_{171}$ & \tiny{$\left(2,1\right)$} & $444$ \\
30. & 13 & $\left(\frac{3}{2},\frac{5}{8}\right)$ & $273$ & \tiny{(2600,26)} & ${\bf III_{20}}$ & 19 & $\left(\frac{1}{2},\frac{19}{8}\right)$ & $703$ & \tiny{(38,$2^{18}$)} & $D_{19,1}$ & \tiny{$\left(1,2\right)$} & $976$ \\
31. & $\frac{27}{2}$ & $\left(\frac{11}{16},\frac{3}{2}\right)$ & $270$ & \tiny{$(54,2871)$} & ${\bf III_{21}}$ & $\frac{37}{2}$ & $\left(\frac{21}{16},\frac{3}{2}\right)$ & $185$ & \tiny{$(37\cdot2^6,4921)$} & $\text{GHM}_{185}$ & \tiny{$\left(1,1\right)$} & $455$ \\
32. & $\frac{27}{2}$ & $\left(\frac{3}{2},\frac{11}{16}\right)$ & $270$ & \tiny{$(2871,54)$} & ${\bf III_{21}}$ & $\frac{37}{2}$ & $\left(\frac{1}{2},\frac{37}{16}\right)$ & $666$ & \tiny{(37,$2^{18}$)} & $B_{18,1}$ & \tiny{$\left(1,1\right)$} & $936$ \\
33. & $\frac{68}{5}$ & $\left(\frac{2}{5},\frac{9}{5}\right)$ & $374$ & \tiny{(119,12138)} & \sout{${\bf III_{23}}$} & $\frac{92}{5}$ & $\left(\frac{8}{5},\frac{6}{5}\right)$ & $92$ & \tiny{$(7475,1196)$} & $\mathbf{III_{37}}$ & \tiny{$\left(\frac15,1\right)$} & $466$ \\
34. & $\frac{68}{5}$ & $\left(\frac{7}{5},\frac{4}{5}\right)$ & $136$ & \tiny{$(1700,119)$} & $\mathbf{III_{22}}$ & $\frac{92}{5}$ & $\left(\frac{3}{5},\frac{11}{5}\right)$ & $690$ & \tiny{$(299,178802)$} & \sout{${\bf III_{38}}$} & \tiny{$\left(\frac25,1\right)$} & $826$ \\
35. & 14 & $\left(\frac{7}{4},\frac{1}{2}\right)$ & $378$ & \tiny{($2^{13}$,28)} & $D_{14,1}$ & 18 & $\left(\frac{1}{4},\frac{5}{2}\right)$ & $598$ & \tiny{$(25,221\cdot2^{10})$} & \sout{${\bf III_{36}}$} & \tiny{$\left(\frac{1}{4},1\right)$} & $976$ \\
36. & $\frac{29}{2}$ & $\left(\frac{13}{16},\frac{3}{2}\right)$ & $261$ & \tiny{$(116,3393)$} & ${\bf III_{26}}$ & $\frac{35}{2}$ & $\left(\frac{19}{16},\frac{3}{2}\right)$ & $210$ & \tiny{$(35\cdot2^5,4655)$} & $\text{GHM}_{210}$ & \tiny{$\left(1,1\right)$} & $471$ \\
37. & $\frac{29}{2}$ & $\left(\frac{3}{2},\frac{13}{16}\right)$ & $261$ & \tiny{$(3393,116)$} & ${\bf III_{26}}$ & $\frac{35}{2}$ & $\left(\frac{1}{2},\frac{35}{16}\right)$ & $595$ & \tiny{(35,$2^{17}$)} & $B_{17,1}$ & \tiny{$\left(1,1\right)$} & $856$ \\
38. & $\frac{76}{5}$ & $\left(\frac{9}{5},\frac{3}{5}\right)$ & $437$ & \tiny{$(11875,57)$} & ${\bf III_{28}}$ & $\frac{84}{5}$ & $\left(\frac{1}{5},\frac{12}{5}\right)$ & $534$ & \tiny{(33,55924)} & \sout{${\bf III_{34}}$} & \tiny{$\left(\frac{2}{25},1\right)$} & $971$ \\
39. & $\frac{31}{2}$ & $\left(\frac{1}{2},\frac{31}{16}\right)$ & $465$ & \tiny{(31,$2^{15}$)} & $B_{15,1}$ & $\frac{33}{2}$ & $\left(\frac{3}{2},\frac{17}{16}\right)$ & $231$ & \tiny{$(4301,528)$} & ${\bf III_{31}}$ & \tiny{$\left(1,1\right)$} & $696$ \\
40. & $\frac{31}{2}$ & $\left(\frac{15}{16},\frac{3}{2}\right)$ & $248$ & \tiny{(248,3875)} & $E_{8,2}$ & $\frac{33}{2}$ & $\left(\frac{17}{16},\frac{3}{2}\right)$ & $231$ & \tiny{$(528,4301)$} & ${\bf III_{31}}$ & \tiny{$\left(1,1\right)$} & $479$ \\
\hline
\hline
\rowcolor{Mywhite}\caption{Inconsistent pairings, $c^\mathcal{H}=32$ with $(n_1,n_2)=(2,3)$ The integer $\cN$ in the last column gives the total number of dimension-1 states in the meromorphic character $j^{\frac23}(j-992+\cN)$.}
\label{T17}
\end{longtable}

\noindent {\bf Conclusion:} From Table \ref{T17} we were not able to newly rule out any solutions from being CFTs, but instead we were able to predict the absence of meromorphic theories with the following values of $\cN$ coupled with a particular factor in their Kac-Moody algebra. This happens when either one solution in the bilinear pair is a WZW theory (or a known RCFT) and the other solution has integral $Y_i$ values. Furthermore, these two solutions also have a nice bilinear pairing, that is, $d_i$s are integral. In addition to the above two conditions, the $\cN$ value of this bilinear pair must be less than or equal to $2016$ which is the dimension of the Kac-Moody algebra $D_{32,1}$. Table \ref{T1799} lists the cases for which meromorphic theories at $c=32$ with given values of $\cN$ and simple factor in their Kac-Moody algebras have been ruled out. 

\setlength\LTleft{145pt}
\setlength\LTright{0pt}
\rowcolors{2}{Mygrey}{Mywhite}
\begin{longtable}{l||c||c}
\hline
\hline
\rowcolor{Mywhite}\# & $\cN$ & Factor \\
\hline
\hline
1. & 131 & $F_{4,1}^{\otimes 2}$ \\
2. & 342 & $A_{4,1}^{\otimes 5}$ \\
3. & 342 &  $A_{9,2}B_{3,1}$ \\
4. & 431 & $D_{8,2}B_{4,1}$ \\
5. & 431 & $C_{6,1}^{\otimes 2}$ \\
6. & 444 & $A_{7,1}^{\otimes 2} D_{5,1}$ \\
7. & 455 & $E_{7,2}F_{4,1}$ \\
8. & 471 & $C_{10,1}$ \\
9. & 479 & $E_{8,2}$ \\
10. & 696 & $B_{15,1}$ \\
11. & 856 & $B_{17,1}$ \\
12. & 936 & $B_{18,1}$ \\
13. & 976 & $D_{19,1}$ \\
14. & 1016 & $B_{19,1}$ \\
15. & 1046 & $E_{6,1}^{\otimes 2}$ \\
16. & 1488 & $D_{10,1}$ \\
17. & 1508 & $F_{4,1}^{\otimes 2}$ \\
\hline
\hline
\rowcolor{Mywhite}\caption{List of meromorphic theories ruled out by Table \ref{T17}}
\label{T1799}
\end{longtable}
As a mild check of these predictions, wherever the algebra listed above is simply laced and of level 1 one can check  from \cite{kervaire1994unimodular} that there are no lattice theories with complete root systems at $c=32$ having these dimensions and subalgebras.

\subsection{Cosets of $c^\cH=40$}

In this subsection we classify all bilinear pairings that add up to a central charge of 40. From \eref{sumbound} this means $n_1+n_2=6$, from which we find the three possibilities $(n_1,n_2)=(1,5),(2,4)$ and $(3,3)$. We consider each one in turn. The meromorphic theory has the character $\chi^\mathcal{H}=j^{2/3}(j-1240+\cN)$ where $\mathcal{N}$ denotes the dimension of the Kac-Moody algebra.

\subsubsection*{$\underline{(n_1,n_2)=(1,5)}$}

As we saw at $c=24,32$, the consistent CFT pairings with $n_1=1$ are all of a standard kind, namely cosets of the $c^{\mathcal H}=40$ meromorphic theory $\mathcal{E}_1{[D_{40,1}]}$, whose Kac-Moody algebra has dimension 3160. 
Thus we have pairings of (i) $B_{r_1,1}$ and $B_{r_2,1}$ with  $r_1 + r_2 = 39,~ 0 \leq r_1, r_2 \leq 39$, (ii) $D_{r_1,1}$ and $D_{r_2,1}$ with $r_1 + r_2 = 40, ~ 1 \leq r_1, r_2 \leq 39$.
The pairing of $D_{20,1}$ is a self-coset relation. 

There are no IVOA-type pairings with $(n_1,n_2)=(1,5)$ so we move on to list the inconsistent pairings.

\subsection*{Comments on Table \ref{T20}}

In row 1 of this table we find ${\bf V_{63}}$ which we have so far been unable to characterise as a CFT or otherwise. It pairs with a consistent theory leading to an invariant at $c=40$ with 5344 currents. This is above the bound of 3160 for a meromorphic theory in this dimension (see Appendix \ref{upperb}), hence this is not a genuine pairing to a meromorphic theory at the level of CFT. We conclude that ${\bf V_{63}}$ is not a CFT. This was actually the last admissible character (other than those of IVOA type) to remain uncharacterised from our original list. 

In row 2 we have an inconsistent pairing, visible from the fractional value of one of the pair $(d_1,d_2)$, which implies the absence of a $c=40$ modular invariant with an algebra of dimension 2584. If the pairing existed then we would have a lattice theory $\cE_1[D_{4,1}^{\otimes 2}D_{32,1}]$. Hence such a theory should not exist. This is a prediction about lattices with complete root systems in 40 dimensions, which we were unable to independently confirm.

The pair in rows 7 and 13, and also ${\bf III_{46}}$ in row 11, have been ruled out by tables \ref{T6}, \ref{T11} and \ref{T14}. All the other entries are self-evidently inconsistent.

\setlength\LTleft{-40pt}
\setlength\LTright{0pt}
\rowcolors{2}{Mygrey}{Mywhite}
\begin{longtable}{l||ccccc||ccccc||cc} 
\hline
\hline
\rowcolor{Mywhite}\# & $c$ & $(h_1,h_2)$ & $m_1$ & \tiny{$(D_1,D_2)$} & $\mathcal{W}$ & $\Tilde{c}$ & $(\Tilde{h}_1,\Tilde{h}_2)$ & $\tilde{m}_1$ & \tiny{$(\Tilde{D}_1,\Tilde{D}_2)$} & $\widetilde{\mathcal{W}}$ & \tiny{$(d_1,d_2)$} & $\mathcal{N}$  \\
\hline
\hline
1. & $4$ & $\left(\frac{1}{3},\frac{2}{3}\right)$ & $16$ & \tiny{(3,9)} & $A_{2,1}^{\otimes 2}$ & 36 & $\left(\frac{2}{3},\frac{13}{3}\right)$ & $3384$  & \tiny{$(324,8\cdot3^{20})$} & ${\bf V_{63}}$ & \tiny{$\left(2,2\right)$} & $5344$ \\
2. & 8 & $\left(\frac{1}{2},1\right)$ & $56$ & \tiny{(8,64)} & $D_{4,1}^{\otimes 2}$ & 32 & $\left(\frac{1}{2},4\right)$ & $2016$ & \tiny{(64,$2^{31}$)} & $D_{32,1}$ & \tiny{$\left(3,\frac{57}{16}\right)$} & $3608$ \\
3. & $\frac{44}{5}$ & $\left(\frac{1}{5},\frac{7}{5}\right)$ & $253$ & \tiny{(11,242)} & \sout{${\bf III_{9}}$} & $\frac{156}{5}$ & $\left(\frac{4}{5},\frac{18}{5}\right)$ & $3612$ & \tiny{(14877,250774426)} & \sout{${\bf III_{62}}$} & \tiny{$\left(\frac{1}{250},1\right)$} & $\frac{1129897}{250}$ \\
4. & 9 & $\left(\frac{1}{8},\frac{3}{2}\right)$ & $261$ & \tiny{$(9,456)$} & \sout{${\bf III_{10}}$} & 31 & $\left(\frac{7}{8},\frac{7}{2}\right)$ & $5239$ & \tiny{$(9269,2295147\cdot2^7)$} & \sout{${\bf III_{61}}$} & \tiny{$\left(\frac{1}{32}, 1\right)$} & $\frac{259421}{32}$ \\
5. & $\frac{19}{2}$ & $\left(\frac{3}{16},\frac{3}{2}\right)$ & $266$ & \tiny{(19,703)} & \sout{${\bf III_{11}}$} & $\frac{61}{2}$ & $\left(\frac{13}{16},\frac{7}{2}\right)$ & $3599$ & \tiny{(47763,264580485)} & \sout{${\bf III_{60}}$} & \tiny{$\left(\frac{1}{512},1\right)$} & $\frac{2886377}{512}$ \\
6. & $10$ & $\left(\frac{1}{4},\frac{3}{2}\right)$ & $700$ & \tiny{$(5,960)$} & \sout{${\bf III_{13}}$} & 30 & $\left(\frac{3}{4},\frac{7}{2}\right)$ & $2778$ & \tiny{$(539,14421\cdot2^{14})$} & \sout{${\bf III_{59}}$} & \tiny{$\left(\frac{1}{2}, 1\right)$} & $\frac{8791}{2}$ \\
7. & 12 & $\left(\frac{1}{3},\frac{5}{3}\right)$ & $318$ & \tiny{$(9,2\cdot3^7)$} & ${\bf V_{18}}$ & 28 & $\left(\frac{2}{3},\frac{10}{3}\right)$ & $1948$ & \tiny{$(225,11\cdot2\cdot3^{14})$} & ${\bf V_{58}}$ & \tiny{$\left(1,1\right)$} & $4291$ \\
8. & $\frac{68}{5}$ & $\left(\frac{2}{5},\frac{9}{5}\right)$ & $374$ & \tiny{(119,12138)} & \sout{${\bf III_{23}}$} & $\frac{132}{5}$ & $\left(\frac{3}{5},\frac{16}{5}\right)$ & $1536$ & \tiny{(2392,47018049)} & \sout{${\bf III_{57}}$} & \tiny{$\left(\frac{1}{125},1\right)$} & $\frac{523398}{125}$ \\
9. & 14 & $\left(\frac{3}{4},\frac{3}{2}\right)$ & $266$ & \tiny{(56,$56^2$)} & $E_{7,1}^{\otimes 2}$ & 26 & $\left(\frac{1}{4},\frac{7}{2}\right)$ & $1118$ & \tiny{$(117,3315\cdot2^{14})$} & \sout{${\bf III_{56}}$} & \tiny{$\left(\frac{1}{16},1\right)$} & $\frac{3587}{2}$ \\
10. & $\frac{84}{5}$ & $\left(\frac{1}{5},\frac{12}{5}\right)$ & $534$ & \tiny{(33,55924)} & \sout{${\bf III_{34}}$} & $\frac{116}{5}$ & $\left(\frac{4}{5},\frac{13}{5}\right)$ & $1711$ & \tiny{(1653,910803)} & \sout{${\bf III_{53}}$} & \tiny{$\left(\frac{1}{125},1\right)$} & $\frac{335174}{125}$ \\
11. & 18 & $\left(\frac{1}{4},\frac{5}{2}\right)$ & $598$ & \tiny{$(25,221\cdot2^{10})$} & \sout{${\bf III_{36}}$} & 22 & $\left(\frac{3}{4},\frac{5}{2}\right)$ & $1298$ & \tiny{$(154,847\cdot2^{10})$} & ${\bf III_{46}}$ & \tiny{$\left(\frac{1}{4},1 \right)$} & $\frac{5717}{2}$ \\
12. & $\frac{92}{5}$ & $\left(\frac{3}{5},\frac{11}{5}\right)$ & $690$ & \tiny{(299,178802)} & \sout{${\bf III_{38}}$} & $\frac{108}{5}$ & $\left(\frac{2}{5},\frac{14}{5}\right)$ & $860$ & \tiny{(833,3015426)} & \sout{${\bf III_{43}}$} & \tiny{$\left(\frac{1}{125},1\right)$} & $\frac{442817}{125}$ \\
13. & 20 & $\left(\frac{1}{3},\frac{8}{3}\right)$ & $728$ & \tiny{$(12,2\cdot3^{12})$} & ${\bf V_{40}}$ & 20 & $\left(\frac{2}{3},\frac{7}{3}\right)$ & $890$ & \tiny{$(135,10\cdot2\cdot3^9)$} & ${\bf V_{41}}$ & \tiny{$\left(1,1\right)$} & $3238$ \\
\hline
\hline
\rowcolor{Mywhite}\caption{Inconsistent pairings, $c^\mathcal{H}=40$ with $(n_1,n_2)=(1,5)$.}
\label{T20}
\end{longtable}

{\bf Conclusion:} From Table \ref{T20} we learn for the first time that ${\bf V_{63}}$ is not a CFT. With this we have completed the characterisation of all admissible solutions appearing in table \ref{T0}. We also predict that there is no meromorphic theory at $c=40$ of the form $\cE_1[D_{4,1}^{\otimes 2}D_{32,1}]$.

\subsubsection*{$\underline{(n_1,n_2)=(2,4)}$}

We move on to pairings with $(n_1,n_2)=(2,4)$. In these cases (as well as the ones to follow with $(n_1,n_2)=(3,3)$), no non-trivial embeddings of Kac-Moody algebras can be involved, as we explained earlier. Thus they are relatively simpler to deal with.

\subsection*{Comments on Table \ref{T21}}

This table is made up entirely of consistent bilinear pairings of known theories. Note that ${\bf III_{50}}$ and ${\bf III_{45}}$ have previously been characterised as CFT in 
\eref{idents}. Hence these pairings are predictions about the existence of meromorphic theories at $c=40$. More details of these predictions can be found in \cite{Das:2022slz}.

\setlength\LTleft{-45pt}
\setlength\LTright{0pt}
\rowcolors{2}{Mygrey}{Mywhite}
\begin{longtable}{l||ccccc||ccccc||cc}
\hline
\hline
\rowcolor{Mywhite}\# & $c$ & $(h_1,h_2)$ & $m_1$ & \tiny{$(D_1,D_2)$} & $\mathcal{W}$ & $\Tilde{c}$ & $(\Tilde{h}_1,\Tilde{h}_2)$ & $\Tilde{m}_1$ & \tiny{$(\Tilde{D}_1,\Tilde{D}_2)$} & $\widetilde{\mathcal{W}}$ & \tiny{$(d_1,d_2)$} & $\mathcal{N}$  \\
\hline
\hline
1. & 12 & $\left(\frac{3}{2},\frac{1}{2}\right)$ & 276 & \tiny{($2^{11}$,24)} & $D_{12,1}$ & 28 & $\left(\frac{1}{2},\frac{7}{2}\right)$ & 1540 & \tiny{(56,$2^{27}$)} & $D_{28,1}$ & \tiny{$(1,1)$} & $1816$ \\
2. & 14 & $\left(\frac{3}{2},\frac{3}{4}\right)$ & 266 & \tiny{($56^2$,56)} & $E_{7,1}^{\otimes 2}$ & 26 & $\left(\frac{1}{2},\frac{13}{4}\right)$ & 1326 & \tiny{(52,$2^{25}$)} & $D_{26,1}$ & \tiny{$\left(1,2\right)$} & $1592$ \\
3. & 15 & $\left(\frac{3}{2},\frac{7}{8}\right)$ & 255 & \tiny{$(3640,120)$} & $\text{GHM}_{255}$ & 25 & $\left(\frac{1}{2},\frac{25}{8}\right)$ & 1225 & \tiny{(50,$2^{24}$)} & $D_{25,1}$ & \tiny{$\left(1,2\right)$} & $1480$ \\
4. & $\frac{31}{2}$ & $\left(\frac{3}{2},\frac{15}{16}\right)$ & 248 & \tiny{(3875,248)} & $E_{8,2}$ & $\frac{49}{2}$ & $\left(\frac{1}{2},\frac{49}{16}\right)$ & 1176 & \tiny{(49,$2^{24}$)} & $B_{24,1}$ & \tiny{$\left(1,1\right)$} & $1424$ \\
5. & $\frac{33}{2}$ & $\left(\frac{1}{2},\frac{33}{16}\right)$ & 528 & \tiny{(33,$2^{16}$)} & $B_{16,1}$ & $\frac{47}{2}$ & $\left(\frac{3}{2},\frac{31}{16}\right)$ & 0 & \tiny{$(4371,47\cdot2^{11})$} & BM & \tiny{$\left(1,1\right)$} & $528$ \\
6. & 17 & $\left(\frac{3}{2},\frac{9}{8}\right)$ & 221 & \tiny{$(561\cdot2^3,544)$} & $\text{GHM}_{221}$ & 23 & $\left(\frac{1}{2},\frac{23}{8}\right)$ & 1035 & \tiny{(46,$2^{22}$)} & $D_{23,1}$ & \tiny{$\left(1,2\right)$} & $1256$ \\
7. & 17 & $\left(\frac{1}{2},\frac{17}{8}\right)$ & 561 & \tiny{(34,$2^{16}$)} & $D_{17,1}$ & 23 & $\left(\frac{3}{2},\frac{15}{8}\right)$ & 23 & \tiny{$(4600,23\cdot2^{11})$} & ${\bf III_{50}}$ & \tiny{$\left(1,2\right)$} & $584$ \\
8. & $\frac{35}{2}$ & $\left(\frac{3}{2},\frac{19}{16}\right)$ & 210 & \tiny{$(4655,35\cdot2^5)$} & $\text{GHM}_{210}$ & $\frac{45}{2}$ & $\left(\frac{1}{2},\frac{45}{16}\right)$ & 990 & \tiny{(45,$2^{22}$)} & $B_{22,1}$ & \tiny{$\left(1,1\right)$} & $1200$ \\
9. & $\frac{35}{2}$ & $\left(\frac{1}{2},\frac{35}{16}\right)$ & 595 & \tiny{(35,$2^{17}$)} & $B_{17,1}$ & $\frac{45}{2}$ & $\left(\frac{3}{2},\frac{29}{16}\right)$ & 45 & \tiny{$(4785,45\cdot2^{10})$} & $\text{GHM}_{45}$ & \tiny{$\left(1,1\right)$} & $640$ \\
10. & 18 & $\left(\frac{1}{2},\frac{9}{4}\right)$ & 630 & \tiny{(36,$2^{17}$)} & $D_{18,1}$ & 22 & $\left(\frac{3}{2},\frac{7}{4}\right)$ & 66 & \tiny{$(77\cdot2^6,11\cdot2^{11})$} & ${\bf III_{45}}$ & \tiny{$\left(1,2\right)$} & $696$ \\
11. & 18 & $\left(\frac{3}{2},\frac{5}{4}\right)$ & 198 & \tiny{$(75\cdot2^6,9\cdot2^7)$} & $\text{GHM}_{198}$ & 22 & $\left(\frac{1}{2},\frac{11}{4}\right)$ & 946 & \tiny{(44,$2^{21}$)} & $D_{22,1}$ & \tiny{$\left(1,2\right)$} & $1144$ \\
12. & $\frac{37}{2}$ & $\left(\frac{1}{2},\frac{37}{16}\right)$ & 666 & \tiny{(37,$2^{18}$)} & $B_{18,1}$ & $\frac{43}{2}$ & $\left(\frac{3}{2},\frac{27}{16}\right)$ & 86 & \tiny{$(5031,43\cdot2^9)$} & $\text{GHM}_{86}$ & \tiny{$\left(1,1\right)$} & $752$ \\
13. & $\frac{37}{2}$ & $\left(\frac{3}{2},\frac{21}{16}\right)$ & 185 & \tiny{$(4921,37\cdot2^6)$} & $\text{GHM}_{185}$ & $\frac{43}{2}$ & $\left(\frac{1}{2},\frac{43}{16}\right)$ & 903 & \tiny{(43,$2^{21}$)} & $B_{21,1}$ & \tiny{$\left(1,1\right)$} & $1088$ \\
14. & 19 & $\left(\frac{1}{2},\frac{19}{8}\right)$ & 703 & \tiny{(38,$2^{18}$)} & $D_{19,1}$ & 21 & $\left(\frac{3}{2},\frac{13}{8}\right)$ & 105 & \tiny{$(5096,21\cdot2^9)$} & $\text{GHM}_{105}$ & \tiny{$\left(1,2\right)$} & $808$ \\
15. & 19 & $\left(\frac{3}{2},\frac{11}{8}\right)$ & 171 & \tiny{$(5016,19\cdot2^7)$} & $\text{GHM}_{171}$ & 21 & $\left(\frac{1}{2},\frac{21}{8}\right)$ & 861 & \tiny{(42,$2^{20}$)} & $D_{21,1}$ & \tiny{$\left(1,2\right)$} & $1032$ \\
16. & $\frac{39}{2}$ & $\left(\frac{1}{2},\frac{39}{16}\right)$ & 741 & \tiny{(39,$2^{19}$)} & $B_{19,1}$ & $\frac{41}{2}$ & $\left(\frac{3}{2},\frac{25}{16}\right)$ & 123 & \tiny{$(5125,41\cdot2^8)$} & $\text{GHM}_{123}$ & \tiny{$\left(1,1\right)$} & $864$ \\
17. & $\frac{39}{2}$ & $\left(\frac{3}{2},\frac{23}{16}\right)$ & 156 & \tiny{$(5083,39\cdot2^7)$} & $\text{GHM}_{156}$ & $\frac{41}{2}$ & $\left(\frac{1}{2},\frac{41}{16}\right)$ & 820 & \tiny{(41,$2^{20}$)} & $B_{20,1}$ & \tiny{$\left(1,1\right)$} & $976$ \\
18. & 20 & $\left(\frac{1}{2},\frac{5}{2}\right)$ & 780 & \tiny{(40,$2^{19}$)} & $D_{20,1}$ & 20 & $\left(\frac{3}{2},\frac{3}{2}\right)$ & 140 & \tiny{(5120,5120)} & $\text{GHM}_{140}$ & \tiny{$\left(1,2\right)$} & 920  \\
\hline
\hline
\rowcolor{Mywhite}\caption{CFT pairings, $c^\mathcal{H}=40$ with $(n_1,n_2)=(2,4)$. The character of the meromorphic theory is $j^{\frac23}(j-1240+\cN)$ with $\cN$ given in the last column of the table.}
\label{T21}
\end{longtable}

\noindent {\bf Conclusion:} Table \ref{T21} gives us predictions for meromorphic theories at $c=40$. We do not go into detail here since we have already presented these predictions in \cite{Das:2022slz}.

\subsection*{Comments on Table \ref{T39999}}

This table has just one pair of admissible characters of IVOA type. Both members have already been identified as such in previous tables.

\setlength\LTleft{-25pt}
\setlength\LTright{0pt}
\rowcolors{2}{Mygrey}{Mywhite}
\begin{longtable}{l||ccccc||ccccc||cc}
\hline
\hline
\rowcolor{Mywhite}\# & $c$ & $(h_1,h_2)$ & $m_1$ & \tiny{$(D_1,D_2)$} & $\mathcal{W}$ & $\Tilde{c}$ & $(\Tilde{h}_1,\Tilde{h}_2)$ & $\Tilde{m}_1$ & \tiny{$(\Tilde{D}_1,\Tilde{D}_2)$} & $\widetilde{\mathcal{W}}$ & \tiny{$(d_1,d_2)$} & $\mathcal{N}$  \\
\hline
\hline
1. & $\frac{124}{7}$ & $\left(\frac{9}{7},\frac{10}{7}\right)$ & 248 & \tiny{(2108,2108)} & ${\bf III_{35}}$ & $\frac{156}{7}$ & $\left(\frac{5}{7},\frac{18}{7}\right)$ & 1248 & \tiny{(130,799500)} & ${\bf III_{48}}$ & \tiny{$\left(1,1\right)$} & $1496$ \\
\hline
\hline
\rowcolor{Mywhite}\caption{IVOA-type pairing, $c^\mathcal{H}=40$ with $(n_1,n_2)=(2,4)$.  The meromorphic character is $j^{\frac23}(j-1240+\cN)$ with $\cN$ given in the last column of the table.}
\label{T39999}
\end{longtable}

\subsection*{Comments on Table \ref{T29999}}

The rows without a type \sout{${\bf III}$} factor are 12--15, 18, 20, 23, 25, 27, 30, 31. All of them contain precisely one member that has been shown not to be a CFT. As a consequence we again get a set of cases for which a meromorphic theory at $c=40$ is ruled out. We list these below. The remaining rows have a type \sout{${\bf III}$} factor that is paired with an affine theory in most cases, and with an inconsistent type ${\bf III}$ solution in the remaining cases. Either way we get no new information from such pairs.

\setlength\LTleft{-40pt}
\setlength\LTright{0pt}
\rowcolors{2}{Mygrey}{Mywhite}
\begin{longtable}{l||ccccc||ccccc||cc}
\hline
\hline
\rowcolor{Mywhite}\# & $c$ & $(h_1,h_2)$ & $m_1$ & \tiny{$(D_1,D_2)$} & $\mathcal{W}$ & $\Tilde{c}$ & $(\Tilde{h}_1,\Tilde{h}_2)$ & $\Tilde{m}_1$ & \tiny{$(\Tilde{D}_1,\Tilde{D}_2)$} & $\widetilde{\mathcal{W}}$ & \tiny{$(d_1,d_2)$} & $\mathcal{N}$  \\
\hline
\hline
1. & $\frac{17}{2}$ & $\left(\frac{3}{2},\frac{1}{16}\right)$ & 255 & \tiny{(221,17)} & \sout{${\bf III_6}$} & $\frac{63}{2}$ & $\left(\frac{1}{2},\frac{63}{16}\right)$ & 1953 & \tiny{(63,$2^{31}$)} & $B_{31,1}$ & \tiny{$\left(1,\frac{1}{16}\right)$} & $2208$ \\
2. & $\frac{44}{5}$ & $\left(\frac{6}{5},\frac{2}{5}\right)$ & 220 & \tiny{$(275,11)$} & 
${\bf III_8}$ & $\frac{156}{5}$ & $\left(\frac{4}{5},\frac{18}{5}\right)$ & 3612 & \tiny{(14877,250774426)} & \sout{${\bf III_{62}}$} & \tiny{$\left(\frac{1}{25},1\right)$} & $3832$ \\
3. & 9 & $\left(\frac{9}{8},\frac{1}{2}\right)$ & 153 & \tiny{(256,18)} & $D_{9,1}$ & 31 & $\left(\frac{7}{8},\frac{7}{2}\right)$ & 5239 & \tiny{$(9269,2295147\cdot2^7)$} & \sout{${\bf III_{61}}$} & \tiny{$\left(\frac{1}{4},1\right)$} & $5392$ \\
4. & 9 & $\left(\frac{3}{2},\frac{1}{8}\right)$ & 261 & \tiny{$(456,9)$} & \sout{${\bf III_{10}}$} & 31 & $\left(\frac{1}{2},\frac{31}{8}\right)$ & 1891 & \tiny{(62,$2^{30}$)} & $D_{31,1}$ & \tiny{$\left(1,\frac{1}{4}\right)$} & $2152$ \\
5. & $\frac{19}{2}$ & $\left(\frac{19}{16},\frac{1}{2}\right)$ & 171 & \tiny{($2^9$,19)} & $B_{9,1}$ & $\frac{61}{2}$ & $\left(\frac{13}{16},\frac{7}{2}\right)$ & 3599 & \tiny{(47763,264580485)} & \sout{${\bf III_{60}}$} & \tiny{$\left(\frac{1}{64},1\right)$} & $3770$ \\
6. & $\frac{19}{2}$ & $\left(\frac{3}{2},\frac{3}{16}\right)$ & 266 & \tiny{(703,19)} & \sout{${\bf III_{11}}$} & $\frac{61}{2}$ & $\left(\frac{1}{2},\frac{61}{16}\right)$ & 1830 & \tiny{(61,$2^{30}$)} & $B_{30,1}$ & \tiny{$\left(1,\frac{1}{8}\right)$} & $2096$ \\
7. & $10$ & $\left(\frac{5}{4},\frac{1}{2}\right)$ & 190 & \tiny{($2^9$,20)} & $D_{10,1}$ & 30 & $\left(\frac{3}{4},\frac{7}{2}\right)$ & 2778 & \tiny{$(539,14421\cdot2^{14})$} & \sout{${\bf III_{59}}$} & \tiny{$\left(1,1\right)$}& $2968$ \\
8. & $10$ & $\left(\frac{3}{2},\frac{1}{4}\right)$ & 270 & \tiny{$(960,5)$} & \sout{${\bf III_{13}}$} & 30 & $\left(\frac{1}{2},\frac{15}{4}\right)$ & 1770 & \tiny{(60,$2^{29}$)} & $D_{30,1}$ & \tiny{$\left(1,1\right)$} & $2040$ \\
9. & $\frac{21}{2}$ & $\left(\frac{3}{2},\frac{5}{16}\right)$ & 273 & \tiny{(1225,21)} & \sout{${\bf III_{14}}$} & $\frac{59}{2}$ & $\left(\frac{1}{2},\frac{59}{16}\right)$ & 1711 & \tiny{(59,$2^{29}$)} & $B_{29,1}$ & \tiny{$\left(1,\frac{1}{4}\right)$} & $1984$ \\
10. & 11 & $\left(\frac{3}{2},\frac{3}{8}\right)$ & 275 & \tiny{$(1496,11)$} & \sout{${\bf III_{15}}$} & 29 & $\left(\frac{1}{2},\frac{29}{8}\right)$ & 1653 & \tiny{(58,$2^{28}$)} & $D_{29,1}$ & \tiny{$\left(1,1\right)$} & $1928$ \\
11. & $\frac{23}{2}$ & $\left(\frac{3}{2},\frac{7}{16}\right)$ & 276 & \tiny{(1771,23)} & \sout{${\bf III_{16}}$} & $\frac{57}{2}$ & $\left(\frac{1}{2},\frac{57}{16}\right)$ & 1596 & \tiny{(57,$2^{28}$)} & $B_{28,1}$ & \tiny{$\left(1,\frac{1}{2}\right)$} & $1872$ \\
12. & 12 & $\left(\frac{4}{3},\frac{2}{3}\right)$ & 156 & \tiny{($27^2$,27)} & $E_{6,1}^{\otimes 2}$ & 28 & $\left(\frac{2}{3},\frac{10}{3}\right)$ & 1948 & \tiny{$(225,11\cdot2\cdot3^{14})$} & ${\bf V_{58}}$ & \tiny{$\left(2,2\right)$} & $2104$ \\
13. & $\frac{25}{2}$ & $\left(\frac{3}{2},\frac{9}{16}\right)$ & 275 & \tiny{(2325,25)} & ${\bf III_{19}}$ & $\frac{55}{2}$ & $\left(\frac{1}{2},\frac{55}{16}\right)$ & 1485 & \tiny{(55,$2^{27}$)} & $B_{27,1}$ & \tiny{$\left(1,1\right)$} & $1760$ \\
14. & 13 & $\left(\frac{3}{2},\frac{5}{8}\right)$ & 273 & \tiny{$(2600,26)$} & ${\bf III_{20}}$ & 27 & $\left(\frac{1}{2},\frac{27}{8}\right)$ & 1431 & \tiny{(54,$2^{26}$)} & $D_{27,1}$ & \tiny{$\left(1,2\right)$} & $1704$ \\
15. & $\frac{27}{2}$ & $\left(\frac{3}{2},\frac{11}{16}\right)$ & 270 & \tiny{$(2871,54)$} & ${\bf III_{21}}$ & $\frac{53}{2}$ & $\left(\frac{1}{2},\frac{53}{16}\right)$ & 1378 & \tiny{(53,$2^{26}$)} & $B_{26,1}$ & \tiny{$\left(1,1\right)$} & $1648$ \\
16. & $\frac{68}{5}$ & $\left(\frac{7}{5},\frac{4}{5}\right)$ & 136 & \tiny{$(1700,119)$} & ${\bf III_{22}}$ & $\frac{132}{5}$ & $\left(\frac{3}{5},\frac{16}{5}\right)$ & 1536 & \tiny{(2392,47018049)} & \sout{${\bf III_{57}}$} & \tiny{$\left(\frac{2}{25},1\right)$} & $1672$ \\
17. & 14 & $\left(\frac{7}{4},\frac{1}{2}\right)$ & 378 & \tiny{($2^{13}$,28)} & $D_{14,1}$ & 26 & $\left(\frac{1}{4},\frac{7}{2}\right)$ & 1118 & \tiny{$(117,3315\cdot2^{14})$} & \sout{${\bf III_{56}}$} & \tiny{$\left(\frac{1}{16},1\right)$} & $1496$ \\
18. & $\frac{29}{2}$ & $\left(\frac{3}{2},\frac{13}{16}\right)$ & 261 & \tiny{$(3393,116)$} & ${\bf III_{26}}$ & $\frac{51}{2}$ & $\left(\frac{1}{2},\frac{51}{16}\right)$ & 1275 & \tiny{(51,$2^{25}$)} & $B_{25,1}$ & \tiny{$\left(1,1\right)$} & $1536$ \\
19. & $\frac{33}{2}$ & $\left(\frac{17}{16},\frac{3}{2}\right)$ & 231 & \tiny{$(528,4301)$} & ${\bf III_{31}}$ & $\frac{47}{2}$ & $\left(\frac{15}{16},\frac{5}{2}\right)$ & 4371 & \tiny{$(4371,1135003)$} & \sout{${\bf III_{55}}$} & \tiny{$\left(\frac12,1\right)$} & $4602$ \\
20. & $\frac{33}{2}$ & $\left(\frac{3}{2},\frac{17}{16}\right)$ & 231 & \tiny{$(4301,528)$} & ${\bf III_{31}}$ & $\frac{47}{2}$ & $\left(\frac{1}{2},\frac{47}{16}\right)$ & 1081 & \tiny{(47,$2^{23}$)} & $B_{23,1}$ & \tiny{$\left(1,1\right)$} & $1312$ \\
21. & $\frac{84}{5}$ & $\left(\frac{6}{5},\frac{7}{5}\right)$ & 336 & \tiny{$(770,1452)$} & ${\bf III_{33}}$ & $\frac{116}{5}$ & $\left(\frac{4}{5},\frac{13}{5}\right)$ & 1711 & \tiny{$(1653,910803)$} & \sout{${\bf III_{53}}$} & \tiny{$\left(\frac15,1\right)$} & $2047$ \\
22. & $\frac{84}{5}$ & $\left(\frac{1}{5},\frac{12}{5}\right)$ & 534 & \tiny{(33,55924)} & \sout{${\bf III_{34}}$} & $\frac{116}{5}$ & $\left(\frac{9}{5},\frac{8}{5}\right)$ & 58 & \tiny{$(27550,4959)$} & ${\bf III_{52}}$ & \tiny{$\left(\frac{2}{25},1\right)$} & $592$ \\
23. & 17 & $\left(\frac{9}{8},\frac{3}{2}\right)$ & 221 & \tiny{$(544,561\cdot2^3)$} & $\text{GHM}_{221}$ & 23 & $\left(\frac{7}{8},\frac{5}{2}\right)$ & 2323 & \tiny{$(575,32683\cdot2^5)$} & ${\bf III_{51}}$ & \tiny{$\left(2,1\right)$} & $2544$ \\
24. & $\frac{35}{2}$ & $\left(\frac{19}{16},\frac{3}{2}\right)$ & 210 & \tiny{$(35\cdot2^5,4655)$} & $\text{GHM}_{210}$ & $\frac{45}{2}$ & $\left(\frac{13}{16},\frac{5}{2}\right)$ & 1640 & \tiny{(1595,956449)} & \sout{${\bf III_{49}}$} & \tiny{$\left(\frac14,1\right)$} & $1850$ \\
25. & 18 & $\left(\frac{5}{4},\frac{3}{2}\right)$ & 198 & \tiny{$(9\cdot2^7,75\cdot2^6)$} & $\text{GHM}_{198}$ & 22 & $\left(\frac{3}{4},\frac{5}{2}\right)$ & 1298 & \tiny{$(154,847\cdot2^{10})$} & ${\bf III_{46}}$ & \tiny{$\left(2,1\right)$} & $1496$ \\
26. & 18 & $\left(\frac{1}{4},\frac{5}{2}\right)$ & 598 & \tiny{$(25,221\cdot2^{10})$} & \sout{${\bf III_{36}}$} & 22 & $\left(\frac{7}{4},\frac{3}{2}\right)$ & 66 & \tiny{$(11\cdot2^{11},77\cdot2^6)$} & ${\bf III_{45}}$ & \tiny{$\left(\frac14,1\right)$} & $664$ \\
27. & $\frac{92}{5}$ & $\left(\frac{6}{5},\frac{8}{5}\right)$ & 92 & \tiny{$(1196,7475)$} & ${\bf III_{37}}$ & $\frac{108}{5}$ & $\left(\frac{4}{5},\frac{12}{5}\right)$ & 1404 & \tiny{$(459,153\cdot5^5)$} & ${\bf III_{44}}$ & \tiny{$\left(1,2\right)$} & $1496$ \\
28. & $\frac{92}{5}$ & $\left(\frac{8}{5},\frac{6}{5}\right)$ & 92 & \tiny{$(7475,1196)$} & ${\bf III_{37}}$ & $\frac{108}{5}$ & $\left(\frac{2}{5},\frac{14}{5}\right)$ & 860 & \tiny{(833,3015426)} & \sout{${\bf III_{43}}$} & \tiny{$\left(\frac{1}{25},1\right)$} & $952$ \\
29. & $\frac{92}{5}$ & $\left(\frac{3}{5},\frac{11}{5}\right)$ & 690 & \tiny{(299,178802)} & \sout{${\bf III_{38}}$} & $\frac{108}{5}$ & $\left(\frac{7}{5},\frac{9}{5}\right)$ & 27 & \tiny{$(2295,42483)$} & ${\bf III_{42}}$ & \tiny{$\left(\frac25,1\right)$} & $717$ \\
30. & 20 & $\left(\frac{4}{3},\frac{5}{3}\right)$ & 80 & \tiny{$(2430,17496)$} & ${\bf V_{39}}$ & 20 & $\left(\frac{2}{3},\frac{7}{3}\right)$ & 890 & \tiny{$(135,20\cdot3^9)$} & ${\bf V_{41}}$ & \tiny{$\left(1,1\right)$} & $970$ \\
31. & 20 & $\left(\frac{5}{3},\frac{4}{3}\right)$ & 80 & \tiny{$(17496,2430)$} & ${\bf V_{39}}$ & 20 & $\left(\frac{1}{3},\frac{8}{3}\right)$ & 728 & \tiny{$(12,2\cdot3^{12})$} & ${\bf V_{40}}$ & \tiny{$\left(1,1\right)$} & $808$ \\
\hline
\hline
\rowcolor{Mywhite}\caption{Inconsistent pairings, $c^\mathcal{H}=40$ with $(n_1,n_2)=(2,4)$.  The meromorphic character is $j^{\frac23}(j-1240+\cN)$ with $\cN$ given in the last column of the table.}
\label{T29999}
\end{longtable}

\noindent {\bf Conclusion:} From Table \ref{T29999} we were able to predict the absence of meromorphic theories with the following values of $\cN < 3160$ coupled with a particular factor in their Kac-Moody algebra:

\setlength\LTleft{125pt}
\setlength\LTright{0pt}
\rowcolors{2}{Mygrey}{Mywhite}
\begin{longtable}{l||c||c}
\hline
\hline
\rowcolor{Mywhite}\# & $\cN$ & Factor \\
\hline
\hline
1. & 808 & $A_{2,1}^{\otimes 10}$ \\
2. & 808 & $A_{5,2}^{\otimes 2}\,C_{2,1}$ \\
3. & 808 & $A_{8,3}$ \\
4. & 970 & $A_{2,1}^{\otimes 10}$ \\
5. & 970 & $A_{5,2}^{\otimes 2}C_{2,1}$ \\
6. & 970 & $A_{8,3}$ \\
7. & 1312 & $B_{23,1}$ \\
8. & 1496 & $E_{6,3}G_{2,1}$ \\
9. & 1496 & $D_{6,1}^{\otimes 3}$ \\
10. & 1496 & $A_{9,1}^{\otimes 2}$ \\
11. & 1536 & $B_{25,1}$ \\
12. & 1648 & $B_{26,1}$ \\
13. & 1704 & $D_{27,1}$ \\
14. & 1760 & $B_{27,1}$ \\
15. & 2104 & $E_{6,1}^{\otimes 2}$ \\
16. & 2544 & $A_{11,1}\,E_{6,1}$ \\
\hline
\hline
\rowcolor{Mywhite}\caption{List of meromorphic theories ruled out by Table \ref{T29999}}
\label{T17989}
\end{longtable}


\subsubsection*{$\underline{(n_1,n_2)=(3,3)}$}

\subsection*{Comments on Table \ref{T22}}

All entries in this table are genuine coset pairs. Several CFTs of GHM type from \cite{Gaberdiel:2016zke} are paired with each other. This includes a self-coset in row 9. Rows 3 and 8 are similar, the theories ${\bf III_{45},V_{39}}$ were not listed in \cite{Gaberdiel:2016zke} but this should count as an oversight as they properly belong in Table 2 of that paper. In row 7 we see a self-pairing of $D_{20,1}$ to a meromorphic theory at $c=40$ without enhancement of Kac-Moody algebra, so the resulting theory can be written $\cE_1[D_{20,1}D_{20,1}]$ (this is to be contrasted with the pairing of the same factors in the $(n_1,n_2)=(1,5)$ case, where the meromorphic theory is $D_{40,1}$).

\setlength\LTleft{-40pt}
\setlength\LTright{0pt}
\rowcolors{2}{Mygrey}{Mywhite}
\begin{longtable}{l||ccccc||ccccc||cc}
\hline
\hline
\rowcolor{Mywhite}\# & $c$ & $(h_1,h_2)$ & $m_1$ & \tiny{$(D_1,D_2)$} & $\mathcal{W}$ & $\Tilde{c}$ & $(\Tilde{h}_1,\Tilde{h}_2)$ & $\tilde{m}_1$ & \tiny{$(\Tilde{D}_1,\Tilde{D}_2)$} & $\widetilde{\mathcal{W}}$ & \tiny{$(d_1,d_2)$} & $\mathcal{N}$  \\
\hline
\hline
1. & 17 & $\left(\frac{9}{8},\frac{3}{2}\right)$ & $221$ & \tiny{$(544,4488)$} & $\text{GHM}_{221}$ & 23 & $\left(\frac{15}{8},\frac{3}{2}\right)$ & $23$ & \tiny{$(23\cdot2^{11},4600)$} & ${\bf III_{50}}$ & \tiny{$\left(2,1\right)$} & $244$ \\
2. & $\frac{35}{2}$ & $\left(\frac{3}{2},\frac{19}{16}\right)$ & $210$ & \tiny{$(4655,35\cdot2^5)$} & $\text{GHM}_{210}$ & $\frac{45}{2}$ & $\left(\frac{3}{2},\frac{29}{16}\right)$ & $45$ & \tiny{$(4785,45\cdot2^{10})$} & $\text{GHM}_{45}$ & \tiny{$\left(1,1\right)$} & $255$ \\
3. & 18 & $\left(\frac{5}{4},\frac{3}{2}\right)$ & $198$ & \tiny{$(9\cdot2^7, 75\cdot2^6)$} & $\text{GHM}_{198}$ & 22 & $\left(\frac{7}{4},\frac{3}{2}\right)$ & $66$ & \tiny{$(11\cdot2^{11},77\cdot2^6)$} & ${\bf III_{45}}$ & \tiny{$\left(2,1\right)$} & $264$ \\
4. & $\frac{37}{2}$ & $\left(\frac{3}{2},\frac{21}{16}\right)$ & $185$ & \tiny{$(4921,37\cdot2^6)$} & $\text{GHM}_{185}$ & $\frac{43}{2}$ & $\left(\frac{3}{2},\frac{27}{16}\right)$ & $86$ & \tiny{$(5031,43\cdot2^9)$} & $\text{GHM}_{86}$ & \tiny{$\left(1,1\right)$} & $271$ \\
5. & 19 & $\left(\frac{3}{2},\frac{11}{8}\right)$ & $171$ & \tiny{$(5016,19\cdot2^7)$} & $\text{GHM}_{171}$ & 21 & $\left(\frac{3}{2},\frac{13}{8}\right)$ & $105$ & \tiny{$(5096,21\cdot2^9)$} & $\text{GHM}_{105}$ & \tiny{$\left(1,2\right)$} & $276$ \\
6. & $\frac{39}{2}$ & $\left(\frac{3}{2},\frac{23}{16}\right)$ & $156$ & \tiny{$(5083,39\cdot2^7)$} & $\text{GHM}_{156}$ & $\frac{41}{2}$ & $\left(\frac{3}{2},\frac{25}{16}\right)$ & $123$ & \tiny{$(5125,41\cdot2^8)$} & $\text{GHM}_{123}$ & \tiny{$\left(1,1\right)$} & $279$ \\
7. & 20 & $\left(\frac{1}{2},\frac{5}{2}\right)$ & $780$ & \tiny{(40,$2^{19}$)} & $D_{20,1}$ & 20 & $\left(\frac{5}{2},\frac{1}{2}\right)$ & $780$ & \tiny{($2^{19},40$)} & $D_{20,1}$ &  \tiny{$(1,1)$} & $1560$ \\
8. & 20 & $\left(\frac{4}{3},\frac{5}{3}\right)$ & $80$ & \tiny{$(2430,17496)$} & ${\bf V_{39}}$ & 20 & $\left(\frac{5}{3},\frac{4}{3}\right)$ & $80$ & \tiny{$(17496,2430)$} & ${\bf V_{39}}$ & \tiny{$\left(1,1\right)$} & $160$ \\
9. & 20 & $\left(\frac{7}{5},\frac{8}{5}\right)$ & $120$ & \tiny{$(4\cdot5^4,13\cdot5^4)$} & $\text{GHM}_{120}$ & 20 & $\left(\frac{8}{5},\frac{7}{5}\right)$ & $120$ & \tiny{$(13\cdot5^4,4\cdot5^4)$} & $\text{GHM}_{120}$ & \tiny{$\left(2,2\right)$} & $240$ \\
\hline
\hline
\rowcolor{Mywhite}\caption{CFT pairings, $c^\mathcal{H}=40$ with $(n_1,n_2)=(3,3)$. $\mathcal{H}$ with $\chi^\mathcal{H}=j^{2/3}(j+\mathcal{N}_0)$ where $\mathcal{N}_0 \geq -1240$.  The meromorphic character is $j^{\frac23}(j-1240+\cN)$ with $\cN$ given in the last column of the table.}
\label{T22}
\end{longtable}

\subsection*{Comments on Table \ref{T2287}}

This table contains four pairings that all involve characters of IVOA type. Seven of these have been encountered before, but one of the solutions in row 1, with $c=\frac{236}{7}$, is appearing here for the first time. This one has been previously noted in \cite{Hampapura:2016mmz} in the context of a study of three-character solutions without Kac-Moody symmetry. Hence we denote this character as HM$(7,2)$.

\setlength\LTleft{-30pt}
\setlength\LTright{0pt}
\rowcolors{2}{Mygrey}{Mywhite}
\begin{longtable}{l||ccccc||ccccc||cc}
\hline
\hline
\rowcolor{Mywhite}\# & $c$ & $(h_1,h_2)$ & $m_1$ & \tiny{$(D_1,D_2)$} & $\mathcal{W}$ & $\Tilde{c}$ & $(\Tilde{h}_1,\Tilde{h}_2)$ & $\tilde{m}_1$ & \tiny{$(\Tilde{D}_1,\Tilde{D}_2)$} & $\widetilde{\mathcal{W}}$ & \tiny{$(d_1,d_2)$} & $\mathcal{N}$  \\
\hline
\hline
1. & $\frac{44}{7}$ & $\left(\frac{4}{7},\frac{5}{7}\right)$ & $88$ &  \tiny{(11,44)} & ${\bf III_{3}}$ & $\frac{236}{7}$ & $\left(\frac{17}{7},\frac{16}{7}\right)$ & $0$ & \tiny{$(848656,715139)$} & $\text{HM}(7,2)$ & \tiny{$\left(1,1\right)$} & $88$ \\
2. & $\frac{116}{7}$ & $\left(\frac{8}{7},\frac{10}{7}\right)$ & $348$ & \tiny{(725,1972)} & ${\bf III_{32}}$ & $\frac{164}{7}$ & $\left(\frac{13}{7},\frac{11}{7}\right)$ & $41$ & \tiny{(50922,4797)} & ${\bf III_{54}}$ & \tiny{$\left(1,1\right)$} & $389$ \\
3. & $\frac{84}{5}$ & $\left(\frac{6}{5},\frac{7}{5}\right)$ & $336$ & \tiny{$(770,1452)$} & ${\bf III_{33}}$ & $\frac{116}{5}$ & $\left(\frac{9}{5},\frac{8}{5}\right)$ & $58$ &  \tiny{$(27550,4959)$} & ${\bf III_{52}}$ & \tiny{$\left(2,1\right)$} & $394$ \\
4. & $\frac{124}{7}$ & $\left(\frac{9}{7},\frac{10}{7}\right)$ & $248$ & \tiny{(2108,2108)} & ${\bf III_{35}}$ & $\frac{156}{7}$ & $\left(\frac{12}{7},\frac{11}{7}\right)$ & $78$ & \tiny{(27170,5070)} & ${\bf III_{47}}$ & \tiny{$\left(1,1\right)$} & $326$ \\
\hline
\hline
\rowcolor{Mywhite}\caption{IVOA-type pairings, $c^\mathcal{H}=40$ with $(n_1,n_2)=(3,3)$.  The meromorphic character is $j^{\frac23}(j-1240+\cN)$ with $\cN$ given in the last column of the table.}
\label{T2287}
\end{longtable}

\noindent {\bf Conclusion:} In Table \ref{T2287} we find seven IVOA-type solutions that were previously discussed above, and one that appears for the first time in this table but has been noted before.

\subsection*{Comments on Table \ref{T2299}}

This table contains 10 pairings. Rows 1--8 have  one inconsistent solution paired with a known CFT, while rows 9 and 10 are self-pairings where both members are known to be inconsistent. As a result, rows 1, 3, 6 and 7 lead to negative predictions for specific types of meromorphic theories at $c=40$, while rows 9 and 10 do not. Meanwhile rows 2, 4, 5 and 8 have one factor with fractional $Y_i$ values, so these also do not lead to negative predictions for meromorphic theories.

\setlength\LTleft{-35pt}
\setlength\LTright{0pt}
\rowcolors{2}{Mygrey}{Mywhite}
\begin{longtable}{l||ccccc||ccccc||cc}
\hline
\hline
\rowcolor{Mywhite}\# & $c$ & $(h_1,h_2)$ & $m_1$ & \tiny{$(D_1,D_2)$} & $\mathcal{W}$ & $\Tilde{c}$ & $(\Tilde{h}_1,\Tilde{h}_2)$ & $\tilde{m}_1$ & \tiny{$(\Tilde{D}_1,\Tilde{D}_2)$} & $\widetilde{\mathcal{W}}$ & \tiny{$(d_1,d_2)$} & $\mathcal{N}$  \\
\hline
\hline
1. & $\frac{33}{2}$ & $\left(\frac{17}{16},\frac{3}{2}\right)$ & $231$ & \tiny{$(528,4301)$} & 
${\bf III_{31}}$ & $\frac{47}{2}$ & $\left(\frac{31}{16},\frac{3}{2}\right)$ & $0$ & \tiny{$(47\cdot2^{11},4371)$} & BM & \tiny{$\left(1,1\right)$} & $231$ \\
2. & $\frac{33}{2}$ & $\left(\frac{1}{2},\frac{33}{16}\right)$ & $528$ & \tiny{(33,$2^{16}$)} & 
$B_{16,1}$ & $\frac{47}{2}$ & $\left(\frac{5}{2},\frac{15}{16}\right)$ & $4371$ & \tiny{(1135003,4371)} & \sout{${\bf III_{55}}$} & \tiny{$\left(1,\frac{1}{2}\right)$} & $4899$ \\
3. & 17 & $\left(\frac{1}{2},\frac{17}{8}\right)$ & $561$ & \tiny{(34,$2^{16}$)} & $D_{17,1}$ & 23 & $\left(\frac{5}{2},\frac{7}{8}\right)$ & $2323$ & \tiny{$(32683\cdot2^5,575)$} & ${\bf III_{51}}$ & \tiny{$\left(1,2\right)$} & $2884$ \\
4. & $\frac{35}{2}$ & $\left(\frac{1}{2},\frac{35}{16}\right)$ & $595$ & \tiny{(35,$2^{17}$)} & $B_{17,1}$ & $\frac{45}{2}$ & $\left(\frac{5}{2},\frac{13}{16}\right)$ & $1640$ & \tiny{(956449,1595)} & \sout{${\bf III_{49}}$} & \tiny{$\left(1,\frac{1}{4}\right)$} & $2235$ \\
5. & 18 & $\left(\frac{1}{4},\frac{5}{2}\right)$ & $598$ & \tiny{$(25,221\cdot2^{10})$} & \sout{${\bf III_{36}}$} & 22 & $\left(\frac{11}{4},\frac{1}{2}\right)$ & $946$ & \tiny{($2^{21}$,44)} & $D_{22,1}$ & \tiny{$\left(\frac{1}{4},1\right)$} & $1544$ \\
6. & 18 & $\left(\frac{1}{2},\frac{9}{4}\right)$ & $630$ & \tiny{(36,$2^{17}$)} & $D_{18,1}$ & 22 & $\left(\frac{5}{2},\frac{3}{4}\right)$ & $1298$ & \tiny{$(847\cdot2^{10},154)$} & ${\bf III_{46}}$ & \tiny{$\left(1,2\right)$} & $1928$ \\
7. & $\frac{92}{5}$ & $\left(\frac{6}{5},\frac{8}{5}\right)$ & $92$ & \tiny{$(1196,7475)$} & ${\bf III_{37}}$ & $\frac{108}{5}$ & $\left(\frac{9}{5},\frac{7}{5}\right)$ & $27$ & \tiny{$(42483,2295)$} & ${\bf III_{42}}$ & \tiny{$\left(1,2\right)$} & $119$ \\
8. & $\frac{92}{5}$ & $\left(\frac{3}{5},\frac{11}{5}\right)$ & $690$ & \tiny{(299,178802)} & \sout{${\bf III_{38}}$} & $\frac{108}{5}$ & $\left(\frac{12}{5},\frac{4}{5}\right)$ & $1404$ & \tiny{$(153\cdot5^5,459)$} & ${\bf III_{44}}$ & \tiny{$\left(\frac25,1\right)$} & $2094$ \\
9. & 20 & $\left(\frac{1}{3},\frac{8}{3}\right)$ & $728$ & \tiny{$(12,2\cdot3^{12})$} & ${\bf V_{40}}$ & 20 & $\left(\frac{8}{3},\frac{1}{3}\right)$ & $728$ & \tiny{$(2\cdot3^{12},12)$} & ${\bf V_{40}}$ & \tiny{$(1,1)$} & $1456$ \\
10. & 20 & $\left(\frac{2}{3},\frac{7}{3}\right)$ & $890$ & \tiny{$(135,10\cdot2\cdot3^9)$} & ${\bf V_{41}}$ & 20 & $\left(\frac{7}{3},\frac{2}{3}\right)$ & $890$ & \tiny{$(10\cdot2\cdot3^9,135)$} & ${\bf V_{41}}$ & \tiny{$(1,1)$} & $1780$ \\
\hline
\hline
\rowcolor{Mywhite}\caption{Inconsistent pairings, $c^\mathcal{H}=40$ with $(n_1,n_2)=(3,3)$. 
 The meromorphic character is $j^{\frac23}(j-1240+\cN)$ with $\cN$ given in the last column of the table.
}
\label{T2299}
\end{longtable}

From Table \ref{T2299} we were able to predict the absence of meromorphic theories with the following values of $\cN < 3160$ coupled with a particular factor in their Kac-Moody algebra:

\setlength\LTleft{175pt}
\setlength\LTright{0pt}
\rowcolors{2}{Mygrey}{Mywhite}
\begin{longtable}{l||c||c}
\hline
\hline
\rowcolor{Mywhite}\# & $\cN$ & Factor \\
\hline
\hline
1. & 119 & $E_{6,3}G_{2,1}$ \\
2. & 231 & $\text{BM}$ \\
3. & 1928 & $D_{18,1}$ \\
4. & 2884 & $D_{17,1}$ \\
\hline
\hline
\rowcolor{Mywhite}\caption{List of meromorphic theories ruled out by Table \ref{T2299}}
\label{T169899}
\end{longtable}

\section{Discussion and conclusions}
\label{discussion}

In this paper we started with a list of 65 admissible characters, listed in Table \ref{T0}, and found all possible bilinear pairings involving them such that the total central charge is $\le 40$. We then examined them for consistency as CFTs.  
24 of these were ruled out as CFTs at the outset since they do not have integer multiplicities $Y_1,Y_2$. We then studied the remaining 41 through their bilinear pairings to modular invariants, and were able to classify all of them into three groups: (i) 6 consistent CFTs, for which we have found the Kac-Moody algebra, (ii) 20 candidates for Intermediate Vertex Operator Algebras, whose fusion rules are not all positive, (iii) 15 admissible characters that cannot correspond to any CFT. 

\subsection{Our results}

Table \ref{cft} lists the cases that have been classified as CFTs. We see that in some cases there are multiple CFTs corresponding to a single set of admissible characters, as was already seen in \cite{Gaberdiel:2016zke} for the two-character case. All entries of this table were identified by \cite{Duan:2022ltz}.

\setlength\LTleft{40pt}
\setlength\LTright{0pt}
\rowcolors{2}{Mygrey}{Mywhite}
\begin{longtable}{l||ccc||cl}
\hline
\hline
\rowcolor{Mywhite}\# & $c$ & $(h_1,h_2)$ & $m_1$ & $\cW$ & Chiral Algebra \\
\hline
\hline
1. & $\frac{12}{5}$ & $(\frac{1}{5},\frac{3}{5})$ & $6$ & ${\bf III_2}$ & $\mathcal{E}_3[A_{1,8}]$ \\
2. & $\frac{68}{5}$ & $(\frac{4}{5},\frac{7}{5})$ & $136$ & ${\bf III_{22}}$ & $\mathcal{E}_3[C_{8,1}]$ \\
3. & $\frac{92}{5}$ & $(\frac{6}{5},\frac{8}{5})$ & $92$ & ${\bf III_{37}}$ & $\mathcal{E}_3[E_{6,3}G_{2,1}]$ \\
4. & $20$ & $(\frac{4}{3},\frac{5}{3})$ & $80$ & ${\bf V_{39}}$ & $\cE_3[A_{2,1}^{\otimes 10}],~\cE_3[A_{5,2}^{\otimes 2}C_{2,1}],~\cE_3[A_{8,3}]$ \\
5. & $22$ & $(\frac{3}{2},\frac{7}{4})$ & $66$ & ${\bf III_{45}}$ & $\mathcal{E}_3[A_{1,1}^{\otimes 22}],~
\mathcal{E}_3[A_{3,2}^{\otimes 4}A_{1,1}^{\otimes 2}],~
\mathcal{E}_3[A_{5,3}D_{4,3}A_{1,1}]$,\\
\rowcolor{Mywhite} &  &  &  &  & $\mathcal{E}_3[A_{7,4}A_{1,1}],
~\mathcal{E}_3[D_{5,4}C_{3,2}],~
\mathcal{E}_3[D_{6,5}]$ \\
\rowcolor{Mygrey}6. & $23$ & $(\frac{3}{2},\frac{15}{8})$ & $23$ & ${\bf III_{50}}$ & $\mathcal{E}_3[D_{1,1}^{\otimes 23}]$ \\
\hline
\hline
\rowcolor{Mywhite}\caption{Consistent CFTs}
\label{cft}
\end{longtable}

Next we list the cases that were in our Table \ref{T0}, other than those already eliminated at the outset, which cannot be identified as consistent CFTs. These fall into two classes: the first are those of IVOA type: ${\bf III_1, III_3, III_4, III_5, III_7, III_8, III_{12}, III_{24}, III_{25}, III_{27}, III_{28}, III_{29}}$, ${\bf III_{30}, III_{32}, III_{33}, III_{35}, III_{47}, III_{48}, III_{52}, III_{54}}$, while the second are inconsistent in the sense that they cannot be CFT: ${\bf III_{17}, V_{18}, III_{19}, III_{20}, III_{21}, III_{26}, III_{31}, V_{40}, V_{41}, III_{42}}$,\\ ${\bf III_{44}, III_{46}, III_{51}, V_{58}, V_{63}}$ (recall that the $(c,h_1,h_2)$ and $m_1$ values of these are listed in Table \ref{T0}). From the inconsistent list, the ten type-${\bf III}$ solutions were first discovered as admissible characters in \cite{franc2020classification} while the five type-${\bf V}$ solutions are among the seven that were newly found last year in \cite{Kaidi:2021ent, Das:2021uvd, Bae:2021mej}.

Our work once more highlights the intimate relation between general RCFT and meromorphic CFT. We see that this relation, when properly applied, allows us to rule in and also rule out characters from being CFT, and likewise gives positive and negative predictions for the existence of meromorphic theories.

While we have not aspired to mathematical rigour in this work, we believe our conclusions can and should be tested at a more formal and rigorous level. Basic properties of Modular Tensor Categories (MTC) at low numbers of primaries \cite{rowell2009classification} lead us to believe that whenever two admissible characters pair up and both are known CFTs, the pair is also a CFT -- but technically this is only known up to 4 primaries and a few of our examples have more primaries than that, despite having only three characters. There are also possible subtleties about linear equivalence vs equivalence of embeddings, as well as about possibly inequivalent embeddings in different simple factors of the same algebra. Such questions were addressed in \cite{Mukhi:2022bte} where the focus was on a rigorous classification for exactly two primaries in a range of central charge. Something similar can surely be attempted for three primaries (rather than three characters) in a more rigorous fashion than was done here using the MTC data for theories with three simple objects.

On the other hand, a positive aspect of the present approach based on MLDE and bilinear pairing of $q$-series is that the classification of pairings is explicit and exhaustive, and does not rely on mathematically subtle questions. Also it raises intriguing questions about admissible characters that are not CFT -- we do not know why they nevertheless exhibit bilinear pairings, and what this teaches us. This point may be of interest to the community studying vector-valued modular forms. 

\Comment{
\subsection{Ising}

\begin{enumerate}
    \item\label{it1} In all the cases above (except one) $\mathcal{M}(4,3)$ always appears in the following way,
    \begin{align}
        \frac{D_{c^{\mathcal{H}},1}}{B_{c^{\mathcal{H}-1},1}} \cong \mathcal{M}(4,3). \label{ising0}
    \end{align}
    This can be explained from the following maximal special embedding: $D_{c^{\mathcal{H}},1}\overset{m}{\underset{S}{\longrightarrow}}B_{c^{\mathcal{H}-1},1}$ (where $S$ above the arrow stands for special maximal embedding) which implies that the $B_{c^{\mathcal{H}-1},1}$ has trivial commutant inside $D_{c^{\mathcal{H}},1}$ (as was seen in the case of a specific example in Eq.(\ref{no_singlet})). 
    
    Here, the numerator theory should be thought of as, $\mathcal{E}[D_{c^{\mathcal{H}}}]$ being interpreted as a $1$-character meromorphic theory. For example, one can readily see that $\mathcal{E}[D_{8,1}] = j^{1/3}$.
    
    \item\label{it2} In one particular case (as we will see below), $\mathcal{M}(4,3)$ appears due to the case of diagonal embedding of the form,
    \begin{align}
        &A_{1,2} \hookrightarrow A_{1,1}\otimes A_{1,1} \nonumber\\
        \text{leading to,} & \, \, \frac{A_{1,1}\otimes A_{1,1}}{A_{1,2}} \cong \mathcal{M}(4,3) \nonumber\\
        &E_{8,2} \hookrightarrow E_{8,1}\otimes E_{8,1} \nonumber\\
        \text{leading to,} & \, \, \frac{E_{8,1}\otimes E_{8,1}}{E_{8,2}} \cong \mathcal{M}(4,3) \nonumber
    \end{align}
\end{enumerate} 

\subsection{Baby Monster}
We see that BM always appears in the following way. Consider $c^{\mathcal{H}}=8\left(m+3\right)$ (where $m\in\mathbb{N}\cup\{0\}$). Then for $(n_1,n_2)=(2,m+2)$ we always have, $B_{8m,1}\leftrightarrow \text{BM}$ (where at the level of characters, by an abuse of notation, we have identified $B_{0,1}\cong \mathcal{M}(4,3)$).

Now if we consider $\mathcal{E}[B_{8m,1}\otimes \text{BM}]$ then this would be a $1$-character $c^{\mathcal{H}}=8\left(m+3\right)$ meromorphic theory with the following identity character,
\begin{align}
    \chi^{c^\mathcal{H}}_0 = \chi_0 + \chi_2 + 2048 \, \chi_{m+2}, \label{BM}
\end{align}
where $\chi_2$ and $\chi_{m+2}$ are two of the characters of the $9$-character theory $B_{8m,1}\otimes \text{BM}$ with conformal dimensions as $2$ and $m+2$ respectively.

We have checked that Eq.(\ref{BM}) holds for $c^{\mathcal{H}}=24,32$ and $40$ and we conjecture that it should hold in general for $c^{\mathcal{H}}=8\left(m+3\right)$. For instance, we have,
\begin{align}
    &\chi_0^{c \, = \, 24} = j \\
    &\chi_0^{c \, = \, 32} = j^{1/3}\left(j-856\right) \\
    &\chi_0^{c \, = \, 40} = j^{2/3}\left(j-712\right) \\
\end{align}
where $d_m := \text{dim}(B_{8m,1}) = 128 m^2 + 8m$. Similar conjectures would exist for other infinite lists whose first members appear in Schellekens list.
}

\subsection{Complete list of unitary (3,0) CFTs, except $c=8,16$}

In this section we tabulate the complete list of unitary $(3,0)$ CFTs (except at $c=8,16$). Here IVOA-type solutions are excluded since properly speaking they are not strict CFTs. The last column \#(primaries) denote the number of primaries of the given theory $\cW$.

\setlength\LTleft{30pt}
\setlength\LTright{0pt}
\rowcolors{2}{Mygrey}{Mywhite}
\begin{longtable}{l||ccc||ll||c}
\hline
\hline
\rowcolor{Mywhite}\# & $c$ & $(h_1,h_2)$ & $m_1$ & $\cW$ & Chiral Algebra & \#(primaries) \\
\hline
\hline
1. & $\frac{2r+1}{2}$ & $(\frac{1}{2},\frac{2r+1}{16})$ & $2r^2+r$ & ${\bf I}$ & $B_{r,1}$ & 3 \\
2. & $r$ & $(\frac{1}{2},\frac{r}{8})$ & $2r^2-r$ & ${\bf I}$ & $D_{r,1}$ $(r\neq 8,16)$ & 4 \\
\hline
\hline
3. & $\frac{12}{5}$ & $(\frac{1}{5},\frac{3}{5})$ & $6$ & ${\bf III_2}$ & $\mathcal{E}_3[A_{1,8}]$ & 4 \\
4. & $4$ & $(\frac{2}{5},\frac{3}{5})$ & $24$ & ${\bf I}$ & $A_{4,1}$ & 5 \\
5. & $\frac{28}{5}$ & $(\frac{2}{5},\frac{4}{5})$ & $28$ & ${\bf I}$ & $G_{2,1}^{\otimes 2}$ & 4 \\
6. & $\frac{52}{5}$ & $(\frac{3}{5},\frac{6}{5})$ & $104$ & ${\bf I}$ & $F_{2,1}^{\otimes 2}$ & 4 \\
7. & $12$ & $(\frac{2}{3},\frac{4}{3})$ & $156$ & ${\bf I}$ & $E_{6,1}^{\otimes 2}$ & 9 \\
8. & $\frac{68}{5}$ & $(\frac{4}{5},\frac{7}{5})$ & $136$ & ${\bf III_{22}}$ & $\mathcal{E}_3[C_{8,1}]$ & 4 \\
9. & $14$ & $(\frac{3}{4},\frac{3}{2})$ & $266$ & ${\bf I}$ & $E_{7,1}^{\otimes 2}$ & 4 \\
10. & $15$ & $(\frac{7}{8},\frac{3}{2})$ & $255$ & $\text{GHM}_{255}$ & $\mathcal{E}_3[A_{15,1}]$ & 4 \\
11. & $\frac{31}{2}$ & $(\frac{15}{16},\frac{3}{2})$ & $248$ & ${\bf I}$ & $E_{8,2}$ & 3 \\
12. & $17$ & $(\frac{9}{8},\frac{3}{2})$ & $221$ & $\text{GHM}_{221}$ & $\mathcal{E}_3[A_{11,1}E_{6,1}]$ & 4 \\
13. & $\frac{35}{2}$ & $(\frac{19}{16},\frac{3}{2})$ & $210$ & $\text{GHM}_{210}$ & $\mathcal{E}_3[C_{10,1}]$ & 3 \\
14. & $18$ & $(\frac{5}{4},\frac{3}{2})$ & $198$ & $\text{GHM}_{198}$ & $\mathcal{E}_3[D_{6,1}^{\otimes 3}]$ & 4 \\
\rowcolor{Mygrey} 15. &  &  &  &  & $\mathcal{E}_3[A_{9,1}^{\otimes 2}]$ & 4 \\
\rowcolor{Mywhite} 16. & $\frac{92}{5}$ & $(\frac{6}{5},\frac{8}{5})$ & $92$ & ${\bf III_{37}}$ & $\mathcal{E}_3[E_{6,3}G_{2,1}]$ & 4 \\
\rowcolor{Mygrey} 17. & $\frac{37}{2}$ & $(\frac{21}{16},\frac{3}{2})$ & $185$ & $\text{GHM}_{185}$ & $\mathcal{E}_3[E_{7,2}F_{4,1}]$ & 3 \\
\rowcolor{Mywhite} 18. & $19$ & $(\frac{11}{8},\frac{3}{2})$ & $171$ & $\text{GHM}_{171}$ & $\mathcal{E}_3[A_{7,1}^{\otimes 2}D_{5,1}]$ & 4 \\
\rowcolor{Mygrey} 19. & $\frac{39}{2}$ & $(\frac{23}{16},\frac{3}{2})$ & $156$ & $\text{GHM}_{156}$ & $\mathcal{E}_3[B_{4,1}D_{8,2}]$ & 3 \\
\rowcolor{Mygrey} 20. &  &  &  &  & $\mathcal{E}_3[C_{6,1}^{\otimes 2}]$ & 3 \\
\rowcolor{Mywhite} 21. & $20$ & $(\frac{4}{3},\frac{5}{3})$ & $80$ & ${\bf V_{39}}$ & $\cE_3[A_{2,1}^{\otimes 10}]$ & 9 \\
\rowcolor{Mywhite} 22. &  &  &  &  & $\cE_3[A_{5,2}^{\otimes 2}C_{2,1}]$ & 9 \\
\rowcolor{Mywhite} 23. &  &  &  &  & $\cE_3[A_{8,3}]$ & 9 \\
\rowcolor{Mygrey}24. & $20$ & $(\frac{7}{5},\frac{8}{5})$ & $120$ & $\text{GHM}_{120}$ & $\cE_3[A_{4,1}^{\otimes 5}]$ & 5 \\
\rowcolor{Mygrey} 25. &  &  &  &  & $\cE_3[A_{9,2}B_{3,1}]$ & 5 \\
\rowcolor{Mywhite} 26. & $\frac{41}{2}$ & $(\frac{3}{2},\frac{25}{16})$ & $123$ & $\text{GHM}_{123}$ & $\mathcal{E}_3[D_{6,2}C_{4,1}B_{3,1}]$ & 3 \\
\rowcolor{Mywhite} 27. &  &  &  &  & $\mathcal{E}_3[A_{9,2}A_{4,1}]$ & 3 \\
\rowcolor{Mygrey} 28. & $21$ & $(\frac{3}{2},\frac{13}{8})$ & $105$ & $\text{GHM}_{105}$ & $\cE_3[A_{3,1}^{\otimes 7}]$ & 4 \\
\rowcolor{Mygrey} 29. &  &  &  &  & $\cE_3[A_{3,1}D_{5,2}^{\otimes 2}]$ & 4 \\
\rowcolor{Mygrey} 30. &  &  &  &  & $\cE_3[A_{7,2}C_{3,1}^{\otimes 2}]$ & 4 \\
\rowcolor{Mygrey} 31. &  &  &  &  & $\cE_3[D_{7,3}G_{2,1}]$ & 4 \\
\rowcolor{Mygrey} 32. &  &  &  &  & $\cE_3[C_{7,2}]$ & 4 \\
\rowcolor{Mywhite} 33. & $\frac{43}{2}$ & $(\frac{3}{2},\frac{27}{16})$ & $86$ & $\text{GHM}_{86}$ & $\mathcal{E}_3[C_{2,1}^{\otimes 3}D_{4,2}^{\otimes 2}]$ & 3 \\
\rowcolor{Mywhite} 34. &  &  &  &  & $\mathcal{E}_3[A_{5,2}^{\otimes 2}A_{2,1}^{\otimes 2}]$ & 3 \\
\rowcolor{Mywhite} 35. &  &  &  &  & $\cE_3[A_{2,1}E_{6,4}]$ & 3 \\
\rowcolor{Mygrey} 36. & $22$ & $(\frac{3}{2},\frac{7}{4})$ & $66$ & ${\bf III_{45}}$ & $\mathcal{E}_3[A_{1,1}^{\otimes 22}]$, & 4 \\
\rowcolor{Mygrey} 37. &  &  &  &  & $\mathcal{E}_3[A_{3,2}^{\otimes 4}A_{1,1}^{\otimes 2}]$ & 4 \\
\rowcolor{Mygrey} 38. &  &  &  &  & $\mathcal{E}_3[A_{5,3}D_{4,3}A_{1,1}]$ & 4 \\
\rowcolor{Mygrey} 39. &  &  &  &  & $\mathcal{E}_3[A_{7,4}A_{1,1}]$ & 4 \\
\rowcolor{Mygrey} 40. &  &  &  &  & $\mathcal{E}_3[D_{5,4}C_{3,2}]$ & 4 \\
\rowcolor{Mygrey} 41. &  &  &  &  & $\mathcal{E}_3[D_{6,5}]$ & 4 \\
\rowcolor{Mywhite} 42. & $\frac{45}{2}$ & $(\frac{3}{2},\frac{29}{16})$ & $45$ & $\text{GHM}_{45}$ & $\mathcal{E}_3[A_{1,2}^{\otimes 15}]$ & 3 \\
\rowcolor{Mywhite} 43. &  &  &  &  & $\mathcal{E}_3[A_{3,4}^{\otimes 3}]$ & 3 \\
\rowcolor{Mywhite} 44. &  &  &  &  & $\cE_3[A_{5,6}C_{2,3}]$ & 3 \\
\rowcolor{Mywhite} 45. &  &  &  &  & $\cE_3[D_{5,8}]$ & 3 \\
\rowcolor{Mygrey} 46. & $23$ & $(\frac{3}{2},\frac{15}{8})$ & $23$ & ${\bf III_{50}}$ & $\mathcal{E}_3[D_{1,1}^{\otimes 23}]$ & 4 \\
\rowcolor{Mywhite} 47. & $\frac{47}{2}$ & $(\frac{3}{2},\frac{31}{16})$ & $0$ & ${\bf IV}$ & $\text{Baby Monster}$ & 3 \\
\hline
\hline
\rowcolor{Mywhite}\caption{Complete list of unitary $(3,0)$ CFTs (except unitary CFTs at $c=8,16$)}
\label{complete30}
\end{longtable}

Finally, Table \ref{incomplete30} lists four  theories at $c=8,16$ that are well-understood. The first of these is the tensor product of an affine theory with itself, the second and third are affine theories and the fourth is more subtle as it is a three-character extension of the fourth power of an affine theory. The first and the last theories  have three characters but $16$ primaries each. A more complete study of the infinite set of cases at $c=8,16$ is left for future work.

\setlength\LTleft{40pt}
\setlength\LTright{0pt}
\rowcolors{2}{Mygrey}{Mywhite}
\begin{longtable}{l||ccc||ll||c}
\hline
\hline
\rowcolor{Mywhite}\# & $c$ & $(h_1,h_2)$ & $m_1$ & $\cW$ & Chiral Algebra & \#(primaries) \\
\hline
\hline
1. & $8$ & $\left(\frac{1}{2},1\right)$ & 56 & ${\bf I}$ & $D_{4,1}^{\otimes 2}$ & 16 \\
2. & $8$ & $\left(\frac{1}{2},1\right)$ & 120 & ${\bf I}$ & $D_{8,1}$ & 4 \\
3. & $16$ & $\left(\frac{1}{2},2\right)$ & 496 & ${\bf I}$ & $D_{16,1}$ & 4 \\
4. & $16$ & $\left(\frac{3}{2},1\right)$ & 112 & ${\bf III''}(m_1 = 112)$ & $\cE_3[D_{4,1}^{\otimes 4}]$ & 16 \\
\hline
\hline
\rowcolor{Mywhite}\caption{4 unitary $(3,0)$ CFTs at $c=8,16$}
\label{incomplete30}
\end{longtable}

\begin{appendix}

\section*{Acknowledgements}

AD would like to thank Nabil Iqbal, Napat Poovuttikul, Daniele Dorigoni, T.V. Karthik and Madalena Lemos for useful discussions on Lie algebras. He expresses his gratitude to Jagannath Santara for helpful discussions on fusion coefficients. He would also like to thank Jishu Das and Naveen Umasankar for insightful discussions on MLDEs. He would also like to express his gratitude to Sigma Samhita for her immense help in the type setting of the tables required for this work. CNG thanks J. Santara for helpful discussions and collaboration on related projects. SM gratefully acknowledges the hospitality of the Institute for Advanced Study, Princeton, where this work was completed with generous support from NSF Grant PHY-2207584, and the hospitality of the Isaac Newton Institute for Mathematical Sciences, Cambridge during the programme  New Connections in Number Theory and Physics, where early work on this paper was undertaken with support from EPSRC grant no EP/R014604/1.  He is grateful to Brandon Rayhaun for several useful discussions (including suggesting the proof in Appendix D) and to Sahand Seifnashri for patient explanations of Modular Tensor Categories.

\section{Computations of some embedding indices}\label{c1}
\subsection*{Example 1: $\mathbf{F_4\rightarrow A_1^{(a)}\times G_2^{(b)}}$}
Here we consider $F_4\rightarrow A_1^{(a)}\times G_2^{(b)}$ (which is a maximal \textit{S} type embedding). We shall compute $a$ and $b$ which are embedding indices. For the above embedding consider the following branching,
\begin{align}
    &\mathbf{52} = \mathbf{(3,1)} \oplus \mathbf{(5,7)} \oplus \mathbf{(1,14)} \\
\text{now}, \qquad   &\mathscr{L}^{F_4}(\mathbf{52}) = 18 \nonumber\\
    &\mathscr{L}^{A_1}_{\text{net}} = 1 \times \mathscr{L}^{A_1}(\mathbf{3}) + 7 \times \mathscr{L}^{A_1}(\mathbf{5}) + 14 \times \mathscr{L}^{A_1}(\mathbf{1}) = 1 \times 4 + 7 \times 20 + 14 \times 0 = 144 \nonumber\\
    &\mathscr{L}^{G_2}_{\text{net}} = 3 \times \mathscr{L}^{G_2}(\mathbf{1}) + 5 \times \mathscr{L}^{G_2}(\mathbf{7}) + 1\times \mathscr{L}^{G_2}(\mathbf{14})  = 3 \times 0 + 5 \times 2 + 1 \times 8 = 18 \nonumber\\
\end{align}
where $\mathscr{L}^{\mathfrak{g}}(\text{irrep})$ denotes the Dynkin index of the corresponding irrep of the Lie algebra $\mathfrak{g}$ in question, $\mathscr{L}^{A_1}_{\text{net}}$ denotes the net Dynkin index computed from the above branching and $\mathscr{L}^{\mathfrak{g}}_{\text{net}}$ has a similar meaning for the corresponding Lie algebra $\mathfrak{g}$.
\begin{align}
\text{thus}, \qquad &a = \frac{\mathscr{L}^{A_1}_{\text{net}}}{\mathscr{L}^{F_4}(\mathbf{52})} = \frac{144}{18} = 8 \nonumber\\
\text{and}, \qquad &b = \frac{\mathscr{L}^{G_2}_{\text{net}}}{\mathscr{L}^{F_4}(\mathbf{26})} = \frac{18}{18} = 1
\end{align}
Hence, we have: $F_4\rightarrow A_1^{(8)}\times G_2^{(1)}$ and in the affine case we would have: $\hat{F}_{4,1}\rightarrow \hat{A}_{1,8} \times \hat{G}_{2,1}$.

\subsection*{Example 2: $\mathbf{E_8\to A_1^{(a)}\times G_2^{(b)} \times G_2^{(c)}}$}\label{step3G2_FM}
Consider the following embedding, $E_8\to A_1^{(a)}\times G_2^{(b)} \times G_2^{(c)}$ (non-maximal)[$E_8\to {\bf III_2}\times G_2^{(1)} \times G_2^{(1)}$]. \\

To understand the above non-maximal embedding let us first understand the maximal embeddings from which the above can be obtained,
\begin{align}
    &E_8\overset{m}{\longrightarrow} G_2^{(1)}\times F_4^{(1)} \\
\text{furthermore}, \qquad &F_4 \overset{m}{\longrightarrow} A_1^{(8)}\times G_2^{(1)} \\
\text{implying}, \qquad &E_8\overset{n-m}{\longrightarrow} G_2^{(a)}\times A_1^{(b)}\times G_2^{(c)}.
\end{align}
From the first embedding consider the following branching rule,
\begin{align}
    \mathbf{248} = \mathbf{(14,1)} \oplus \mathbf{(7,26)} \oplus \mathbf{(1,52)}
\end{align}
Now let us employ the second embedding to write the above branching rule as,
\begin{align}
    \mathbf{(14,1)} \oplus \mathbf{(7,26)} \oplus \mathbf{(1,52)} &= \mathbf{(14,1,1)} \oplus \mathbf{(7,((5,1)\oplus(3,7)))} \oplus \mathbf{(1,((3,1)\oplus(5,7)\oplus(1,14)))} \nonumber\\
    &= \mathbf{(14,1,1)} \oplus \mathbf{(7,5,1)} \oplus \mathbf{(7,3,7)} \oplus \mathbf{(1,3,1)} \oplus \mathbf{(1,5,7)} \oplus \mathbf{(1,1,14)},
\end{align}
where in the second equality we are just expanding the first equality and considering that the numbers inside a parenthesis have to be multiplied.

Now let us compute the embedding indices $a,b,c$,
\begin{align}
    &\mathscr{L}^{E_8}(\mathbf{248}) = 60 \nonumber\\
    &\mathscr{L}^{G_2}_{\text{net}} = 0 + 0 + 42 + 0 + 10 + 8 =60 \nonumber\\
    &\mathscr{L}^{A_1}_{\text{net}} = 1 \times 14 \times \mathscr{L}^{A_1}(\mathbf{1}) + 1 \times 7 \times \mathscr{L}^{A_1}(\mathbf{5}) + 7 \times 7 \times \mathscr{L}^{A_1}(\mathbf{3}) + 1 \times 1 \times \mathscr{L}^{A_1}(\mathbf{3}) \nonumber\\
    &+ 7 \times 1 \times \mathscr{L}^{A_1}(\mathbf{5}) + 14 \times 1 \times \mathscr{L}^{A_1}(\mathbf{1}) = 0 + 140 + 196 + 4 + 140 + 0 = 480 \nonumber\\
   &\mathscr{L}^{G_2}_{\text{net}} = 8 + 10 + 42 + 0 + 0 + 0 =60 \nonumber\\
\text{thus}, \qquad &a = \frac{\mathscr{L}^{G_2}_{\text{net}}}{\mathscr{L}^{E_8}(\mathbf{248})} = \frac{60}{60} = 1 \nonumber\\
\text{and}, \qquad &b = \frac{\mathscr{L}^{A_1}_{\text{net}}}{\mathscr{L}^{E_8}(\mathbf{248})} = \frac{480}{60} = 8 \nonumber\\
\text{and}, \qquad &c = \frac{\mathscr{L}^{G_2}_{\text{net}}}{\mathscr{L}^{E_8}(\mathbf{248})} = \frac{60}{60} = 1
\end{align}
Hence we have: $E_8\overset{n-m}{\longrightarrow} G_2^{(1)}\times A_1^{(8)} \times G_2^{(1)}$ which implies that in the affine case we would get: $\hat{E}_{8,1}\overset{n-m}{\longrightarrow} \hat{G}_{2,1}\times \hat{A}_{1,8} \times \hat{G}_{2,1}$.

\subsection*{Example 3: Non-maximal embedding}
Here let us try to give an example of $\text{MMS theory}\overset{c^\mathcal{H}=8}{\underset{n_1=1}{\longleftrightarrow}}\text{MMS theory}$, where a non-maximal embedding is involved. Consider, 
\begin{align}
&E_8\overset{m}{\longrightarrow} D_8^{(a)} \overset{m}{\longrightarrow} D_4^{(b)}\times D_4^{(c)} \\
\text{implying}, \qquad &E_8\overset{n-m}{\longrightarrow} D_4^{(r)}\times D_4^{(s)},
\end{align}
where $m$ stands for a maximal embedding and $n-m$ stands for a non-maximal embedding.\\

Let us compute the embedding indices, $a, b, c, r, s$, for the above three embeddings. 
\subsubsection*{$\mathbf{E_8 \to D_8^{(a)}}$}
Now let us consider the embedding $E_8 \to D_8^{(a)}$ (maximal and \textit{R} type). For the above embedding consider the following branching,
\begin{align}
    &\mathbf{248} = \mathbf{120} \oplus \mathbf{128} \\
\text{now}, \qquad   &\mathscr{L}^{E_8}(\mathbf{248}) = 60 \nonumber\\
    &\mathscr{L}^{D_8}(\mathbf{120}) = 28 \nonumber\\
    &\mathscr{L}^{D_8}(\mathbf{128}) = 32 \nonumber\\
\text{thus}, \qquad &a = \frac{\mathscr{L}^{D_8}(\mathbf{120}) + \mathscr{L}^{D_8}(\mathbf{128})}{\mathscr{L}^{E_8}(\mathbf{248})} = \frac{60}{60} = 1
\end{align}
Hence, we have: $E_8\rightarrow D_8^{(1)}$.

\subsubsection*{$\mathbf{E_8\overset{m}{\longrightarrow} D_8^{(a)}}$}
Let us consider $D_8 \to D_4^{(b)}\times D_4^{(c)}$ (maximal and {\it R} type),
\begin{align}
    &\mathbf{120} = \mathbf{(8_v,8_v)} \oplus \mathbf{(28,1)} \oplus \mathbf{(1,28)} \\
\text{now}, \qquad   &\mathscr{L}^{D_8}(\mathbf{120}) = 28 \nonumber\\
    &\mathscr{L}^{D_4}_{\text{net}} = 8 \times \mathscr{L}^{D_4}(\mathbf{8_v}) + 1 \times \mathscr{L}^{D_4}(\mathbf{28}) + 28 \times \mathscr{L}^{D_4}(\mathbf{1}) = 8 \times 2 + 1 \times 12 + 28 \times 0 = 28 \nonumber\\
   &\mathscr{L}^{D_4}_{\text{net}} = 8 \times \mathscr{L}^{D_4}(\mathbf{8_v}) + 28 \times \mathscr{L}^{D_4}(\mathbf{1}) + 1 \times \mathscr{L}^{D_4}(\mathbf{28}) = 8 \times 2 + 28 \times 0 + 1 \times 12 = 28 \nonumber\\
\text{thus}, \qquad &b = \frac{\mathscr{L}^{D_4}_{\text{net}}}{\mathscr{L}^{D_8}(\mathbf{120})} = \frac{28}{28} = 1 \nonumber\\
\text{and}, \qquad &c = \frac{\mathscr{L}^{D_4}_{\text{net}}}{\mathscr{L}^{D_8}(\mathbf{120})} = \frac{28}{28} = 1
\end{align}
Hence, we have: $D_8 \overset{m}{\longrightarrow} D_4^{(1)}\times D_4^{(1)}$.

\subsubsection*{$\mathbf{E_8\to D_4^{(r)}\times D_4^{(s)}}$}
Let us consider $E_8\to D_4^{(r)}\times D_4^{(s)}$ (non-maximal),
\begin{align}
    &\mathbf{248} = \mathbf{120} \oplus \mathbf{128} = \mathbf{(8_v,8_v)} \oplus \mathbf{(28,1)} \oplus \mathbf{(1,28)} \oplus \mathbf{(8_c,8_s)} \oplus \mathbf{(8_s,8_c)} \\
\text{now}, \qquad   &\mathscr{L}^{E_8}(\mathbf{248}) = 60 \nonumber\\
    &\mathscr{L}^{D_4}_{\text{net}} = 8 \times \mathscr{L}^{D_4}(\mathbf{8_v}) + 1 \times \mathscr{L}^{D_4}(\mathbf{28}) + 28 \times \mathscr{L}^{D_4}(\mathbf{1}) + 8 \times \mathscr{L}^{D_4}(\mathbf{8_c}) + 8 \times \mathscr{L}^{D_4}(\mathbf{8_s}) \nonumber\\
    &= 8 \times 2 + 1 \times 12 + 28 \times 0 + 8 \times 2 + 8 \times 2 = 60 \nonumber\\
   &\mathscr{L}^{D_4}_{\text{net}} = 8 \times \mathscr{L}^{D_4}(\mathbf{8_v}) + 28 \times \mathscr{L}^{D_4}(\mathbf{1}) + 1 \times \mathscr{L}^{D_4}(\mathbf{28}) + 8 \times \mathscr{L}^{D_4}(\mathbf{8_s}) + 8 \times \mathscr{L}^{D_4}(\mathbf{8_c}) \nonumber\\
   &= 8 \times 2 + 28 \times 0 + 1 \times 12 + 8 \times 2 + 8 \times 2 = 60 \nonumber\\
\text{thus}, \qquad &r = \frac{\mathscr{L}^{D_4}_{\text{net}}}{\mathscr{L}^{E_8}(\mathbf{248})} = \frac{60}{60} = 1 \nonumber\\
\text{and}, \qquad &s = \frac{\mathscr{L}^{D_4}_{\text{net}}}{\mathscr{L}^{E_8}(\mathbf{248})} = \frac{60}{60} = 1
\end{align}
Hence, we have: $E_8\overset{n-m}{\longrightarrow} D_4^{(1)}\times D_4^{(1)}$. This implies that, $D_4$ as a sub-algebra of $E_8$ has commutant $D_4$ inside $E_8$. This is the statement that was made in \cite{Gaberdiel:2016zke}.

\subsection*{Example 4: $\mathbf{D_{16}\rightarrow D_1^{(a)}\times A_{15}^{(b)}}$}
Here we consider $D_{16}\rightarrow D_1^{(a)}\times A_{15}^{(b)}$ (which is a maximal \textit{R} type embedding). For the above embedding consider the following branching,
\begin{align}
    &\mathbf{496} = \mathbf{(1,255)} \oplus \mathbf{(1,1)} + \mathbf{(1,120)} \oplus \mathbf{(1,120)} \\
\text{now}, \qquad   &\mathscr{L}^{D_{16}}(\mathbf{496}) = 60 \nonumber\\
    &\mathscr{L}^{D_1}_{\text{net}} = 255 \times \mathscr{L}^{D_1}(\mathbf{1}) + 1 \times \mathscr{L}^{D_1}(\mathbf{1}) + 2 \times 120 \times \mathscr{L}^{D_1}(\mathbf{1}) = 255 \times 0 + 1 \times 0 + 240 \times \frac{1}{4} = 60 \nonumber\\
    &\mathscr{L}^{A_{15}}_{\text{net}} = 1 \times \mathscr{L}^{A_{15}}(\mathbf{255}) + 1 \times \mathscr{L}^{A_{15}}(\mathbf{1}) + 2\times 1\times \mathscr{L}^{A_{15}}(\mathbf{120})  = 1 \times 32 + 1 \times 0 + 2\times 1 \times 14 = 60 \nonumber\\
\end{align}
Hence, we get, for the embedding indices, $a, b$,
\begin{align}
&a = \frac{\mathscr{L}^{D_1}_{\text{net}}}{\mathscr{L}^{D_{16}}(\mathbf{496})} = \frac{60}{60} = 1 \nonumber\\
\text{and}, \qquad &b = \frac{\mathscr{L}^{A_{15}}_{\text{net}}}{\mathscr{L}^{D_{16}}(\mathbf{496})} = \frac{60}{60} = 1
\end{align}
Hence, we have: $D_{16}\rightarrow D_1^{(1)}\times A_{15}^{(1)}$ and in the affine case we would have: $\hat{D}_{16,1}\rightarrow \hat{D}_{1,1} \times \hat{A}_{15,1}$.

\subsection*{Example 5: $\mathbf{D_{16}\rightarrow A_1^{(a)}\times C_8^{(b)}}$}
Here we consider $D_{16}\rightarrow A_1^{(a)}\times C_8^{(b)}$ (which is a maximal \textit{S} type embedding). For the above embedding consider the following branching,
\begin{align}
    &\mathbf{496} = \mathbf{(1,136)} \oplus \mathbf{(3,1)} \oplus \mathbf{(3,119)} \\
\text{now}, \qquad   &\mathscr{L}^{D_{16}}(\mathbf{496}) = 60 \nonumber\\
    &\mathscr{L}^{A_1}_{\text{net}} = 136 \times \mathscr{L}^{A_1}(\mathbf{1}) + 1 \times \mathscr{L}^{A_1}(\mathbf{3}) + 119 \times \mathscr{L}^{A_1}(\mathbf{3}) = 136 \times 0 + 1 \times 4 + 119 \times 4 = 480 \nonumber\\
    &\mathscr{L}^{C_8}_{\text{net}} = 1 \times \mathscr{L}^{C_8}(\mathbf{136}) + 3 \times \mathscr{L}^{C_8}(\mathbf{1}) + 3\times \mathscr{L}^{C_8}(\mathbf{119})  = 1 \times 18 + 3 \times 0 + 3 \times 14 = 60 \nonumber\\
\end{align}
Hence, we get, for the embedding indices, $a, b$,
\begin{align}
&a = \frac{\mathscr{L}^{A_1}_{\text{net}}}{\mathscr{L}^{D_{16}}(\mathbf{496})} = \frac{480}{60} = 8 \nonumber\\
\text{and}, \qquad &b = \frac{\mathscr{L}^{C_8}_{\text{net}}}{\mathscr{L}^{D_{16}}(\mathbf{496})} = \frac{60}{60} = 1
\end{align}
Hence, we have: $D_{16}\rightarrow A_1^{(8)}\times C_8^{(1)}$ and in the affine case we would have: $\hat{D}_{16,1}\rightarrow \hat{A}_{1,8} \times \hat{C}_{8,1}$.

\section{Infinite family of $\mathbf{c=8}$ and $\mathbf{c=16}$ for category ${\bf III}$ solutions} \label{Inhomo}
In this appendix, we briefly summarise the results of Sec. 2.3 of \cite{Das:2021uvd} which explains, from an MLDE perspective, why there are an infinite family of $c=8$ and $c=16$ for category ${\bf III}$ solutions (this fact was previously noted in \cite{Mason:c8c16}). It is shown in \cite{Das:2021uvd} that the identity character $\chi_0$ can be written in terms of the other two characters, $\chi_1,\chi_2$ as,
\bea \label{228}
\chi_0(q) = q^{\frac{1}{2} - \alpha_1 - \alpha_2}\sum\limits_{n=0}^{\infty} \left[a_{0,n} + A_1\, q^{-\frac{1}{2} + 2 \alpha_1 + \alpha_2}\,  a_{1,n}  + A_2\, q^{-\frac{1}{2} +  \alpha_1 + 2 \alpha_2}\,  a_{3,n}   \right] q^n.
\eea
where $a_{i,n}$ are the Fourier coefficients in the q-series of the character $\chi_i(q)$ and $i=0,1,2$.

Now say the values of $\alpha_1$ and $\alpha_2$, for which admissible solutions, exist are such that $-\frac{1}{2} + 2 \alpha_1 + \alpha_2$ and $-\frac{1}{2} +  \alpha_1 + 2 \alpha_2$  are not non-negative integers, then to get admissible solution for $\chi_0$, we have to set $A_1$ and $A_2$ to be zero. This is what happens in most examples. However, one can imagine the following situation. 

(i) If  $-\frac{1}{2} + 2 \alpha_1 + \alpha_2$ is a non-negative integer, then $A_1$ isn't required to vanish. $A_1$ can take any positive integral value and we would have an admissible solution for $\chi_0$.
\begin{align} \label{229}
&-\frac{1}{2} + 2 \alpha_1 + \alpha_2 \in \mathbb{Z}_{\geq 0}, \qquad  A_1 \in \mathbb{Z}_{\geq 0}\nonumber \\ 
&\chi_1(q) = q^{\frac{1}{2} - \alpha_2 - \alpha_3}\sum\limits_{n=0}^{\infty} \left[a_{0,n} + A_1\, q^{-\frac{1}{2} + 2 \alpha_1 + \alpha_2}\,  a_{1,n}    \right] q^n.
\end{align}
We thus have an infinite number of admissible  character solutions, parametrised by $A_1$, in  \eqref{229}. All members of this infinite family have the same indices and hence the same $c$, $h_1$, $h_2$ and also they have the same Wronskian. However, they are different solutions as in they differ in  the identity character.

(ii) If  $-\frac{1}{2} + \alpha_1 + 2 \alpha_2$ is a non-negative integer, then $A_2$ isn't required to vanish. $A_2$ can take any positive integral value and we would have an admissible solution for $\chi_0$.
\begin{align} \label{230}
&-\frac{1}{2} +  \alpha_1 + 2 \alpha_2 \in \mathbb{Z}_{\geq 0}, \qquad  A_2 \in \mathbb{Z}_{\geq 0}\nonumber \\ 
&\chi_1(q) = q^{\frac{1}{2} - \alpha_1 - \alpha_2}\sum\limits_{n=0}^{\infty} \left[a_{0,n} + A_2\, q^{-\frac{1}{2} +  \alpha_1 + 2 \alpha_2}\,  a_{2,n}    \right] q^n.
\end{align}
We thus have an infinite number of admissible  character solutions, parametrised by $A_2$, in  \eqref{230}. All members of this infinite family have the same indices and hence the same $c$, $h_1$, $h_2$ and also they have the same Wronskian. However, they are different solutions as in they differ in  the identity character.

So, in the study of admissible solutions to $\mathbf{[3,0]}$ MLDEs, one encounters the above two infinite families of CFTs where each family has the same $c$, $h_1$, $h_2$ values, one following \eqref{229} and another following \eqref{230}. 

Note that, $D_{4,1}^{\otimes 2}$ is a part of the infinite family of $c=8$ solutions with $m_1=56$, $D_{8,1}$ is a part of the infinite family of $c=8$ solutions with $m_1=120$ and $D_{4,1}^{\otimes 4}$ is a part of the infinite family of $c=16$ solution with $m_1=112$. The key point to note here is that, in the notation of \cite{Das:2021uvd}, $D_{4,1}^{\otimes 2}$, $D_{8,1}$ and $D_{4,1}^{\otimes 4}$ are the only three solutions which belong to both category ${\bf I}$ and ${\bf III}$.

\section{Upper bound on $\cN$ for meromorphic theories}

\label{upperb}

Here we prove the following bound: for any meromorphic CFT with $c=8N$, the dimension $\cN$ of its Kac-Moody algebra is bounded above by $8N(16N-1)$. This bound is saturated by the meromorphic theory $\cE_1[D_{8N}]$ \footnote{We are grateful to Brandon Rayhaun for suggesting this line of argument.}. 

To show this, let us first consider meromorphic theories with a ``complete'' Kac-Moody algebra with simple factors, i.e. theories whose entire central charge comes from non-Abelian Kac-Moody factors. This holds for 69 of 71 theories at $c=24$, and additional examples come from lattice theories with ``complete root systems'' at higher values of $c$ such as Kervaire lattices in 32d \cite{kervaire1994unimodular}. In this situation we have:
\be
c=\sum_a c_a,\qquad 
c_a =\frac{k_a\,{\rm dim}\,\cG_a}{k_a+g_a}
\ee
where $k_a$ is the level, $g_a$ is the dual Coxeter number and ${\rm dim}\,\cG_a$ is the dimension of the $a$'th simple factor. The sum runs over all the simple factors.

Next, we note that simply-laced algebras $\cG_a$ satisfy the inequality, ${\rm rank}\,\cG_a\le c_a\le {\rm dim}\,\cG_a$ where the first inequality is saturated at $k_a=1$ and the second as $k_a\to\infty$. In fact, as one can easily check, the same inequality is satisfied by non-simply-laced algebras, except that the lower bound becomes strict and is never saturated. 

Meanwhile, the total dimension of the Kac-Moody algebra is:
\be
\cN=\sum_a {\rm dim}\,\cG_a
\ee
Our problem then is to maximise $\cN$ keeping $c$ fixed.  

Now we further restrict to complete Kac-Moody algebras with just one simple factor. Using standard formulae for the dimensions $\cN_r$ and dual Coxeter numbers of the classical compact Lie algebras $(A_r,B_r,C_r,D_r)$ we find:
\be
\begin{split}
c(A_{r,k}) &= \frac{kr(r+2)}{k+r+1} \\[2mm]
c(B_{r,k}) &= \frac{kr(2r+1)}{k+2r-1} \\[2mm]
c(C_{r,k}) &= \frac{kr(2r+1)}{k+r+1} \\[2mm]
c(D_{r,k}) &= \frac{kr(2r-1)}{k+2r-2}
\end{split}
\ee
It follows that, at fixed central charge, $r$ decreases as $k$ increases. Thus to maximise the rank in each family (which maximises the dimension, which is monotonic in the rank) we must take $k=1$, which gives the simpler formulae:
\be
\begin{split}
c(A_{r,1}) &= r\\
c(B_{r,1}) &= r+\shalf \\
c(C_{r,1}) &= \frac{r(2r+1)}{r+2} \\
c(D_{r,1}) &= r
\end{split}
\ee
Notice that $c(B_{r,1}),c(C_{r,1})$ are non-integral for all $r\ge 2$. From the above, the dimension of the algebra at fixed $c$ is:
\be
\begin{split}
A_{r,1}:~\cN&=c(c+2)\\
B_{r,1}:~ \cN &= c(2c-1)\\
C_{r,1}:~ \cN &= \frac14\Big(7c + c^2 + c \sqrt{1 + 14 c + c^2}\Big) \\
D_{r,1}:~ \cN &= c(2c-1)
\end{split}
\ee

It is easy to verify that for any fixed $c\ge 8$, the common value of $\cN$ for $B_r,D_r$ is the largest in the above set. However since $B_r$ has non-integral central charge it cannot be a complete simple factor. Therefore $D_{r,1}$ has the largest possible dimension among simple algebras at a fixed integral central charge. Moreover there is indeed a meromorphic theory with Kac-Moody algebra $D_{8N,1}$ for every $r$, corresponding to the modular invariant extension $\cE_1[D_{8N,1}]$ (for $N=1$ this is $E_{8,1}$, while for all $N\ge 2$ the extension does not enhance the Kac-Moody algebra).  

Now we can go on to the general case: direct sums of  Kac-Moody algebras, including exceptional as well as Abelian algebras, at arbitrary levels. We also allow meromorphic theories where the Kac-Moody algebra is not complete (for example the algebra could contain minimal model or higher-spin modules). We argue that all these generalisations lower the dimension of the Kac-Moody algebra, proving the bound. First, when we take direct sums, the sum  of dimensions of the factors is always less than the dimension of a simple algebra of the same $c$. Since there are finitely many exceptional algebras one can also verify explicitly that none of them ``wins'' over $D_{8N,1}$. Also for Abelian algebras the dimensions are always smaller than those of non-Abelian algebras of comparable central charge. Second, raising the level of any factor raises its central charge without changing its dimension, and therefore lowers its dimension for fixed central charge. Finally  if the Kac-Moody algebra is not complete, its dimension will be smaller than that of a complete algebra with the same $c$. This then proves the result.

\end{appendix}

\bibliographystyle{JHEP}

\bibliography{3char}

\end{document}